\documentclass{article}
\usepackage[a4paper]{geometry}
\usepackage{authblk} 
\usepackage{color}
\usepackage{refcount}
\usepackage{graphicx}
\usepackage{slashed}
\usepackage{braket,amsmath,amssymb}
\usepackage[export]{adjustbox}
\usepackage{subcaption}
\usepackage{graphics,psfrag,empheq,float}
\usepackage[colorlinks=true,linkcolor=black]{hyperref}
\usepackage{cite}
\usepackage{wasysym}
\usepackage[normalem]{ulem} 
\def\Xint#1{\mathchoice
   {\XXint\displaystyle\textstyle{#1}}%
   {\XXint\textstyle\scriptstyle{#1}}%
   {\XXint\scriptstyle\scriptscriptstyle{#1}}%
   {\XXint\scriptscriptstyle\scriptscriptstyle{#1}}%
   \!\int}
\def\XXint#1#2#3{{\setbox0=\hbox{$#1{#2#3}{\int}$}
     \vcenter{\hbox{$#2#3$}}\kern-.5\wd0}}

\def\dashint{\Xint-}

\newcommand{\cV}{{\cal V}}

\newcommand{\hvp}{{\hat{\boldsymbol{p}}}}
\newcommand{\vp}{{\boldsymbol{p}}}
\newcommand{\vq}{{\boldsymbol{q}}}

\newcommand{\vu}{{\boldsymbol{u}}}
\newcommand{\vv}{{\boldsymbol{v}}}

\newcommand{\vk}{{\boldsymbol{k}}}
\newcommand{\vz}{\boldsymbol{z}}

\newcommand{\bg}{\begin{align}}
\newcommand{\eeg}{\end{align}}
\newcommand{\be}{\begin{equation}}
\newcommand{\ee}{\end{equation}}
\newcommand{\ba}{\begin{eqnarray}}
\newcommand{\ea}{\end{eqnarray}}

\newcommand{\nn}{\nonumber}

\newcommand{\ve}{\varepsilon}

\newcommand{\la}{\langle}
\newcommand{\ra}{\rangle}

\newcommand{\ep}{\epsilon}
\newcommand{\fa}{\alpha}

\newcommand{\vc}{{\rm vp}}

\newcommand{\jo}[1]{\textcolor{black}{{{#1}}}}

\begin{document}





\title{Unitarization of electron scattering with an external potential at NLO in QED}

\author[a]{J. A.  Oller\thanks{oller@um.es}}
\author[b]{Marcela Pel\'aez\thanks{mpelaez@fing.edu.uy}}
\affil[a]{\it Departamento de F\'{\i}sica, Universidad de Murcia, E-30071 Murcia,  Spain}
\affil[b]{{\it 
  Instituto de F\'{\i}sica, Facultad de Ingenier\'{\i}a, Universidad de la Rep\'ublica,} \authorcr {\it J. H. y Reissig 565, 11000 Montevideo, Uruguay } }

\maketitle

\begin{abstract}
We have calculated the one-loop scattering amplitude of an electron by an external Coulomb potential in QED  free of infrared divergences. This feature is achieved by applying the Faddeev-Kulish formalism, which implies a redefinition of both the asymptotic electronic states and of the $S$ matrix.  Additionally, we have also derived the infrared-finite one-loop partial-wave amplitudes for this process by applying a recent method in the literature. Next, these partial-wave amplitudes are  unitarized based on analyticity and unitarity by employing three different methods of unitarization: the algebraic $N/D$ method, the Inverse Amplitude Method and the first-iterated $N/D$ method. Then, we have studied several partial waves both for physical momentum and for complex ones  to look for bound-state poles. The binding momentum for the fundamental bound state in $S$ wave is discussed with special detail.  This is a wide-ranging method for calculating nonperturbative partial-wave amplitudes for infinite-range interactions that could be applied to many other systems. 
\end{abstract}
\newpage
\tableofcontents

\bigskip
\section{Introduction}
\def\theequation{\arabic{section}.\arabic{equation}}
\setcounter{equation}{0}

The scattering of particles by a Coulomb field is a fundamental topic because it represents the interaction between charged particles, such as electrons and nuclei. In nonrelativistic scenarios, the scattering can be computed by solving the Schrödinger equation, yielding closed-form solutions for the scattering amplitude or differential cross section. As it is well known, the  cross section for nonrelativistic electron scattering off a central Coulomb potential is described by the classical Rutherford formula. The wave function for the scattering of a Dirac electron by a Coulomb potential was obtained in 
 Ref.~\cite{Mott1929TheSO} as a series of spherical harmonics. Based on this calculation, the lowest-order correction to the Rutherford cross section for a Dirac electron (positron) scattering off a heavy, point-like target with charge Z was first derived, though the  correct result had to await until Dalitz's calculation in Ref.~\cite{Dalitz:1951ah} by evaluating the second Born approximation. 

To obtain more precise and accurate predictions, it is necessary to consider higher-order corrections in QED, typically calculated as a power series expansion in $\alpha=e^2/4\pi$, the fine structure constant, with $e$ the electron charge. Higher-order perturbation theory for this problem has been explored in several articles, such as \cite{Mott1929TheSO,McKinley:1948zz,Dalitz:1951ah,Yennie:1954zz,Baier:1972jm}. Additionally, two-loop corrections of the $g$-factor in a nonperturbative Coulomb field are discussed in \cite{Sikora:2018zda}. Similar two-loop calculations for processes like M\"oller scattering \cite{Delto:2023kqv,Banerjee:2021qvi}, Bhabha scattering \cite{Bonciani:2005im,Fleischer:2006ht}, and electron-muon scattering \cite{Engel:2022kde} can be found. Of particular note, Refs.~\cite{Khalil:2017cqm,Khalil:2017ons} compute the high-energy elastic scattering cross section for an electron off a classical point source to next-to-leading order (NLO) accuracy. They employ the MS renormalization scheme to manage ultraviolet divergences, while the Kinoshita-Lee-Nauenberg theorem addresses infrared singularities at the level of the cross section. 

 In this work, we are interested in the scattering of an electron by an external Coulomb potential, as an example of an infinite-range potential. Our central objects are the partial-wave amplitudes (PWAs), first evaluated perturbatively in QED up to NLO, and then unitarized by implementing non-perturbative unitarization techniques \cite{Oller:2020guq,Blas:2020dyg,Blas:2020och,Oller:2022tmo}. These results are also applied to computing the bound states of an electron in a Coulomb potential. By considering corrections to one loop in QED, in addition to the once-time iterated potential, already provided by the second Born approximation of the Dirac electron \cite{Dalitz:1951ah}, we take into account the effects of the particle self-energy and vertex-modified corrections, like the anomalous magnetic moment. The latter ones  modify the energy of the bound state compared to the  prediction based on the Dirac equation. 

 Despite the lack of convergence of a partial-wave expansion for the Coulomb potential, they are often used when considering hadron and nuclear physics. For instance, Coulomb effects \cite{landau.170517.1} 
are  important since they typically contribute to the low-energy cross section near the threshold of a reaction involving charged particles. Furthermore, the study of the modifications of Coulomb atoms due to strong interactions is also an important technique for hadron physics. And vice versa, an accurateness calculation of the hadronic spectrum requires to take into account electromagnetic effects. 

This article builds upon the work of Ref.~\cite{Oller:2022tmo}, which investigates nonrelativistic Coulomb scattering as a benchmark for the unitarization method of Refs.~\cite{Blas:2020dyg, Blas:2020och} for the case of infinite-range potentials, due to the null mass of photons and gravitons. It was also shown as a clear counterexample that the unitarization method used in Ref.~\cite{Delgado:2022qnh} is unsuitable for unitarizing Coulomb scattering, because the impact of infrared divergences on scattering amplitudes is not properly taken into account.  Ref.~\cite{Oller:2022tmo} shows that the exact Coulomb $S$-wave cannot be approximated using the approach described in Ref.~\cite{Delgado:2022qnh}, because it is typically mismatched by an order of magnitude at least. Furthermore,  it lacks of bound states with non-zero binding momentum.

Our present study extends the work of \cite{Oller:2022tmo} by unitarizing the scattering amplitude of an electron by an external Coulomb potential that we calculate up to NLO in perturbative QED.  The perturbative QED scattering amplitudes are affected by two types of infinities, infrared (IR) and ultraviolet (UV) divergences. The latter can be treated according to the general theory of renormalization, that we apply here to end with UV-finite scattering amplitudes. For the IR divergences one has to account for two main effects in order to finally have  scattering amplitudes free of IR divergences. At this point it is important to differentiate between the standard formalism to treat IR divergences at the level of transition rates (cross sections) \cite{Bloch:1937pw, Weinberg:1965nx}, and the less used  Faddeev-Kulish  (FK) formalism \cite{Kulish:1970ut}, which allows to obtain  finite $S$-matrix elements. In recent years, the developing of a well-defined $S$-matrix theory for QED, gravity as well as for  QED with massless charged particles and in Yang-Mills theory has experienced  striking advances in 
different directions \cite{Ware:2013zja,Choi:2017bna,Choi:2017ylo,Prabhu:2022zcr,Prabhu:2024zwl,Prabhu:2024lmg,Campiglia:2016hvg,Campiglia:2014yka,Ashtekar:2018lor,Hirai:2020kzx,Himwich:2020rro,Hannesdottir:2019umk}.

On the one hand, there are IR divergences at the one-loop level that in the formalism based on the finiteness of transition rates requires to account for the radiation of low-energy photons. In the FK formalism the IR divergences of this type are taken into account by a redefinition of the asymptotic electronic  states 
based on introducing a proper 
cloud of real   soft photons. Such a definition of  asymptotic electron states (without hard photons) was firstly proposed in the literature by Chung in Ref.~\cite{Chung:1965zza}. This step to cure IR divergences was not needed in Refs.~\cite{Blas:2020dyg,Blas:2020och,Oller:2022tmo} because only tree-level amplitudes were unitarized. Therefore, they are accounted here for the first time in order to unitarize infinite range interactions.

On the other hand,  there is still the issue of the well-known divergent Coulomb phase that already appears in the nonrelativistic case \cite{Dalitz:1951ah,Weinberg:1965nx,Kulish:1970ut}. To account for this type of divergence one needs to redefine the $S$-matrix operator itself  by a phase factor depending on the initial and final states \cite{Kulish:1970ut}. The partial-wave projection of this new $S$ matrix is considered, and in terms of it one can define IR-finite PWAs, as firstly derived in Refs.~\cite{Blas:2020dyg,Blas:2020och,Oller:2022tmo}. \jo{ We also emphasize the connection between the  FK formalism \cite{Kulish:1970ut} with Weinberg's one \cite{Weinberg:1965nx} in order to extract the unitary and IR-finite $S$ matrix, which has the advantage of further clarifying our approach.}

By applying this scheme of work we start by calculating perturbatively the IR-finite PWAs up to NLO in QED. Next, we unitarize them by employing three different unitarization methods: the algebraic $N/D$ method, $g$-method for brevity \cite{Oller:1998zr,Oller:1999me}, the Inverse Amplitude Method (IAM) \cite{Lehmann:1972kv,Dobado:1989qm,Oller:1997ng}, and the first-iterated $N/D$ method \cite{Oller:2019opk,Oller:2019rej,Oller:2020guq}. 
Several  PWAs are calculate with these unitarization methods, and we compare between them and with the perturbative PWAs from NLO QED. The resulting PWAs from the different methods are typically in agreement in a broad range of momentum, $m\alpha\lesssim p \lesssim m/\alpha$. In the lower end nonperturbative effects become very important, since one is close to the hydrogen atom, whose reproduction can only be accomplished after unitarization, so that the perturbative QED PWAs depart from the unitarized ones. For the upper limit, it occurs that terms suppressed by higher powers of $\alpha$ 
become more important than those with a lower power due to their momentum dependence. This implies that the standard IAM method can fail at such high momenta (as it does, in fact, for some partial waves), similarly as it is deficient in meson-meson scattering around an Adler zero, and variants of the method should be used \cite{GomezNicola:2007qj,Salas-Bernardez:2020hua,Escudero-Pedrosa:2020rwb}. 

We also pay special attention to the bound-state equations that results from the three unitarization methods, and their application to calculate the binding momenta. We argue  that because at NLO the perturbative PWAs have a left-hand cut (LHC)  for $p^2<0$ in the complex $p^2$ plane, it is only the first-iterated $N/D$ method that is robust enough to deal with this situation. The special difficulty stems from the fact that the pole associated to a bound state in the  PWA lies on top of the LHC.  Importantly, the application of  our unitarization scheme offers a novel procedure to study bound states in QED \cite{Pineda:1998kn,Haidar:2019kcp}, that is improvable order by order as the perturbative  PWAs are calculated at higher orders in QED.   Further application to higher orders and to different systems would be valuable in the future.

The article is organized as follows. After this introduction we elaborate in Sec.~\ref{sec.240322.1} on the perturbative calculation of QED scattering amplitudes up to NLO, with more technical details and a compendium of formulas given in  appendix~\ref{sec.240322.2}. In order to calculate the IR-finite PWAs of Sec.~\ref{sec.240322.4} we first discuss the removal of these divergences  in Sec.~\ref{sec.240322.3}, with its associated appendix~\ref{sec.240413.2}. The unitarization methods are introduced in Sec.~\ref{sec.240128.1}, with Sec.~\ref{sec.240128.5}  dedicated to their application for studying scattering amplitudes. The binding equation is the object of study of Sec.~\ref{sec.231221.2}, and its detailed application to the ground state is discussed in Sec.~\ref{sec.240322.5}. Finally,  concluding remarks and a brief outlook are given in Sec.~\ref{sec.240322.6}. The appendix~\ref{App:PWA} is added to give  more technical details for  deriving the final formula used to calculate PWAs.

\section{Standard Diagrams}
\label{sec.240322.1}
\def\theequation{\arabic{section}.\arabic{equation}}
\setcounter{equation}{0}

The calculation up to NLO in QED of the scattering amplitude of an electron by an external potential produced by an infinitely heavy particle of charge $e'$ can be obtained from QED Lagrangian:
\begin{align}
\mathcal{L}_{\text{QED}} = \bar{\psi}(i\gamma^\mu \partial_\mu - m_0)\psi - \frac{1}{4}F^{\mu\nu}F_{\mu\nu} - e_0\bar{\psi}\gamma^\mu A_\mu\psi
\end{align}
where $m_0$ and $e_0$ represent the bare mass and electron charge, respectively. 

As one loop diagrams are plagued with UV divergences it is necessary to introduce
the renormalization factors defined as:
\begin{align}
\psi&=\sqrt{Z_2}\psi^R~,\\
A_\mu&=\sqrt{Z_3}A_\mu^R~,\nonumber\\
e_0&=Z_e e~,\nonumber\\
m_0&=Z_m m~.\nn
\end{align}
The Lagrangian expressed in term of counterterms takes the form (now the constants and fields are the renormalized ones. The latter are expressed again as $\psi$ and $A_\mu$ for ease of writing), 
\begin{align}
\mathcal{L}_{\text{QED}} &= \bar{\psi}(i\gamma^\mu \partial_\mu - m)\psi - \frac{1}{4}F^{\mu\nu}F_{\mu\nu} - e\bar{\psi}\gamma^\mu A_\mu\psi \\
&+\delta Z_2\bar{\psi}(i\gamma^\mu \partial_\mu)\psi-(Z_2  Z_m-1)\bar{\psi} m\psi
- \delta Z_3\frac{1}{4}F^{\mu\nu}F_{\mu\nu} - \delta Z_1 e\bar{\psi}\gamma^\mu A_\mu\psi\nonumber~,
\end{align}
where $\delta Z_2=Z_2-1$, $\delta Z_3=Z_3-1$ and $\delta Z_1=Z_1-1$ being $Z_1=Z_2Z_e\sqrt{Z_3}$. 

\begin{figure}[ht!]
   \centering
 \setkeys{Gin}{width=3cm}

\subfloat[]{\includegraphics[width=0.2\textwidth, valign=c]{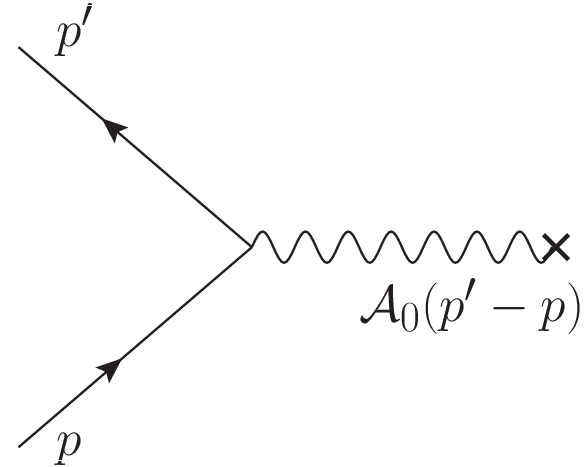}}
\qquad 
\subfloat[]{\includegraphics[width=0.2\textwidth, valign=c]{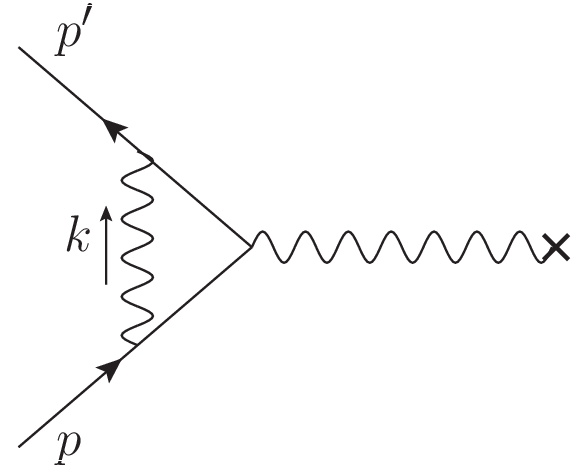}}
\qquad 
\subfloat[]{\includegraphics[width=0.2\textwidth, valign=c]{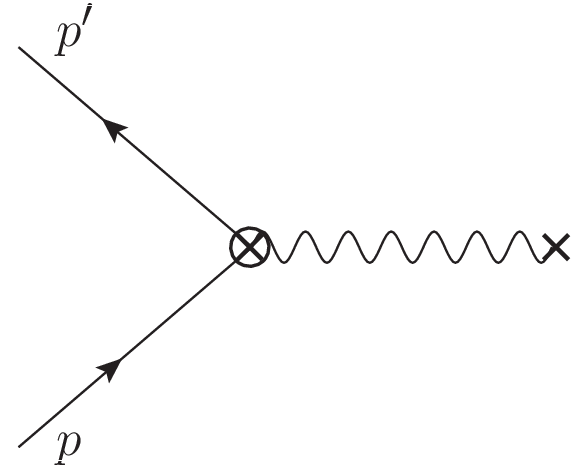}}

\subfloat[]{\includegraphics[width=0.2\textwidth, valign=c]{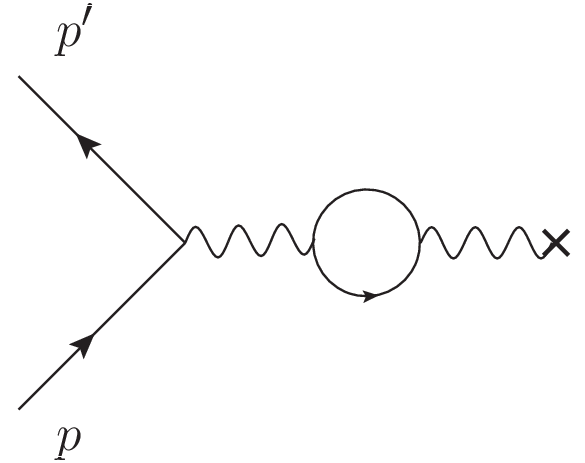}}
\qquad 
\subfloat[]{\includegraphics[width=0.2\textwidth, valign=c]{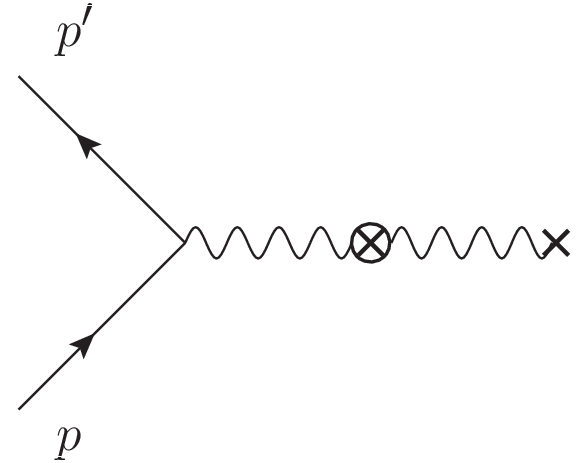}}
\qquad 
\subfloat[]{\includegraphics[width=0.18\textwidth, valign=c]{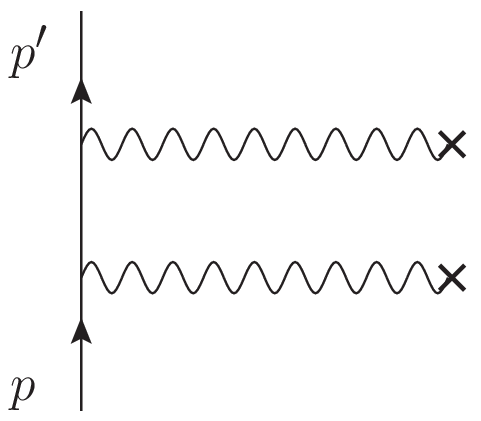}}

\caption{{\small All the one-loop diagrams contributing to the electron scattering by a Coulomb potential.}\label{Fig:todos}}
    \end{figure}

With the QED Lagrangian we can compute Feynman diagrams up to one loop, or equivalently up to ${\cal O}(\alpha^2)$, by  following the standard techniques \cite{Peskin:1995ev,Passarino:1978jh}. The first part of the Lagrangian gives the standard Feynman rules  in terms of the renormalized mass and charge, while the other terms give additional rules for the counterterms.
%
%
In our chosen renormalization scheme, $Z_2$ also absorbs the IR divergences of the electron self-energy and is renormalized to ensure that the electron propagator has a residue equal to one. In this scenario, loop corrections to external legs are nullified with the counterterm of the external leg, allowing us to disregard diagrams modifying external legs.\footnote{This procedure is also followed for the two-loop calculation of Babbha scattering in Ref.~\cite{Fleischer:2006ht}.} The counterterm $Z_3$ is similarly fixed by requiring that the residue of the photon propagator is $-g_{\alpha\beta}$, where $g_{\alpha\beta}={\text{diag}}(1,-1,-1,-1)$ is the Minkowski metric. The set of Feynman diagrams at LO, diagram (a), and NLO in QED, diagrams (b)--(f), are shown in Fig.~\ref{Fig:todos}. 
The vertex-modification diagram [diagram (b)] encompasses both IR and UV divergences, which are absorbed in $Z_1$. As it is well known, the Ward identity dictates that $Z_1=Z_2$, and therefore $\frac{1}{2}\delta Z_3+\delta Z_e=0$ is fixed by the definition of the charge $\bar u(\boldsymbol{p})\Lambda^\mu(p,p)u(\boldsymbol{p})=\bar u(\boldsymbol{p})\gamma^\mu u(\boldsymbol{p})$. A detailed account of our procedure for the calculation of the renormalized scattering amplitude for the scattering of an electron by an external potential is given in the appendix~\ref{App:QED}.


\subsection{Leading-order amplitude}
The leading order (LO) amplitude of the electron scattering by an external potential due to a charge $e'$, represented by diagram (a) of Fig.\ref{Fig:todos}, is
\begin{align}
\label{240316.3}
A_0&=- e e' \,\bar{u}_{\sigma'}(\vp') \slashed{A}(\vq^2) u_\sigma(\vp)~,
\end{align}
where $e$ is the charge of the electron ($e<0$), $u_\sigma(\vp)$ is the Dirac spinor with linear momentum $\vp$ and third component of spin $\sigma$ along the $z$ axis (the Dirac spinors are always associated to on-shell electrons). 
For the case of an infinitely heavy charged point particle one has the Coulomb potential $\displaystyle{A_\mu(\vq^2)=\frac{\delta_\mu^0}{\vq^2+\lambda^2}}$,  where a vanishing small photon mass $\lambda$ has been included in order to handle IR divergences.

\subsection{Vacuum polarization contribution}

We calculate the correction arising from the diagrams (d) and (e) of Fig.~\ref{Fig:todos}, properly renormalized. The final result is
%
%

\begin{align}
\label{Eq:A1}
A_{1}&=-(e e') \bar{u}_{\sigma'}\left(\vp^{\prime}\right) 
  \slashed{A}(\vq^2) u_\sigma(\vp)\left\{1+\frac{\alpha}{9 \pi t}\left[3\left(t+2 m^{2}\right) \sigma\left(t\right) \ln \frac{\sigma\left(t\right)+1}{\sigma\left(t\right)-1}-5 t-12 m^{2}\right]\right\}~,
\end{align}
which has no IR divergences. In the expression, the function $\sigma\left(t\right)=\sqrt{1-4m^2/t}$ and $t=q^2$,  where $q=p'-p$.

\subsection{Vertex correction}

%

The diagrams (b) and (c) of Fig.~\ref{Fig:todos} account for this effect.  The former diagram is usually written using two form factors:
\begin{align}
\bar{u}_{\sigma'}\left(\boldsymbol{p}^{\prime}\right) (-ie\Gamma^\mu) u(\boldsymbol{p})=-ie\bar{u}_{\sigma'}\left(\boldsymbol{p}^{\prime}\right) \left( F_1(q)\gamma^\mu+\frac{i\sigma^{\mu\nu}}{2m}q_\nu F_2(q)\right) u_\sigma(\boldsymbol{p})
\end{align}
with $\sigma^{\mu\nu}=\frac{i}{2}[\gamma^\mu,\gamma^\nu]$ and $q$ the transfer momentum $q=p'-p$. Standard techniques for computing one-loop Feynman diagrams are used in order to obtain the final expression \cite{Passarino:1978jh}. The renormalized vertex is expressed in terms of the operator 
$-i e \Lambda^\mu(p',p)\equiv -i e \left(\gamma^\mu+\Gamma^\mu(p',p)+\delta Z_1 \gamma^\mu\right)$. 
%
%
%
%
In order to renormalize, we impose the following renormalization condition:
\begin{align}
\bar{u}_{\sigma'}\left(\boldsymbol{p}^{\prime}\right)\Lambda^\mu(p',p)u_\sigma(\boldsymbol{p})\Big|_{\slashed{p'}=\slashed{p}=m}=\bar{u}_{\sigma'}\left(\boldsymbol{p}^{\prime}\right)\gamma^\mu u_\sigma(\boldsymbol{p})\Big|_{\slashed{p'}=\slashed{p}=m}
\end{align}
This gives the definition of the $\delta Z_1$ renormalization factor, being $\delta Z_1=-ie F_1(0)$. 
%
Finally, the vertex contribution can be summarized as :

\begin{align}
\label{Eq:A2}
A_{2} & =-(e e') A^{\mu}\left(\boldsymbol{q}^{2}\right) \bar{u}_{\sigma'}\left(\boldsymbol{p}^{\prime}\right)\left[\frac{i \alpha m\left(p^{\prime}-p\right)^{\nu} \sigma_{\mu \nu}}{2 \pi\left(t-4 m^{2}\right)} \sigma(t) \ln \frac{\sigma(t)-1}{\sigma(t)+1}\right. \nonumber\\
& +\frac{\alpha \gamma_{\mu}}{4 \pi t \sigma(t)}\left\{4\left(t \sigma(t)+\left(t-2 m^{2}\right) \ln \frac{\sigma(t)-1}{\sigma(t)+1}\right) \ln \frac{m}{\lambda}\right. \nonumber\\
& -4 t \sigma(t)+\left(8 m^{2}-3 t+\left(t-2 m^{2}\right) \ln \left(4-\frac{t}{m^{2}}\right)\right) \ln \frac{\sigma(t)-1}{\sigma(t)+1} \nonumber\\
& \left.\left.+2\left(2 m^{2}-t\right)\left(\operatorname{Li}_{2}\left(\frac{1}{2}-\frac{1}{2 \sigma(t)}\right)-\operatorname{Li}_{2}\left(\frac{1}{2}+\frac{1}{2 \sigma(t)}\right)\right)\right\}\right] u_\sigma(\boldsymbol{p})\,,
\end{align}

and it has an infrared divergent term.   
   
%

\subsection{The iterated diagram}
\label{sec.240318.1}

The diagram (f) of Fig.\ref{Fig:todos} for a Coulomb  external potential 
was first correctly calculated in Ref.~\cite{Dalitz:1951ah}.  We have also calculated it independently and the final expression for $\lambda\to 0$ is,

\begin{align}
\label{Eq:A3}
A_{3} & =-(e e')^{2} \bar{u}_{\sigma'}\left(\vp^{\prime}\right)
\left(E \gamma^{0}
\left[
\frac{1-\csc \frac{\theta}{2}}{32 p^{3} \cos ^{2} \frac{\theta}{2}}
+i \frac{\ln \frac{2 p}{\lambda}}{16 \pi p^{3} \cos ^{2} \frac{\theta}{2}}
-i \frac{ 1+\cos ^{2} \frac{\theta}{2}}{32 \pi p^3 \sin ^{2} \frac{\theta}{2}\,\cos ^{2} \frac{\theta}{2}\,\sqrt{1+\frac{\lambda^2}{p^2\sin^2\frac{\theta}{2}}} }
\ln\frac{\sqrt{1+\frac{\lambda^2}{p^2\sin^2\theta/2}}+1}{\sqrt{1+\frac{\lambda^2}{p^2\sin^2\theta/2}}-1}\right]\right. \nonumber\\
& \left.+m\left[-\frac{1-\csc \frac{\theta}{2}}{32 p^{3} \cos ^{2} \frac{\theta}{2}}+i \frac{\ln  \sin \frac{\theta}{2}}{16 \pi p^3 \cos ^{2} \frac{\theta}{2}}\right]\right) u_\sigma(\vp),
\end{align}

Let us notice that $A_3$ is affected by IR divergences. Extra $\lambda$ dependence is kept in the denominators and argument of the log as directly given by the loop integrals. They will be needed when performing the partial-wave projection of this amplitude, which requires an angular  integration with $\theta\in[0,\pi]$. Note that for $\theta\to 0$ the factor $\sqrt{1+\frac{\lambda^2}{p^2\sin^2\theta/2}}$ cannot be expanded in powers of $\lambda$.

\section{Infrared divergences}
\label{sec.240322.3}
\def\theequation{\arabic{section}.\arabic{equation}}
\setcounter{equation}{0}

When computing scattering amplitudes in gauge theories, one encounters infrared (IR) divergences linked to massless gauge fields \cite{Weinberg:1995mt}. Infrared divergences are a common issue in quantum electrodynamics (QED), arising due to the massless character of the photon, which results in the long-range nature of electromagnetic interactions. These divergences occur when considering the  radiative contributions due to soft photons, i.e. photons whose energy and momenta are much smaller than the typical energies and masses in the process,  when integrating over all possible photon momenta in loop diagrams. Furthermore, the radiation of real soft photons also lead to the same type of divergences when integrating over their phase space. In the calculation of observables it is well known that these divergences cancel between each other to all orders in perturbation theory leading to transitions rates free of IR divergences \cite{Bloch:1937pw,Yennie:1961ad,Weinberg:1965nx}. 

However, this approach does not resolve the issue of IR divergences appearing in transition amplitudes, c.f.  Eqs.(\ref{Eq:A2})-(\ref{Eq:A3}). These divergences stem from the methodology of constructing a quantum field theory from free fields, assuming that incoming and outgoing states do not interact in the asymptotic limit (far before and after a scattering event). Conceptually one must take into account that  in quantum electrodynamics (QED) interactions are significant even in the asymptotic limit due to the infinite range of massless force-carrying bosons \cite{Kulish:1970ut}. By considering these asymptotic interactions, one can identify an  operator that transforms free Fock states into asymptotically interacting Faddeev-Kulish (FK) states \cite{Kulish:1970ut}\footnote{See, e.g. \cite{Gaharia:2019xlh} for a pedagogical summary.}. This type of states was first introduced by Chung in Ref.~\cite{Chung:1965zza}. For asymptotic states not containing hard photons but only charge particles, the Chung's and FK states are equivalent. A further point is that Ref.~\cite{Kulish:1970ut} also considered the case when hard photons are part of asymptotic states, so that FK states should be employed then. Heuristically, these FK states could be pictured as consisting of charged particles and hard photons surrounded by a cloud of real soft photons, making all scattering processes infrared finite at the $S$-matrix level. The full FK procedure implies not only a redefinition of the asymptotic states but also a redefinition of the $S$-matrix itself, getting rid of a generalized Coulomb phase that does not affect the transition rates. Directly solving the problem of infrared divergences at the level of the $S$ matrix allows one to  perform meaningful partial-wave expansions of QED scattering amplitudes, unitarize them by  employing techniques stemming from general principles of $S$-matrix theory (for instance, commonly used in hadron physics \cite{Oller:2000ma,Oller:2019opk,Albaladejo:2010tj,Oller:2020guq}), and, in terms of them, to study bound states and resonances. Previous studies along these lines are Refs.~\cite{Blas:2020dyg,Blas:2020och,Oller:2022tmo}.

\subsection{Redefinition of the asymptotic electron states}
\label{sec.240316.1}

In order to introduce the FK states we need some preliminary notation. 
The photon annihilation/creation operators $a(\vk)/a(\vk)^\dagger$, correspond to 
\begin{align}
\label{240311.3}
a^{(h)}(\vk)= \varepsilon^{(h)}_\mu a^\mu(\vk)~,    \end{align}
 where $\varepsilon^{(h)}(\vk)\,$ $h=1,2$ are the polarization vectors;  they are transversal,  $\varepsilon^{(h)}(\vk)\cdot k=0$, and orthogonal $\varepsilon^{(h)}(\vk)^*\cdot\varepsilon^{(h')}(\vk)=-\delta_{hh'}$. The $a_\mu(\vk)$ and $a_\nu^\dagger(\vk')$ satisfy the usual creation and annihilation commutation relations
\begin{align}
\label{240310.2}
[a_\mu(\vk),a_\nu(\vk')^\dagger]&=-g_{\mu \nu} 2k^0(2\pi)^3\delta(\vk-\vk')~,\\
[a_\mu(\vk),a_\nu(\vk')]&=[a_\mu(\vk)^\dagger,a_\nu(\vk')^\dagger]=0~.\nn
\end{align}
In turn, the annihilation/creation operators for electrons are $c_\sigma(\vp)/c_\sigma(\vp)^\dagger$, with $\sigma=\pm 1/2$ the third component of the electron spin at rest.

The FK {\it initial} state may be expressed as \cite{Chung:1965zza,Kulish:1970ut}
\begin{align}
\ket{\vp,\sigma}_i=e^{R_f(p_i)}c_\sigma(\vp)^{\dagger}\ket{0}
\end{align}
where $R_f$ takes the form:
\begin{align}
\label{240310.1}
R_f(p)=\sum_{h=1}^2\int^\Lambda \frac{d^3 k}{(2\pi)^3 2k^0} \left(S_i^{(h)}(\vk) a^{(h)}(\vk)^\dagger-S_i^{(h)}(\vk)^* a^{(h)}(\vk)\right)~,
\end{align}
\jo{with $k_0=\sqrt{\vk^2+\lambda^2}$.}  
The operator $\exp(R_f(p))$ of Eq.~\eqref{240310.1} can be interpreted as generating a surrounding cloud of real soft photons around the external electrons. The integral over $\vk$ is cut at $|\vk|<\Lambda$ to preserve the soft character of the photons involved, being $\Lambda$ much smaller than the characteristic momentum of the scattering process and the electron mass.  The function $S^{(h)}_i(\vk)$ is  derived from the asymptotic potential, and in the case of QED \cite{Kulish:1970ut} it can be written as \cite{Chung:1965zza,Kulish:1970ut} 

\begin{align}
\label{240311.2}
S_i^{(h)}(\vk)&=-e  \frac{p\cdot \varepsilon^{(h)}(\vk)}{p\cdot k}~,
\end{align}
where $p$ is the momentum of the incoming electron. Similarly, we define the analogous factors $S_f^{(h)}(\vk)$ associated to the final electronic state of momentum $p'$. Let us notice that $R_f(p)$ respects the Gupta-Bleuler condition because it only involves transversal polarization vectors \cite{Kulish:1970ut}.

\begin{figure}[H]
\centering
   \centering

\subfloat[]{\includegraphics[width=0.2\textwidth, valign=c]{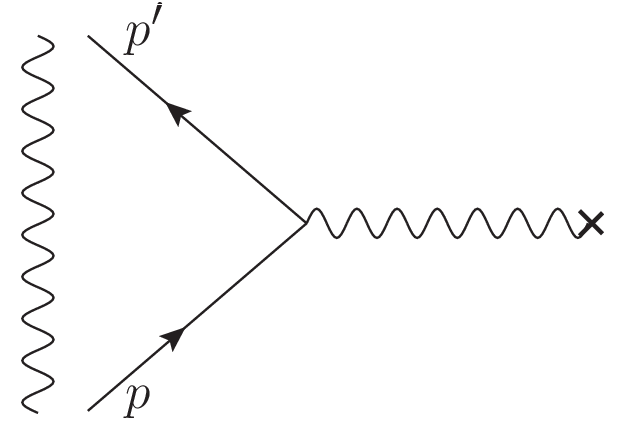}}

\subfloat[]{\includegraphics[width=0.18\textwidth, valign=c]{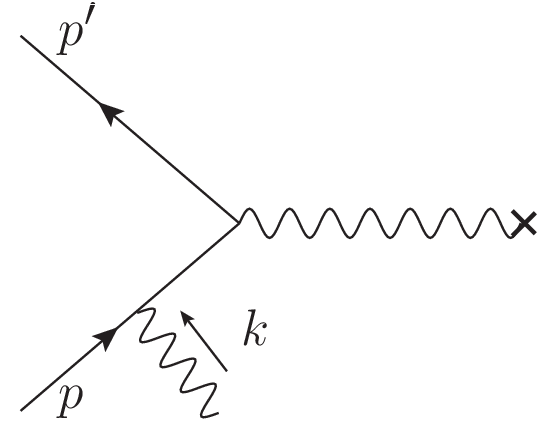}}
\qquad 
\subfloat[]{\includegraphics[width=0.18\textwidth, valign=c]{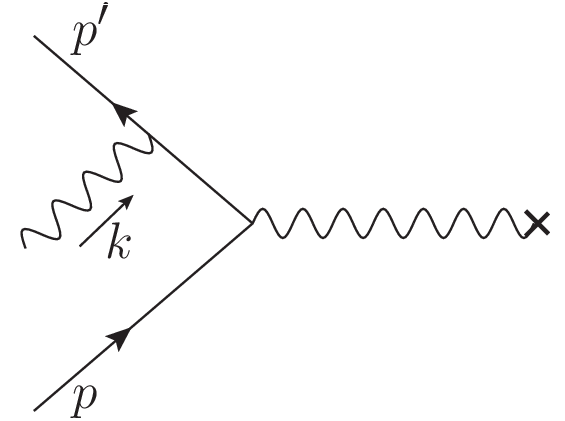}}

\subfloat[]{\includegraphics[width=0.18\textwidth, valign=c]{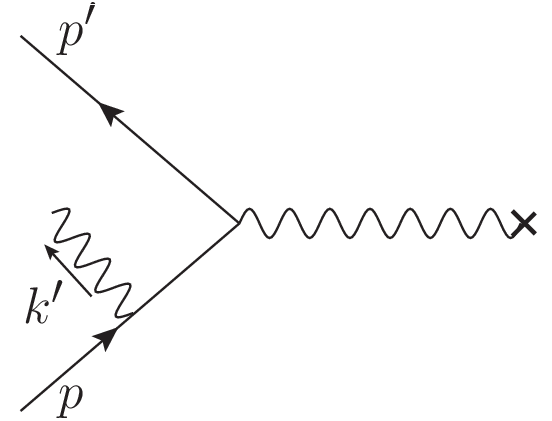}}
\qquad 
\subfloat[]{\includegraphics[width=0.18\textwidth, valign=c]{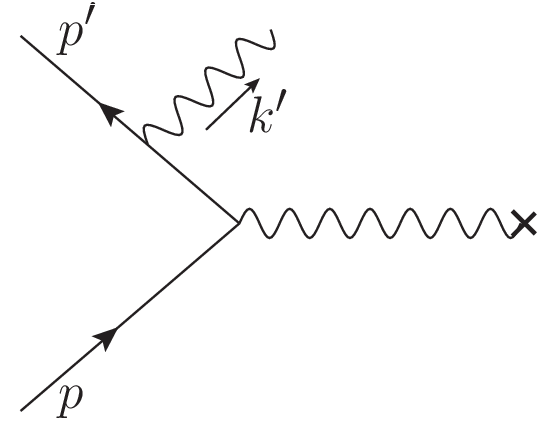}}
     \caption{\label{fig.240316.1} {\small In this figure we show the Chung's diagrams contribution to the second-order scattering amplitude. The depicted photons not corresponding to the external potential are real. Diagram (a): Soft photons from the clouds of the initial and final electrons propagate without interacting. Diagrams (b) and (c): Soft photons from the initial electron cloud are absorbed by the initial and final electronic lines, respectively. Diagrams (d) and (e): Soft photons are emitted from the initial and final electrons, respectively, to the cloud of soft photons of the final state.}}
\end{figure}

 Next, we move to the right the annihilation operators in Eq.~\eqref{240310.1}, by making use of the commutation relations in Eq.~\eqref{240310.2}. Then, one can rewrite the states $\ket{\vp,\sigma}_i$ as
\begin{align}
\ket{\vp,\sigma}_i=\exp\left(-\frac{1}{2}\sum_{h=1}^2 \int^\Lambda \frac{d^3 k}{(2\pi)^3 2k^0} |S_i^{(h)}(\vk)|^2 \right) 
\exp\left( \sum_{h=1}^2\int^\Lambda \frac{d^3k}{(2\pi)^3 2k^0} S_i^{(h)}(\vk)a^{(h)}(\vk)^\dagger \right )
c_\sigma(\vp)^\dagger\ket{0}~.
\end{align}
In our present calculation up to one loop in QED we only need to keep terms up to ${\cal O}(e^2)$ in the definition of $\ket{\vp,\sigma}_i$. Therefore, 
\begin{align}
\label{240316.1}
\ket{\vp,\sigma}_i&= \left(1-\frac{1}{2}\sum_{h=1}^2 \int^\Lambda \frac{d^3 k}{(2\pi)^3 2k^0} |S_i^{(h)}(\vk)|^2 
+
 \sum_{h=1}^2\int^\Lambda \frac{d^3k}{(2\pi)^3 2k^0} S_i^{(h)}(\vk)a^{(h)}(\vk)^\dagger+{\cal O}(e^2)\right)c_\sigma(\vp)^\dagger\ket{0}
 \end{align}
Analogously, the final FK state is
\begin{align}
\label{240316.2}
_f\!\bra{\vp',\sigma'}=\bra{0}c_{\sigma'}(\vp')
\left(1-\frac{1}{2}\sum_{h=1}^2 \int^\Lambda \frac{d^3 k}{(2\pi)^3 2k^0} |S_f^{(h)}(\vk)|^2 
+ \sum_{h=1}^2\int^\Lambda \frac{d^3k}{(2\pi)^3 2k^0} S_f^{(h)}(\vk)^*a^{(h)}(\vk)+{\cal O}(e^2)\right)~.
\end{align}
The redefinition of the asymptotic states in Eqs.~\eqref{240316.1} and \eqref{240316.2}  generates new contributions to the scattering amplitude. The  corresponding diagrams up to ${\cal O}(e^2)$  are shown in Fig.~\ref{fig.240316.1}. There, diagrams (b) and (c) correspond to the absorption of real soft photons sourced by the initial-electron cloud, while in diagrams (e) and (f) real soft photons are emitted to the final-electron   cloud. Non-interacting real soft photons that propagate from the initial- to the final-electron clouds are shown in Diagram (a). 

We refer to the diagrams in Fig.~\ref{fig.240316.1}  as Chung's diagrams, as they were introduced by him for the first time in \cite{Chung:1965zza}. In this reference,  Chung studied in detail the scattering of a charged particle by an external field. It is explicitly shown that the IR divergences from these diagrams cancel the IR divergences from the vertex-modification diagram. The cancellation of IR divergences was shown also in Ref.~\cite{Chung:1965zza} to all orders in perturbation theory.\footnote{No iteration diagrams were considered in Ref.~\cite{Chung:1965zza}.} Calculations are done including a small photon mass $\lambda$ in order to  explicitly handle infrared divergences. Other approaches without a photon mass can be found in \cite{Zwanziger:1974jz}.



To conclude, in order to take into account the FK states, together with the standard diagrams in Fig.~\ref{Fig:todos}, we must consider all the possible interactions involving real photons in the clouds of the electronic states in Fig.~\ref{fig.240316.1}. The explicit calculation of the latter diagrams give
\begin{align}
\label{240316.1b}
A_{\text{4}}&=-A_0\frac{e^2}{2}\int^\Lambda\frac{d^3k}{(2\pi)^3 2 k^0}\left(\frac{p}{p\cdot k}-\frac{p'}{p'\cdot k}\right)^2~.
\end{align}
For the evaluation of the integral we proceed as in Section V of Ref.~\cite{Weinberg:1965nx}, 
and write,  
\begin{align}
\label{240316.4}
A_{4}&=-A_0\frac{e^2}{2 (p^0)^2}
\int_0^\Lambda \frac{|\vk|^2 d|\vk|}{(2\pi)^3 2(k^0)^3}
\int d\Omega_{\vu} \left(\frac{p}{1-\vv\cdot\vu}-\frac{p'}{1-\vv'\cdot\vu}\right)^2~,
\end{align}
where $\vu=\vk/|\vk|$, $\vv=\vp/p^0$, and $\vv'=\vp'/p^0$   (let us recall that $p^0={p'\, }\!^{0}$).  In this way, all the IR divergences are comprised in the first integral over the modulus of $\vk$, while the angular integration is just finite. The former gives
\begin{align}
\label{240316.5}
\int_0^\Lambda\frac{|\vk|^2 d|\vk|}{(\vk^2+\lambda^2)^{3/2}}=\ln\frac{2\Lambda}{\lambda}-1+{\cal O}(\lambda^2)~.
\end{align}
After performing the integration of the solid angle, the result  from the  Chung's diagrams of Fig.~\ref{fig.240316.1} is
\begin{align}
\label{240316.6}
A_{4}&=
\frac{A_0 e^2}{4 \pi^{2}}\left(-1+\frac{2 m^{2}-t}{t \sigma(t)} \ln \frac{\sigma(t)-1}{\sigma(t)+1}\right)\left(\ln \frac{2 \Lambda}{\lambda}-1\right)
\end{align}
As said above, when added to the vertex-modification diagram (b) in Fig.~\ref{Fig:todos} the IR divergences mutually cancel.  The final finite sum of amplitudes of the vertex-modification  and Chung's diagrams, $A_2$ and $A_4$, respectively, gives:
\begin{align}
  \label{230901.1}
  A_{2}+A_4&=-(ee')A^\mu(\vq^2) \bar{u}_{\sigma'}(\vp')\Bigg[
    \frac{i \alpha m (p'-p)^\nu\sigma_{\mu\nu}}{2\pi(t-4m^2)}\sigma(t)\ln\frac{\sigma(t)-1}{\sigma(t)+1}\\
    & +\frac{\alpha \gamma_\mu}{4\pi t \sigma(t)}\Bigg\{4
    \bigg(t\sigma(t)+(t-2m^2)\ln\frac{\sigma(t)-1}{\sigma(t)+1}
    \bigg)\bigg(\ln\frac{m}{2\Lambda}+1\bigg)\nn\\
    &-4t\sigma(t)+\bigg(8m^2-3t+(t-2m^2)\ln(4-\frac{t}{m^2})\bigg)\ln\frac{\sigma(t)-1}{\sigma(t)+1}\nn\\
    &+2(2m^2-t)\bigg({\rm Li}_2(\frac{1}{2}-\frac{1}{2\sigma(t)})
    -{\rm Li}_2(\frac{1}{2}+\frac{1}{2\sigma(t)})\bigg)
    \Bigg\}
    \Bigg]u_{\sigma}(\vp)  ~. \nn
\end{align}

+

\subsection{Redefinition of the $S$ matrix}
\label{sec.240316.2}

The $S$ matrix is standardly defined as the transition amplitude between asymptotic states, describing the scattering of particles in the far past to the far future. However, when dealing with QED, the presence of soft photons leads to infrared divergences, which pose a challenge to the standard $S$-matrix approach. 
We have discussed in the previous section the redefinition of the asymptotic states. But, on top of that, it is still necessary to take care of a divergent phase factor affecting the $S$ matrix, which was first established by Dalitz in Ref.~\cite{Dalitz:1951ah}. There, the Born series for the scattering of a Dirac electron by an external Coulomb potential of an electron is studied up to second order, the diagram (f) in Fig.~\ref{Fig:todos}. It was conjectured that the IR divergent terms affecting this diagram can be resumed in a factor phase 
\begin{align}
\label{240317.1}
\exp\left(i\frac{e e' \jo{E(p)}}{2\pi \jo{p}}\ln\frac{\lambda}{\Lambda}\right)~.
\end{align}
Let us note that there is no effect on the cross section of this factor, since it has unit modulus. The Born series was studied up to third order in the same reference for the  nonrelativistic case, and the expansion of this phase factor also accounts for its IR divergences. 
The validity to all orders of this conjecture was demonstrated by Weinberg \cite{Weinberg:1965nx}.

Faddeev and Kulish  addressed this issue in Ref.~\cite{Kulish:1970ut} and,  in addition to redefining the asymptotic states, as discussed in Sec.~\ref{sec.240316.1}, a related redefinition of the $S$ matrix was derived which removes this type of phase factors. The general form of this phase factor is \cite{Weinberg:1965nx,Kulish:1970ut}
\begin{align}
\label{240317.2}
S_{W;f,i}&=S_f \cdot S_i~,\\
S_i&=\!\!\!\!\!\!\!\!\prod_{\,\,\,\,{\small \begin{tabular}{l}$n\neq m$ \cr $n,m\in \ket{i}$\end{tabular}}} \displaystyle{\!\!\!\!\!\!\!\!\exp\left(i\frac{e_ne_m }{4\pi\beta_{nm}}\ln\frac{\lambda}{\Lambda}\right)}~,\nonumber\\
S_f&=\!\!\!\!\!\!\!\!\prod_{\,\,\,\,{\small \begin{tabular}{l}$n\neq m$ \cr $n,m\in \ket{f}$\end{tabular}}} \displaystyle{\!\!\!\!\!\!\!\!\exp\left(i\frac{e_ne_m }{4\pi\beta_{nm}}\ln\frac{\lambda}{\Lambda}\right)}~.\nonumber
\end{align}
Where  $\ket{i}$ and $\ket{f}$ refer to the initial and final states, respectively, and $\beta_{nm}$ is the Lorentz-invariant relative velocity between the particles $n$ and $m$,
\begin{align}
\label{240317.3}
\beta_{nm}=\sqrt{1-\frac{m_n^2 m_m^2}{(p_n\cdot p_m)^2}}~,
\end{align}
with $m_n$ and $m_m$ the masses of the particles. We notice that  Ref.~\cite{Kulish:1970ut} explicitly states that the natural numbers $n$ and $m$ labelling the charged states in the initial (final) state must be different, as we do here, but this is not indicated in \cite{Weinberg:1965nx}. This fact is  a source of confusion from this latter reference, because the exponent of the exponential becomes infinite for $\beta_{nn}=0$. Indeed,    the real part of the same loop for $n=m$, both belonging to the initial or final state, must be accounted for so as to end with IR-finite transition rates (this loop has a zero imaginary part if calculated explicitly).  In the appendix \ref{sec.240413.2} we show that the loop integral of Ref.~\cite{Weinberg:1965nx} called $J_{nm}$, whose imaginary part gives rise to the phases in Eq.~\eqref{240317.2}, is purely real for $n=m$. These diagrams correspond to the contributions from the wave-function renormalization of the electron. 

The $S$ matrix for the transition $\alpha\to\beta$ free of IR divergences is then \cite{Kulish:1970ut}
\begin{align}
\label{240317.4}
S_{\beta,\alpha}&=\widetilde{S}_{\beta,\alpha} S_{W;\beta,\alpha}^{-1}~,
\end{align}
with $\widetilde{S}_{\beta\alpha}$ the $S$ matrix that results from the QED calculation with the asymptotic states dressed by the cloud of real soft photons, c.f. Sec.~\ref{sec.240316.1}.  
In our present study we employ the IR free $S$ matrix $S_{\vp's',\vp s}$. The correct phase factor is Eq.~\eqref{240317.1}, and it  can be obtained from Eq.~\eqref{240317.2} by taking that the source of the external potential is a infinitely heavy particle of charge $e'$, so that $\beta=p/E(p)$.  

The IR divergent contribution of the iterated diagram in Eq.~\eqref{Eq:A3} for $\theta\neq 0$ is
\begin{align}
\label{240317.5}
i(ee')^2\bar{u}_{\sigma'}(\vp')\gamma^0
u_\sigma(\vp)\frac{E(p)\ln 2p/\lambda}{8\pi p^3\sin^2\frac{\theta}{2}}~.
\end{align}
The IR divergent term proportional to $\ln\lambda$ is equal to  $A_0$, Eq.~\eqref{240316.3}, times the phase factor in Eq.~\eqref{240317.1} expanded up to ${\cal O}(ee')$ and removing the one \cite{Dalitz:1951ah}.

\section{Partial-wave amplitudes}
\label{sec.240322.4}
\def\theequation{\arabic{section}.\arabic{equation}}
\setcounter{equation}{0}   

The  expansion of a scattering amplitude in  PWAs is familiar from elementary 
Quantum Mechanics. This is a technique that typically simplifies the analysis of a particle interaction with a potential of finite range, as unitarity is imposed more simply. The key idea is to take advantage of rotational symmetry and decompose the scattering amplitude into PWAs, each characterized by having a well defined total-angular momentum quantum number $J$.

In our case, the interaction is of infinite range, which makes that the PWA expansion for the Coulomb interaction  does not converge.\footnote{Despite this lack of convergence, the PWA expansion is typically used in nuclear physics to disentangle the effects of Coulomb scattering because of the interference with strong interactions \cite{landau.170517.1}. The point here is to evaluate separately those low-$\ell$ PWAs affected by the strong interactions, and then to resum the rest of Coulomb PWAs by knowing the Coulomb scattering amplitude without strong interactions.} In this case, the Lehmann ellipse collapses because the coalescence of the LHC and RHC to a point at $|\vp|=0$ \cite{Giddings:2009gj,Lehmann:1958ita,martin.200705.1,Kang:1962}. Nonetheless, as stressed in Ref.~\cite{Blas:2020dyg},  due to rotational invariance the angular momentum operator $\boldsymbol{J}$ and the Hamiltonian $H$ commute, $[H,\boldsymbol{J}]=0$, in this case as well. Therefore, it is meaningful to study scattering in the basis of states that simultaneously diagonalize $\boldsymbol{J}$ and $H$, whose scattering amplitudes  are the PWAs. 

Additionally, the unitarity requirement of the $S$ matrix can be expressed in a particularly simple way via PWAs, which is one of their main appeals. Related to this, the study of PWAs also allows one to directly address the spectrum of the theory and find bound states and resonances with the selected quantum numbers of the PWAs coupled. These points will be treated in detail below in Secs.~\ref{sec.240128.1}-\ref{sec.231221.2}. 

We project the scattering amplitudes for an electron scattered by a external potential in the so-called $\ell s J$ basis. Every state of this basis is characterized by having orbital angular momentum $\ell$,  spin $s$,  and angular momentum $J$. It is denoted by $|J\mu,\ell s\ra$, with $\mu$ the third component  of the total angular momentum $\boldsymbol{J}$ along the $z$ axis. Although it is not indicated in the set of quantum numbers inside a ket, a  $\ell s J$ state has well defined modulus of momentum $|\vp|$. Of course, because of rotational invariance and the Wigner-Eckart theorem the PWAs do not depend on $\mu$.

The partial-wave expansion in the $\ell s J$ scheme is developed in detail in Ref.~\cite{Oller:2019rej}, to which we refer for more details, and here we only review briefly the main points. 
The one-particle states with \jo{define} momentum are defined as  
\begin{align}
\label{240320.2}
|\vp,\sigma s\ra&=U(\vp)|\boldsymbol{0},\sigma s\ra~,
\end{align}
with the standard Lorentz transformation directly taking the state at rest to its final momentum $\vp$. It can be decomposed as
\begin{align}
\label{240320.3}
U(\vp)=R(\hvp)B_z(p)R(\hvp)^\dagger~,
\end{align}
with $B_z(p)$ a Lorentz boost along the $z$ axis of velocity $-\vp/E$, and the rotation $R(\hvp)$ takes the $z$ axis to $\vp$. In terms of two rotations around the $y$ and $z$ axis, $R_y$ and $R_z$, it can be expressed as $R(\hvp)=R_z(\phi)R_y(\theta)$, with $\theta$ and $\phi$ the polar and azimuthal angles of $\vp$, in this order. In passing, we mention that if $U(\vp)=R(\hvp)B_z(p)$ were used in Eq.~\eqref{240320.2} the state would have helicity $\sigma$.

Within our normalization the one particle states have the following Lorentz invariant scalar product,
\begin{align}
  \label{231221.1}
  \la \vp',\sigma'|\vp,\sigma\ra&=(2\pi)^32E(p)\delta(\vp'-\vp)\delta_{\sigma'\sigma}~,\\
  E(p)&=\sqrt{\vp^2+m^2}~.\nn
\end{align}

 Let us stress that the composition $\boldsymbol{{s}}+\boldsymbol{\ell}=\boldsymbol{J}$ is compatible with a relativistic formalism. This is so because the states
\begin{align}
\label{240320.1}
|\ell m, \sigma s\rangle=\frac{1}{\sqrt{4 \pi}} \int d \hat{\vp} \,Y_{\ell}^m(\hat{\vp})|\vp,\sigma s\rangle~,
\end{align}
transform under rotations as the direct product of the representations of angular momentum $\ell$ and $s$ \cite{Oller:2019rej}. 
Therefore, the states with definite  angular momentum are
\begin{align}
\label{240320.4}
|J \mu, \ell s\rangle & =\sum_{\sigma, m} C(m \sigma \mu \mid \ell s J)\left|\ell m, \sigma s\right\rangle 
  =\frac{1}{\sqrt{4 \pi}} \sum_{m=-\ell}^{\ell} \sum_{\sigma=-s}^{s} C(m \sigma \mu \mid \ell s J) \int d \hat{\vp}\, Y_{\ell}^{m}(\hat{\vp}) \left|\vp, \sigma s\right\rangle~.
\end{align}
where $C(m \sigma \mu \mid l s J)$ refers to Clebsh-Gordon coefficients of the composition of spins $\ell$ and $s$ to give $J$, with third components $m$, $\sigma$ and $\mu$, respectively. 
 The normalization of the $|J\mu,\ell s\rangle$ is a consequence of Eqs.~\eqref{231221.1} and \eqref{240320.4}. It corresponds to
\begin{align}
  \label{231221.2}
\la J'\mu',\ell' s|J\mu,\ell s\ra&=\frac{4\pi^2}{p}\delta(E'-E)\delta_{\ell'\ell}\delta_{\mu'\mu}\delta_{J'J}~.
\end{align}

When computing the PWAs  for scattering processes, we have to project the initial and final states onto $\ell s J$ states. 
This projection involves integrating over the solid angle to extract the angular momentum components of the initial and final states, according to Eq.~\eqref{240320.4}. Employing rotational invariance, one can get rid of the solid angle integration over the initial state and to keep its momentum fixed along the $z$ axis \cite{Oller:2019rej}. One can furthermore take an extra rotation to restrict the solid angle integration over the final state to a single angle integration over $\theta\in[0,\pi]$, by restricting the final momentum to belong to the $xz$ plane. This last step is explained in Appendix~\ref{App:PWA}. Restricting ourselves to the case of interest  for the calculation of the PWAs of an electron scattered by an external potential, $T^{(J,\ell)}$,  the final equation is
\begin{align}
  \label{231216.1}
  T^{(J,\ell)} \equiv \la J \mu,\ell s |T| J \mu,\ell s\ra&=\frac{1}{2}\int_{-1}^{1}d\cos\theta \,P_\ell(\cos\theta) \la \vp_{xz},\sigma'=\frac{1}{2}s |T|p\hat{\vz}, \sigma=\frac{1}{2} s\ra\\
&  +\frac{(-1)^{\ell+\frac{1}{2}-J}}{2(J+\frac{1}{2})}\int_{-1}^{1}d\cos\theta
 \, P_\ell^1(\cos\theta)\, \la \vp_{xz},\sigma'=-\frac{1}{2} s |T|p\hat{\vz}, \sigma=\frac{1}{2} s\ra~,\nn
\end{align}
and $s=1/2$.
Here we are also denoting the PWAs as
$ \la J \mu,\ell s |T| J\mu,\ell s\ra$. In turn, the scattering amplitudes between states of definite linear momentum are designed as  $\la \vp_{xz},\sigma's |T|p\hat{\vz},\sigma s\ra$,  with the momentum of the incoming electron fixed in the $\hat{\vz}$ direction.  
In Eq.~\eqref{231216.1} we have taken into account that since $J$ is conserved because rotation symmetry, it follows that $\displaystyle{\ell=J\pm \frac{1}{2}}$, as $s=1/2$. Since the parity of an electron state of the type $|J\mu,\ell s\ra$ is $(-1)^\ell$ it follows then that $\ell$ is conserved, and PWAs with different $\ell$ do not mix.

 \subsection{Finite PWAs after redefinition of the $S$ matrix and unitarity}

The redefinition of the $S$ matrix in PWAs is a straightforward consequence of Eq.~\eqref{240317.4} because the phase factor $S_W$ does not depend on spin or on the scattering angle (for one-body scattering in the presence of an external potential and also for two-body scattering of a closed system). Thus, it factorizes out in the process of calculating the PWAs, Eq.~\eqref{231216.1}.

We denote by $\bar{S}^{(J,\ell)}\equiv 1+2i\rho \bar{T}^{(J,\ell)}$ the $S$ matrix in PWAs that follows from the direct calculation within perturbative QED up to some order. The phase space factor $\rho$ is fixed by the normalization of the $|J\mu,\ell s\ra$ states, Eq.~\eqref{231221.2}, by definition is 
\begin{align}
\label{240321.11}
\rho=\frac{p}{4\pi}~.
\end{align}
We have indicated by $\bar{T}^{(J,\ell)}$ the PWA between asymptotic electronic states corrected by the soft photon cloud. The PWAs satisfy unitarity, 
\begin{align}
\label{240321.12}
\Im \bar{T}^{(J,\ell)}&=\rho\left|\bar{T}^{(J,\ell)}\right|^2
\end{align}
The redefined IR-finite $S$ matrix in PWAs $S^{(J,\ell)}$ up to ${\cal O}(\alpha^2)$ is 
 \begin{align}
\label{240318.1}
  S^{(J,\ell)}&= \bar S^{(J,\ell)} S_W^{-1}=1+2 i \rho \bar T_1^{(J,\ell)}+\frac{i e e'}{2\pi v}\ln\frac{\lambda}{\Lambda}+2i\rho\bar T_2^{(J,\ell)}-\rho \frac{e e'}{\pi v}\ln\frac{\Lambda}{\lambda}\bar T_1^{(J,\ell)}-\frac{1}{8\pi^2}\frac{(e e')^2}{v^2}\ln^2\frac{\Lambda}{\lambda}+\mathcal{O}(e^3)
 \end{align}
 In this equation the subscripts indicate the order in the calculation within the QED expansion, $\bar{T}^{(J,\ell)}=\sum_n \bar{T}^{(J,\ell)}_n \alpha^n$.
 
The following step was firstly introduced in Ref.~\cite{Blas:2020och} in order to define IR finite PWAs $T^{(J,\ell)}$ out of the IR-finite $S^{(J,\ell)}$. The procedure is based on  identifying the PWAs free of IR divergences $T^{(J,\ell)}$ from the relation
\begin{align}
\label{240318.2}
S^{(J,\ell)}&=1+2i\rho T^{(J,\ell)}~.
    \end{align}
Then,  from Eq.~\eqref{240318.1} it follows that 
 the IR-finite PWAs at leading and next-to-leading order, $T^{(J,\ell)}_1$ and $T^{(J,\ell)}_2$, respectively, are given by
\begin{align}
\label{231216.5}
T_1^{(J,\ell)}&=\bar{T}_1^{(J,\ell)}+\frac{ee' E}{p^2}\ln\frac{\Lambda}{\lambda}~,\\
\label{231216.6}
T_2^{(J,\ell)}&=\bar{T}_2^{(J,\ell)}+i\frac{e e' E}{2\pi p}\ln \frac{\Lambda}{\lambda}\,\bar{T}_1^{(J,\ell)}+i\frac{(ee')^2E^2}{4\pi p^3}\ln^2\frac{\Lambda}{\lambda}~.
\end{align}
This redefinition only affects the imaginary of $T^{(J,\ell)}_2$. 
This fact is in agreement with our NLO calculation of the iteration diagram (f) in Fig.~\ref{Fig:todos},  whose IR divergences affect only its imaginary, cf. Eq.~\eqref{Eq:A3}.

The redefinition of the $S$-matrix elements $\bar{S}^{(J,\ell)}$ to end with $S^{(J,\ell)}$ only involves the phase factor $S_W$ and unitarity of the $S$ matrix is preserved. Therefore, 
\begin{align}
\label{240321.10}
S^{(J,\ell)}{S^{(J,\ell)}}^*=I~.
\end{align}
Taking into account the relation between $S^{(J,\ell)}$ and $T^{(J,\ell)}$, Eq.~\eqref{240318.2}, it also follows the unitarity relation for the PWAs, 
\begin{align}
\label{240321.11}
\Im T^{(J,\ell)}&=\rho \left| T^{(J,\ell)} \right|^2~,~p^2>0~.
\end{align}
This settles the unitarity or right-hand cut (RHC) for physical values of the momentum.

\subsection{Handling different angular-projection integrals}
\label{sec.140413.1}

Let us first notice that for the term with $P_\ell^1(\cos\theta)$ in Eq.~\eqref{231216.1} the angular integration  is straightforwardly convergent because of the extra factor of $\sin\theta$ that is part of the generalized Legendre Polynomials $P_\ell^m(\theta)$ with $m>0$, 
\begin{align}
\label{240318.3}
P_\ell^m(\cos\theta)=(-1)^m(1-\cos^2\theta)^{m/2}\frac{d^m}{d\cos^m\theta}P_\ell(\cos\theta)~.
    \end{align}
For this reason,  when calculating the partial-wave projection of the iteration diagram, for the term with $P_\ell^1(\cos\theta)$  in Eq.~\eqref{231216.1} one may use instead of Eq.~\eqref{Eq:A3} a simpler expression  given by
\begin{align}
\label{240318.4} 
A_{3}  =-(e e')^{2} \bar{u}_{\sigma'}\left(\boldsymbol{p}^{\prime}\right)&\left(E \gamma^{0}\left[\frac{1-\csc \frac{\theta}{2}}{32 p^{3} \cos ^{2} \frac{\theta}{2}}-i \frac{\ln \frac{2 p}{\Lambda}}{8 \pi p^{3} \sin ^{2} \frac{\theta}{2}}-i \frac{\left(1+\cos^2\frac{\theta}{2}\right)\,\ln \left(\sin \frac{\theta}{2}\right) }{16 \pi p^3 \sin ^{2}\frac{\theta}{2}\, \cos ^{2} \frac{\theta}{2}  }\right]\right.\\& \left.
+m\left[-\frac{1-\csc \frac{\theta}{2}}{32 p^{3} \cos ^{2} \frac{\theta}{2}}+i \frac{\ln \sin \frac{\theta}{2}}{16 \pi p^3 \cos ^{2} \frac{\theta}{2}}\right]\right) u_\sigma(p\boldsymbol{z}),\nn
\end{align}
Actually, the term proportional to $\ln 2p/\Lambda$ in the previous equation results by the cancellation of an IR divergent term proportional to $\ln2p/\lambda$ with another term that stems from the redefinition of the $S$ matrix, from the operation,
\begin{align}
\label{240319.1}
S_W^{-1}A_0\to i\frac{ee'E}{2\pi p}\ln\left(\frac{\Lambda}{\lambda}\right)A_0~,
\end{align}
and then  $\ln 2p/\lambda\to \ln 2p/\Lambda$ in Eq.~\eqref{240318.4}.
 
Let us consider now the term $P_\ell(\cos\theta)$ in Eq.~\eqref{231216.1}, which requires a more careful treatment in order to cancel the $\ln\lambda$ terms. The IR divergent angular integrals of $A_3$ occur for  $\theta \to 0$. On the one hand, one has the terms in Eq.~\eqref{Eq:A3} that do not drive to IR divergent integrals, and their angular integrals for the partial-wave projection can be done in a straightforward manner. Namely, these terms are 
\begin{align}
-(e e')^{2} \bar{u}_{\sigma'}\left(\vp^{\prime}\right)
\left(E \gamma^{0}\frac{1-\csc \frac{\theta}{2}}{32 p^{3} \cos ^{2} \frac{\theta}{2}}
+m\left[-\frac{1-\csc \frac{\theta}{2}}{32 p^{3} \cos ^{2} \frac{\theta}{2}}+i \frac{\ln  \sin \frac{\theta}{2}}{16 \pi p^3 \cos ^{2} \frac{\theta}{2}}\right]\right) u_\sigma(p\vz)~.
\end{align}

On the other hand, the terms in $A_3$, Eq.~\eqref{Eq:A3}, that give rise to IR divergent angular integrals  when multiplied by $P_\ell(\cos\theta)$ when taken to Eq.~\eqref{231216.1}, are the last two terms in between the square brackets of this equation,  namely: 
\begin{align}
\label{240319.2}
A_3^{\text{IR}}&=-\frac{i(e e')^{2} u_\frac{1}{2}^\dagger\left(\vp^{\prime}_{xz}\right)u_\frac{1}{2}(p\hat{\vz})
E }{16\pi p^3\cos^2\frac{\theta}{2}}
\left(
 \ln \frac{2 p}{\lambda}
- \frac{ \cos ^{2} \frac{\theta}{4}}{\sin ^{2} \frac{\theta}{2}\,\sqrt{1+\frac{\lambda^2}{p^2\sin^2\frac{\theta}{2}}} }
\ln\frac{\sqrt{1+\frac{\lambda^2}{p^2\sin^2\theta/2}}+1}{\sqrt{1+\frac{\lambda^2}{p^2\sin^2\theta/2}}-1}\right).
\end{align}
The angular integration of this term times $P_\ell(\cos\theta)$ is not algebraic, but we need an algebraic expression in order to handle with the removal of the IR divergence factor $\ln\lambda$ after including terms from the $S$-matrix redefinition. In order to accomplish this purpose, we split the angular integral over $\theta\in[0,\pi]$ in two intervals. The first one is from 0 up to $\delta$, with $1\gg\delta\gg \lambda^2/p^2$, and the other from $[\delta,\pi]$, such that power-law dependence on $\delta$ are neglected since one takes at the end $\delta\to 0^+$.  

For the former integration, one can simplify  the angular integral in Eq.~\eqref{231216.1} as
\begin{align}
\label{240319.3}
\frac{1}{2}\int_0^\delta d\theta \sin\theta A_3^{\text{IR}}P_\ell(\cos\theta)
&=i(ee')^2E^2 \int_0^\delta \frac{\theta d\theta}{2\pi p^3\theta^2\sqrt{1+\frac{4\lambda^2}{p^2\theta^2}}}\ln\frac{1+\sqrt{1+\frac{4\lambda^2}{p^2\theta^2}}}{\frac{2\lambda}{p\theta}}+\ldots=\frac{i(ee')^2E^2}{4\pi p^3}\ln^2\frac{\lambda}{p\delta}+\ldots
\end{align}
where the ellipsis involve terms suppressed by powers of $\delta$. In the previous equation we have taken into the matrix elements in Eq.~\eqref{240629.1} from the Dirac algebra, and that $P_\ell(\cos\theta)\to 1$ for $\theta\ll 1$, so that this result is valid for all partial waves. Notice that the term proportional to  $\ln(2p/\lambda)$ in Eq.~\eqref{240319.2} is not included in Eq.~\eqref{240319.3} because its integral vanishes linearly in $\delta$ for $\delta\to 0$. 

For the integral with $\theta\in[\delta,\pi]$, since $\delta\gg \lambda^2/p^2$, we can expand $A_3^{\text{IR}}$ in powers of $\lambda/p^2$. The integral gets simpler, as
\begin{align}
\label{240319.4}
\frac{1}{2}\int_\delta^\pi d\theta \sin\theta A_3^{\text{IR}}P_\ell(\cos\theta)
&=\frac{i(ee')^2E}{4\pi p^3}\int_\delta^\pi d\theta \cot\frac{\theta}{2}\left(E+m+(E-m)\cos\theta\right)\,P_\ell(\cos\theta)\left(\ln\frac{2p}{\lambda}+\frac{\left(1+\cos^2\frac{\theta}{2}\right)\ln\sin\frac{\theta}{2}}{2\cos^2\frac{\theta}{2}}\right).
\end{align}
Of course, when summing the results of the integrals in Eq.~\eqref{240319.3} and \eqref{240319.4} the dependence on $\delta$ vanishes for $\delta\to 0$.
For instance, for $\ell=0$ the result from the sum of these two contributions is
\begin{align}
\label{240319.5}
A_3^{\text{IR};\ell=0}&=\frac{i(ee')^2 E}{4\pi p^3}\left[E\ln^2\lambda-\left(m+E(-1+\ln 4)+2E\ln p\right)\ln\lambda
+\frac{1}{24}\left(-m(6+\pi^2)+6E+24E\ln^2 2p-24(E-m)\ln 2p\right)\right],
\end{align}
and the $\ln\delta$ dependence has disappeared. 

Additionally, the $\ln\lambda$ dependence present in Eq.~\eqref{240319.3} and \eqref{240319.4} precisely cancels with the one from the $S$-matrix redefinition. Again for illustration let us take $\ell=0$. The IR divergent piece from the $S$-matrix redefinition, cf. Eq.~\eqref{231216.6}, is
\begin{align}
\label{240319.6}
i\frac{e e' E}{2\pi p}\ln \frac{\Lambda}{\lambda}\bar{T}_1^{(\frac{1}{2},0)}+i\frac{(ee')^2E^2}{4\pi p^3}\ln^2\frac{\Lambda}{\lambda}&=
\frac{i(ee')^2E}{4\pi p^3}\left[
-E\ln^2\lambda + \left(m+E\left(-1+\ln4\right)+2E\ln 2p\right)\ln\lambda\right.\nn\\
&\left.+
\left(E-m+E\left(\ln\Lambda-2\ln 2p\right)\right)\ln\Lambda\right]
\end{align}
and its dependence on $\ln\lambda$ cancels with that from $A_3^{\text{IR};\ell=0}$ in  Eq.~\eqref{240319.5}, such that the combination $\ln 2p/\Lambda$ is the one remaining. 

The calculation of the angular integrals in Eq.~\eqref{231216.1} for the partial-wave projection of the NLO QED amplitudes due to the vacuum polarization, $A_1$, and vertex modification, $A_2$, can be done numerically in a straightforward way. Nonetheless, for small momenta $|\vp|\ll m$ one can derive algebraic expressions by making an  expansion in powers of $t$ of $A_1$ and $A_2$ in Eqs.~\eqref{Eq:A1} and \eqref{Eq:A2}, respectively. In terms of this expansion the angular integrals of Eq.~\eqref{231216.1} can be done algebraically. 

For instance, applying this procedure to the vacuum polarization diagram for $\ell=0$ we have
\begin{align}
\label{240319.7}
A_1^{\ell=0}=-ee'\alpha \frac{ \left(E + m\right) \left(8 E^4 + 16 E^3 m - 59 E^2 m^2 - 43 E m^3 + 
    267 m^4 \right)}{2835 \pi m^6 }+{\cal O}\left((p/m)^6\right)~,
\end{align}
and for the vertex-modification diagram the result is
\begin{align}
\label{240319.8}
A_2^{\ell=0}&=
\frac{ee'\fa}{20160 \pi m^6 }  \left(
-96 E^5 + 80 E^3 m^2 + 4216 E m^4 + 
   m^5 (832 - 945 \pi) + 20 E^2 m^3 (68 + 21 \pi) - 
   E^4 m (512 + 105 \pi) \right.\\
   &\left.+ 
   64 (E + m) (8 E^4 + 16 E^3 m - 53 E^2 m^2 - 37 E m^3 + 
      171 m^4) \ln\frac{m}{2\Lambda}\right)+{\cal O}\left((p/m)^6\right)~.\nn
\end{align}

The partial-wave projection of the Born  amplitude  of an electron by a Coulomb potential was already discussed in Refs.~\cite{Blas:2020dyg,Blas:2020och,Oller:2022tmo} for the nonrelativistic case. The situation is analogous for the LO QED scattering amplitude $A_0$, Eq.~\eqref{240316.3}. Its PWA projection according to Eq.~\eqref{231216.1} can be evaluated algebraically. It is IR divergent, but these divergences are cancelled out when adding the ${\cal O}(ee')$ term after the redefinition the $S$ matrix, following  Eq.~\eqref{231216.5}.

\section{Unitarization methods}
\def\theequation{\arabic{section}.\arabic{equation}}
\setcounter{equation}{0}   
\label{sec.240128.1}

Partial wave unitarity in quantum field theory serves as a cornerstone principle, guaranteeing that the combined probability of all conceivable outcomes in a scattering event remains constrained to 1. This foundational concept is essential for upholding the coherence of quantum field theories.


   For a bound states $|p|\sim m\alpha\ll m$, and this makes that the chain of unitarity diagrams comprising the Lippmann-Schwinger or Schr\"odinger equations are very much enhanced because of unitarity,
\begin{align}
\label{231219.1}
\Im T^{(J,\ell)}=\frac{p}{4\pi}|T^{(J,\ell)}|^2~,~p^2>0~,
\end{align}
as deduced in Eq.~\eqref{240321.11}. 
Since $T^{(J,\ell)}$ at leading order in powers of $\alpha$ goes like $p^{-2}$ for $|p|\ll m$, it is clear that then unitarity produces ever increasing powers of $p^{-1}$, which correspondingly require higher powers of $m$ in the numerator so as to keep  right the dimensions of the amplitude resulting. Of course, this is to be expected since the generation of bound states requires non-perturbative physics. This enhancement can also be seen by analyzing the pinch singularity in the unitarity diagrams, similarly as originally discussed in \cite{Weinberg:1991um} for nucleon-nucleon scattering (there one has the Deuteron among other non-perturbative effects). This same effect was also unveiled for $\bar{K}N$ scattering and the $\Lambda(1405)$ together with other non-perturbative phenomena and near resonances in \cite{Oller:1997ti,Oller:2000fj,Oller:2005ig,Albaladejo:2010tj}. 



Next, we briefly summarize the three unitarization methods that we consider along the manuscript: the $g$-method, the Inverse Amplitude Method (IAM) and the first-iterated $N/D$ method.  We apply and compare the results obtained between them. For scattering along physical values of momentum, that is, real and positive ones, the $g$-method and the 1st-iterated $N/D$ method  are mutually compatible. For the IAM this is usually also the case, except for some qualifications at relatively large momentum, as we discuss in detail below.  However, for the study of the fundamental state of the hydrogen atom, due to the overlap between the pole position and the left-hand cut (LHC), only the  first-iterated $N/D$ method is the one with the analytical structure sound enough to be applied in such demanding circumstance.

Common to all these three unitarization methods is to use that for the inverse of a PWA  the unitarity relation in partial waves,  Eq.~\eqref{231219.1}, implies
\begin{align}
  \label{231221.3a}
\Im\frac{1}{T^{(J,\ell)}}&=-\frac{p}{4\pi}~,~p^2>0~.
\end{align}
For more details in the presentation of the different unitarization methods here employed, including extra  references, the interested reader can consult the text book \cite{Oller:2019rej}, and the reviews \cite{Oller:2019opk,Oller:2020guq}.

\subsection{The $g$-method}
\label{sec.240128.2}

Based on Eq.~\eqref{231221.3a} one can isolate the branch-point singularity at $p^2=0$ due to unitarity by writing
\begin{align}
  \label{231221.3}
 \frac{1}{T^{(J,\ell)}(p^2)}&=\frac{1}{V^{(J,\ell)}(p^2)}+g(p^2)~,
\end{align}
with the unitarity loop function $g(p^2)$ having a RHC for $p^2>0$. The branch point singularity at $p^2=0$ also has another source due to the LHC as photons have null mass. For instance, at NLO QED one has a LHC  contribution for $p^2<0$  from iterating the LO QED result along the RHC. This fact poses some extra difficulties that are properly treated in Sec.~\ref{sec.240128.4}.  From Eq.~\eqref{231221.3} it follows that the unitarized PWA then reads
\begin{align}
  \label{240122.1}
  T^{(J,\ell)}(p^2)&=\frac{1}{V^{(J,\ell)}(p^2)^{-1}+g(p^2)}
 =\frac{V^{(J,\ell)}(p^2)}{1+V^{(J,\ell)}(p^2) g(p^2)}~.
  \end{align}

The unitarity loop function, accounting for the propagation of the electron in the scattering process, obeys the following once-subtracted 
 dispersion relation in the complex $p^2$ plane 

\begin{align}
  \label{240121.1}
  g(p^2)&=g(0)-\frac{p^2}{4\pi^2} \int_{0}^\infty \frac{dk^2 k}{(k^2-p^2)k^2}
=g(0)-i\frac{\sqrt{p^2}}{4\pi}
  ~.
\end{align}
Next,  for $p^2\to 0$ we match the unitarity loop function with its nonrelativistic counterpart \cite{Oller:2022tmo}, which requires that $g(0)=0$. Thus, our final formula for $g(p^2)$ is
\begin{align}
  \label{240121.2}
  g(p^2)&=-i\frac{\sqrt{p^2}}{4\pi}~.
\end{align}
As a matter of fact, it then coincides with its nonrelativistic expression for all $p^2$ \cite{Oller:2022tmo}.

Once we have calculated $g(p^2)$ we can apply Eq.~\eqref{231221.3} to calculate $V^{(J,\ell)}(p^2)$ in a series expansion in powers of $\alpha$, \begin{align}
V^{(J,\ell)}(p^2)=\sum_{n=1}  V_n^{(J,\ell)}(p^2)~,
\end{align}
where  $V^{(J,\ell)}_n$ is  ${\cal O}(\alpha^n)$.  To fix them we match the perturbative QED PWA and its unitarized formula order by order in an expansion in powers of $\alpha$. We denote by $T_n^{(J,\ell)}(p^2)$ the ${\cal O}(\alpha^n)$ contribution to the scattering amplitude calculated perturbatively in QED.   Then, 
by expanding Eq.~\eqref{240122.1} up to NLO  one has that
\begin{align}
  \label{231221.6}
  V_1^{(J,\ell)}(p^2)&=T_1^{(J,\ell)}(p^2)~,\\
  \label{231221.7}
V_2^{(J,\ell)}(p^2)&=T_2^{(J,\ell)}(p^2)+g(p^2)V_1^{(J,\ell)}(p^2)^2~.
\end{align}

\subsection{The Inverse Amplitude Method}
\label{sec.240128.3}

The IAM \cite{Lehmann:1972kv,Dobado:1989qm,Dobado:1996ps,Oller:1997ng,Oller:1998hw} can be derived by  an expansion of the inverse of a PWA up to NLO,
\begin{align}
  \label{240129.1}
\frac{1}{T_1^{(J,\ell)}+T_2^{(J,\ell)}}&=\frac{1}{T_1^{(J,\ell)}}-\frac{T_2^{(J,\ell)}}{{T_1^{(J,\ell)}}^2}+{\cal O}(\alpha)~.
\end{align}
Then, this result is  directly inverted such that the unitarized PWA is 
\begin{align}
  \label{240128.iam}
  T^{(J,\ell)}&=\frac{T_1^{(J,\ell)}(p^2)^2}{T_1^{(J,\ell)}(p^2)-T_2^{(J,\ell)}(p^2)}~.
\end{align}
Unitarity is satisfied because of the perturbative unitarity fulfilled by $T_2^{(J,\ell)}(p^2)$, such that
\begin{align}
  \label{240212.2}
  \Im \,T_2^{(J,\ell)}(p^2)=\frac{p}{4\pi} T_1^{(J,\ell)}(p^2)^2 ~,~p^2>0~.
\end{align}

An issue that raises when applying the IAM for $p\gg m$ is that $T_2^{(J,\ell)}$ decreases more slowly for large $p$ than $T_1^{(J,\ell)}$. While the latter does it  as $1/p$, the former is typically ultraviolet enhanced by logarithmic factors of the type $\displaystyle{\frac{\ln^2( p^2/m^2)}{p}}$, due to the dressed-vertex contribution, as well as by other less strong ones as $\displaystyle{\frac{\ln (p^2/m^2)}{p}}$, that stem from the rest of contributions to $T_2^{(J,\ell)}$.  As a result, for large $p\gtrsim {\cal O}(m/\alpha)$, the expansion of $1/(T_1+T_2)=1/T_1-T_2/T_1^2+\ldots$ loses its meaning. This manifests in the appearance of a marked peak when $T_1^{(J,\ell)}-\Re T_2^{(J,\ell)}=0$,  wherever  these perturbative amplitudes have the same sign for large enough $p$. We show below some examples in this respect concerning the PWAS with $(J,\ell)=(\frac{1}{2},0)$ and $(\frac{1}{2},1)$.  However, these peaks do not correspond to any resonance pole in the second Riemann sheet and are artefacts of the approach.

This  behavior is in fact similar to the one that appears when applying the IAM in  Chiral Perturbation Theory for meson-meson scattering at around an Adler zero \cite{Oller:1998hw,GomezNicola:2007qj,Salas-Bernardez:2020hua,Escudero-Pedrosa:2020rwb}. Around this region, the leading $T_1$ crosses zero and the subleading $T_2$ is then larger, such that $T_1-T_2$ becomes zero at some point in those energies. This translates in a deficient behavior of the IAM in a region around the Adler zero until $|T_1|$ becomes clearly larger again than $|T_2|$.  However, there is a difference between this situation and our problem at hand on Coulomb scattering. Now, $|T_1|<|T_2|$ remains so for large enough $p$ up to infinity, as just discussed, so that a proper behavior for the unitarized PWA by employing  IAM cannot be restored for large momenta when $T_1$ and $\Re T_2$ have the same sign.

\subsection{First-iterated $N/D$ method}
\label{sec.240128.4}

To end having unitarized PWAs with the right analytical properties, we apply the first iterated $N/D$ method \cite{Oller:2019opk,Oller:2019rej,Oller:2020guq}. The issue with the $g$-method and IAM is that they drive to structures in which terms with RHC and LHC are directly multiplied.\footnote{For instance, these are not the right analytical properties of a triangular loop diagram \cite{Eden:1966dnq}.}  These deficiencies manifest when looking for the poles in the unitarized amplitudes following these methods. In particular, complex poles with momentum having positive imaginary part and nonvanishing real part are found in the first or physical Riemann sheet, which is against the due properties of a PWA, cf. Eq.~\eqref{240211.2} below. 
 The right secular equation for looking for poles in the PWAs should only have RHC. In the $N/D$ method they correspond to zeros of the $D$ function, while in the general theory of integral equations for the Lippmann-Schwinger equation one has to look for the zeroes of the Fredholm determinant \cite{Tricomi85,Hernandez:1984zzb}.

In the first-iterated $N/D$ method the function $\cV^{(J,\ell)}(p^2)$, having only LHC and to be fixed below, is taken as the $N$ function. Then, the unitarized PWA and the resulting $D$ function, $D^{(J,\ell)}(p^2)$, are given by the expressions, 
\begin{align}
  \label{240122.3}
  T^{(J,\ell)}(p^2)&=\frac{\cV^{(J,\ell)}(p^2)}{D^{(J,\ell)}(p^2)}~,\\
  \label{240122.2}
D^{(J,\ell)}(p^2)&=1-\frac{1}{4\pi^2}\int_{0}^\infty dk^2\frac{k \cV^{(J,\ell)}(k^2)}{k^2-p^2}~.
\end{align}
 Notice that $D^{(J,\ell)}(p^2)$ has only RHC.  Equation \eqref{240122.2} resembles to the denominator in the last term to the right  of Eq.~\eqref{240122.1}, the difference is that now the analogous to the unitarization kernel $V^{(J,\ell)}(p^2)$ is integrated. 

The function $\cV^{(J,\ell)}(p^2)$ is expressed as a series expansion in powers of $\alpha$ as
\begin{align}
  \label{240128.1}
\cV^{(J,\ell)}(p^2)&=\sum_{n=1}^\infty \cV^{(J,\ell)}_n(p^2)\alpha^n~,
\end{align}
where the $\cV_n^{(J,\ell)}$ are determined 
by requiring the matching of Eq.~\eqref{240122.3} order by order with the perturbative calculation in QED of the PWA. Therefore,
\begin{align}
  \label{240128.2}
\cV_1^{(J,\ell)}(p^2)&=T_1^{(J,\ell)}(p^2)~,\\
  \label{240128.3}
\cV_2^{(J,\ell)}(p^2)&=T_2^{(J,\ell)}(p^2)-\frac{T_1^{(J,\ell)}(p^2)}{4\pi^2}\int_{0}^\infty dk^2\frac{k T_1^{(J,\ell)}(k^2)}{k^2-p^2}~.
\end{align}
We notice that because of perturbative unitarity, cf. Eq.~\eqref{240212.2},  $\cV_2^{(J,\ell)}$ is real for $p^2>0$, cancelling mutually the imaginary parts in $T_2^{(J,\ell)}$ and the integral in Eq.~\eqref{240128.3}. 

At this stage Eq.~\eqref{240128.3} is rather formal since further discussions are necessary because the RHC integration for calculating $D^{(J,\ell)}(p^2)$ is divergent and, at least, one subtraction is needed to make it convergent. 
Thus, the unitarization of the LO QED amplitude takes the form:
\begin{align}
  \label{240131.4}
  T^{(J,\ell)}(p^2)&=\frac{\cV_1^{(J,\ell)}(p^2)}{D_{{\rm LO}}^{(J,\ell)}(p^2)}~,\\
  \label{240131.4b}
D^{(J,\ell)}_{{\rm LO}}(p^2)&=1+C_{\rm LO}^{(J,\ell)}-\frac{p^2+p_0^2}{4\pi^2}\int_0^\infty dk^2\frac{k \cV_1^{(J,\ell)}(k^2)}{(k^2-p^2)(k^2+p_0^2)}~.
\end{align}
Within the iterated $N/D$ method we can keep fixed to 1 the constant term for $p\to 0$ in $D^{(J,\ell)}(p^2)$. Then, we remove  the extra constant term arising from the real part of the integral for $p\to 0$ by properly choosing the constant $C_{\rm LO}^{(J,\ell)}$,
    \begin{align}
      C_{\rm LO}&=\lim_{p^2\to 0}\frac{p_0^2+p^2}{4\pi^2}\dashint_0^\infty dk^2\frac{k \cV_1^{(J,\ell)}(k^2)}{(k^2-p^2)(k^2+p_0^2)}~,
    \end{align}
with $\dashint$ denoting the Cauchy principal value of the integral. As a result the $p_0^2$ dependence in $D_{\rm LO}^{(J,\ell)}(p^2)$ is fully removed.     
This is a legitimate process 
because this reshuffling in the subtraction constant is finally reabsorbed in the LHC function $\cV^{(J,\ell)}$ during the matching procedure in order to reproduce the perturbative QED PWA calculated in higher orders. Our results are more accurate the higher the order in which $T^{(J,\ell)}$ is calculated in QED. 

     This treatment, namely, to keep 1 the constant term in $D^{(J,\ell)}$ for $p\to 0$, is also applied to the contributions of $D^{(J,\ell)}(p^2)$ stemming from the integration of $\cV_2^{(J,\ell)}(p^2)$.  Once this constant term is removed there is only $p_0^2$ dependence left from the RHC integration involving the real part of $T_2^{(J,\ell)}$ that stems from the QED diagram representing the first iteration of LO QED amplitude, diagram (f)  in Fig.~\ref{Fig:todos}. These are pure relativistic terms which are zero in the nonrelativistic limit \cite{Oller:2022tmo}, where the NLO QED amplitude is purely imaginary.\footnote{Notice that if the leading nonrelativistic contribution to the first iteration amplitude is ${\cal O}(\alpha^2 p^{-3})$ (as required by unitarity) then its relativistic correction is expected to be an order $p^2/m^2$ higher, so that it would start at ${\cal O}(\alpha^2 p^{-1})$, as it is indeed the case.} Now, the point we want to comment upon here is that  this term, $\cV_{\rm it}^{(J,1/2)}(p^2)$, is given by $\alpha^2\pi^2C_{J\ell}/p$, with $C_{J\ell}$ a PWA dependent constant ($C_{\frac{1}{2}0}=1$), so that its  RHC integration contributing to $D^{(J,\ell)}$ is
\begin{align}
\label{240213.1}
\delta D_{\rm it}^{(J,\ell)}(p^2)&=  -\frac{(p^2+p_0^2)}{4\pi^2}\int_0^\infty dk^2\frac{k \cV_{\rm it}^{(J,\ell)}(k^2)}{(k^2-p^2)(k^2+p_0^2)}=\frac{\alpha^2 C_{J\ell}}{4}\ln \frac{-p^2}{p_0^2}~.
\end{align}
For such a case we take $p_0^2=m^2$, since any change of subtraction point $p_0^2$ just adds a constant that is again removed by the redefinition of $\cV^{(J,\ell)}$ at higher orders.\footnote{The factor in front of $\ln(-p^2/p_0^2)$ in Eq.~\eqref{240213.1} is fixed by unitarity.} This value for the subtraction point is taken because the scale that settles the relativistic expansion is the electron mass $m$. 
 We show the variation in the results due to this $p_0^2$ dependence left by considering values of $p_0^2\lesssim m^2$, namely, for $p_0^2=m^2/4$ and $m^2$. It is remarkable that for scattering this variation will be shown to be very small, or even  negligible for momenta clearly below or above $p=m$.

A remark is in order for the unitarization  of the other terms in $\cV_2^{(J,\ell)}(p^2)$  that arise from the vacuum polarization ($\cV_{\rm vc}^{(J,\ell)}$), the dressed vertex ($\cV_{\rm vx}^{(J,\ell)}$), and the principal value part of the integration present in the right-hand side of Eq.~\eqref{240128.3}  ($\cV_{D}^{(J,\ell)}$). An interesting fact  is that  these functions are smooth in the threshold region, and admit a Taylor expansion in powers of $p^2$ around threshold, with higher orders suppressed by negative powers of  $m^2$. Then, for $|p^2|\ll m^2$ the integration
\begin{align}
  \label{240131.1}
  -\frac{p^2}{4\pi^2}\int_0^\infty dk^2\frac{k\left[\cV_{\rm vc}^{(J,\ell)}(k^2)+\cV_{\rm vx}^{(J,\ell)}(k^2)+ \cV_{\rm D}^{(J,\ell)}(k^2)\right]}{(k^2-p^2)k^2}&\approx
  -\left[\cV_{\rm vc}^{(J,\ell)}(0)+\cV_{\rm vx}^{(J,\ell)}(0)+\cV_{D}^{(J,\ell)}(0)\right]\frac{p^2}{4\pi^2}\int_0^\infty dk^2\frac{k}{k^2(k^2-p^2)} \\
&  \to \left[\cV_{\rm vc}^{(J,\ell)}(0)+\cV_{\rm vx}^{(J,\ell)}(0)+\cV_D^{(J,\ell)}(0)\right]g(p^2)~.\nn
\end{align}
This makes that for the study of the contributions of these terms 
 to the binding momenta ($p_b$)  of the hydrogen atom one gets the same results either from the first-iterated $N/D$ method or from the $g$-method.

 Then, we have the following final expression for these contributions to $D^{(J,\ell)}(p^2)$, 
\begin{align}
  \label{240131.2}
\delta D_{\rm vc}^{(J,\ell)}+\delta D_{\rm vx}^{(J,\ell)}+\delta D_{D}^{(J,\ell)}&=-\frac{p^2}{4\pi^2}\int_0^\infty dk^2\frac{k\left[\cV_{\rm vc}^{(J,\ell)}(k^2)+\cV_{\rm vx}^{(J,\ell)}(k^2)+\cV_{\rm D}^{(J,\ell)}(k^2)\right]}{(k^2-p^2)k^2}~.
\end{align}

All in all, the unitarized PWA obtained from the first-iterated $N/D$ method applied to the perturbative QED amplitude calculated up to  NLO is
\begin{align}
  \label{240131.6}
T^{(J,\ell)}(p^2)&=\frac{\cV_1^{(J,\ell)}(p^2)+\cV_2^{(J,\ell)}(p^2)}{D_{{\rm LO}}^{(J,\ell)}(p^2)+\delta D_{{\rm it}}^{(J,\ell)}(p^2)+\delta D_{{\rm vc}}^{(J,\ell)}(p^2)+\delta D_{{\rm vx}}^{(J,\ell)}(p^2)+\delta D_{D}^{(J,\ell)}(p^2)}~,
\end{align}
where
\begin{align}
  \label{240131.7}
  \cV_{2}{(J,\ell)}(p^2)&=T_2^{(J,\ell)}
 -\cV_1^{(J,\ell)}\left\{\frac{p^2+p_0^2}{4\pi^2}\int_0^\infty dk^2\frac{k \cV_1^{(J,\ell)}(k^2)}{(k^2-p^2)(k^2+p_0^2)}
  -\displaystyle{\lim_{q^2\to 0}}\frac{q^2+p_0^2}{4\pi^2}\dashint_0^\infty dk^2\frac{k \cV_1^{(J,\ell)}(k^2)}{(k^2-q^2)(k^2+p_0^2)} \right\}~.
\end{align}

The function $D^{(J,\ell)}(p^2)$ is $D_{\rm LO}^{(J,\ell)}+D_{{\rm NLO}}^{(J,\ell)}$, with 
\begin{align}
  \label{240210.2}
D_{{\rm NLO}}^{(J,\ell)}(p^2)&=\delta D_{{\rm it}}^{(J,\ell)}(p^2)+\delta D_{{\rm vc}}^{(J,\ell)}(p^2)+\delta D_{{\rm vx}}^{(J,\ell)}(p^2)+\delta D_{D}^{(J,\ell)}(p^2)~.
\end{align}



To simplify the writing we introduce the following notation: The unitarized PWAs obtained by applying Eqs.~\eqref{240122.1}, \eqref{240128.iam}, and \eqref{240122.3} 
are denoted by $T^{(J,\ell)}_{\rm{on}}(p^2)$, $T^{(J,\ell)}_{\rm{IAM}}(p^2)$ and $T^{(J,\ell)}_{\rm{nd}}(p^2)$, respectively. When needed, the order of the perturbative QED PWA unitarized is indicated as a subscript preceded by a semicolon. 

\section{Scattering in partial waves}
\label{sec.240128.5}
\def\theequation{\arabic{section}.\arabic{equation}}
\setcounter{equation}{0}   

\jo{Nominally, the} scale $\Lambda$ separates ``soft'' photons from ``hard'' ones, since the resummation of soft photons in Ref.~\cite{Weinberg:1965nx} was undertaken for $\Lambda$  smaller (though not necessarily much smaller) than the typical momentum in the transition process. Following Refs.~\cite{Oller:2022tmo,Blas:2020dyg} we write it as 
\begin{align}
  \label{240723.1}
  \Lambda=\frac{2p}{a}
  ~.
\end{align}
For the case of Coulomb scattering  $\ln a=\gamma_E$. For  the nonrelativistic case this value can be derived rigorously \cite{Blas:2020dyg,Oller:2022tmo}, for instance, by comparison  with the known solution for Coulomb scattering  obtained in Quantum Mechanics  \cite{Kang:1962}
\begin{align}
  \label{220710.1}
  S^{(\ell)}_{\rm nr}(p)&=\frac{\Gamma(\ell+1-i \gamma)}{\Gamma(\ell+1+i \gamma)}
  \,,\\
  T^{(\ell)}_{\rm nr}(p)&=\frac{2\pi}{i  p}\left(S^{(\ell)}_{\rm nr}(p)-1\right)~,\nn
\end{align}
with $\gamma=-ee'/(4\pi p)$. 
By comparing with Eq.~\eqref{240122.1}  the exact expression for $V^{(\ell)}(p)$ to all orders can be deduced
\begin{align}
  \label{240123.1}
  V^{(\ell)}_{\rm nr}(p)&=-i\frac{4\pi}{p}\frac{\Gamma(1+\ell-i\gamma)-\Gamma(1+\ell+i\gamma)}{\Gamma(1+\ell-i\gamma)+\Gamma(1+\ell+i\gamma)}~.
\end{align}
 where we have suppressed the superscript $J$ because in the nonrelativistic case $V^{(J,\ell)}$ only depends on $\ell$.  
Its perturbative expansion in powers of $\gamma$ gives
\begin{align}
  \label{220710.4}
V^{(\ell)}_{\rm nr}(p)&=\frac{me e'}{p^2}\psi_0(1+\ell)+{\cal O}(\alpha^3)~,
\end{align}
where $\psi_n(z)=d^{n+1}\log\Gamma(z)/dz^{n+1}$ and $\alpha$ is the fine structure constant.\footnote{There is an explicit formula for $\psi_0(n)=-\gamma_E+\theta(n-1)\sum_{k=1}^{n-1}1/k$\,.} In particular for $\ell=0$ we find $ V_{\rm nr}^{(0)}(p)=-e e'm\gamma_E/p^2+{\cal O}(\alpha^3)$. 
Applying the formalism of Sec.~\ref{sec.240322.4} we end with the expression for $\ell=0$ and $J=1/2$, $V^{(1/2,0)}=-ee' m \ln a/p^2+{\cal O}(\alpha^3,p^2/m^2)$, so  it follows that $\ln a=\gamma_E$. 
If we proceed similarly for the $P$ waves we have from the expansion of Eq.~\eqref{240123.1} that $V_{\rm nr}^{(1)}= -ee'(-1+\gamma_E)m/p^2+{\cal O}(\alpha^3)$, while our calculation of the LO PWAs in the nonrelativistic limit ($\ell=1$ and $J=3/2$ o $1/2$) gives  $V^{(J,1)}=-ee'(-1+\ln a)m/p^2+{\cal O}(\alpha^3,p^2/m^2)$, and the same result  $\ln a=\gamma_E$ is obtained.

This result can be also derived in terms of the scattering amplitude calculated within the small-angle expansion in QED. The scattering amplitude of two charged spinless particles was calculated  in  Ref.~\cite{Bazhanov:1977fa} up to NLO in the small scale expansion (or expansion in $t^{1/2}$), and to all orders in the fine-structure constant. The resulting scattering amplitude obtained in this reference (its equations (21) and (29)) has the extra phase factor $(4p^2/\lambda^2)^{-iee'/4\pi v}\exp(iee'\gamma_E/2\pi v)$,\footnote{The factor $4p^2$ comes from $-t=4p^2 \sin^2\frac{\theta}{2}$, appearing within the global factor $(-t/\lambda^2)^{-iee'/2\pi v}$ in the expression for the scattering amplitude in several equations in Ref.~\cite{Bazhanov:1977fa}.} where $v$ is the velocity in the laboratory frame.  More conveniently for our purposes, this factor can be recast as
\begin{align}
  \exp\left( i\frac{ee'E(p)}{2\pi p} (\gamma_E+\ln \frac{\lambda}{2p})\right)
  =\exp\left( i\frac{ee'E(p)}{2\pi p}\ln\frac{e^{\gamma_E} \lambda}{2p}\right)\,.
\end{align}
Comparing it with Eq.~\eqref{240317.1}, we immediately read that $\Lambda=2p/a$ with $a=\exp\gamma_E$, as we wanted to show. Notice also that this result is valid at once for all the $(J,\ell)$ PWAs, which justifies why above we got it for $S$ and $P$ waves.

\subsection{Discussion}
\label{sec.240930.1}

The value for $\Lambda$ given in Eq.~\eqref{240723.1} is then used for the calculation of Eq.~\eqref{230901.1}, corresponding to $A_2+A_4$. It is also used for the evaluation of the PWAs as detailed in Sec.~\ref{sec.140413.1}, by accounting for the redefinition of the $S$ matrix that allows one to calculate the partial-wave decomposition for the iteration diagram $A_3$ given in Eq.~\eqref{Eq:A3}. In deducing the expressions for $A_4$ and $S_W$ in Eqs.~\eqref{240316.1b} and \eqref{240317.1}, respectively, one typically uses approximations that are valid in the limit of small photon momenta in the involved loops, as we have showed in Sec.~\ref{sec.240316.1} for $A_4$, and detailed in Ref.~\cite{Weinberg:1965nx} for $S_W$. For $A_4$ the momentum flowing to the external field $A_\mu$ in the diagrams (b)--(d) of Fig.~\ref{fig.240316.1}  is actually $\vq- \vk$ for the diagrams (b)--(c) and $\vq+\vk$ for (d) and (e),  with $\vk$ the photon momentum integrated to calculate $R_f$, Eq.~\eqref{240310.1}. From this point of view, the factorization of $A_0$, taking $A_\mu(\vq^2)$ out of the integration in $\vk$ as in Eq.~\eqref{240316.6} for $A_4$, is strictly  valid when $\Lambda^2\ll \vp^2$. At this point, let us stress  that our calculation for $A_4$ in Sec.~\ref{sec.240316.1} follows the lines of the original Chung's calculation, such that the same approximations as in Ref.~\cite{Chung:1965zza}  are done, which obtains the same result.  Similarly, the expression for $S_W$ worked out by Weinberg in Ref.~\cite{Weinberg:1965nx}, and independently redone by us, implies to neglect in the photon propagator factors of $k$ in its numerator and factors of $k^2$ in its denominator, assuming 
that $\Lambda^2 \ll \vp^2$. 

Indeed, our calculated $A_4$, following the Chung's formalism \cite{Chung:1965zza}, and Weinberg's result in Ref.~\cite{Weinberg:1965nx} are in a closer connection than its separate exposition up to now could resemble. 
It is a  matter of fact that  $A_4$ in Eq.~\eqref{240316.6} can be obtained more straightforwardly by applying the formalism of Ref.~\cite{Weinberg:1965nx}. Following this reference, and taking into account also Eq.~\eqref{240317.1} for the imaginary part of the exponent,  the relation is 
\begin{align}
  \label{240930.1}
  S&=\widetilde{S}_F\exp\left(-\Phi-i\frac{ee'E(p)}{2\pi p}\ln\frac{\lambda}{\Lambda}\right)\,.
\end{align}
 This equations  is obtained in Ref.~\cite{Weinberg:1965nx} by resumming the exchange of virtual soft photons to all orders in powers of the electric charge. Here $\widetilde{S}_F$ is the IR-divergent $S$-matrix obtained by directly applying the QED rules without dressing the asymptotic states a la Chung or FK. Its evaluation to NLO is discussed in Sec.~\ref{sec.240322.1}.   
 The loop function $\Phi$, corresponding to the exponent in Eq.~(2.11) of Ref.~\cite{Weinberg:1965nx}, is given by
\begin{align}
\label{240927.2}
\Phi&=\frac{ie^2}{2}\int^\Lambda\frac{d^4k}{(2\pi)^4(k^2-\lambda^2+i\ep)}\left(\frac{m^2}{(p\cdot k)(p \cdot k)}-\frac{p\cdot p'}{(p\cdot k)(p'\cdot k)}\right)\\
&=-\frac{e^2}{4\pi^2}\left(-1+\frac{2m^2-t}{t\sigma(t)}\ln\frac{\sigma(t)-1}{\sigma(t)+t}\right)\left(\ln\frac{2\Lambda}{\lambda}-1\right)\,.\nn
\end{align}
This exponential factor, when included in Eq.~\eqref{240930.1}, gives   the  ${\cal O}(e^4)$ contribution $ -A_0 \Phi$ that is the same as $A_4$ in Eq.~\eqref{240316.6}.

As a result, we can reinterpret the results in Sec.~\ref{sec.240316.1}, obtained in terms of Chung's states, as stemming from $\exp(-\Phi)$ in Eq.~\eqref{240930.1}.  In this way, the factorization of $A_0$ in Eq.~\eqref{240316.6} is not an issue, it arises by construction.  One should stress that this equation provides  a definition of a new  $S$ matrix that is unitary and free of infrared divergences independently of the explicit numerical value used for $\Lambda$. This is a 
crucial point in order to apply our approach based on unitarization. This unitarity character has been explicitly checked by us up to NLO for any $\Lambda$, calculating the PWAs following Sec.~\ref{sec.240322.4}. Importantly, the relation between the IR-divergent and IR-finite $S$-matrices in Eq.~\eqref{240930.1} is set through the exponential factor $\displaystyle{\exp\left(-\Phi-i\frac{ee'E(p)}{2\pi p}\ln\frac{\lambda}{\Lambda}\right)}$, which has no poles. Therefore, we can use the new unitary and IR-finite $S$-matrix to search for its poles (corresponding to bound states, resonances and virtual states) and, in this way, study the related spectroscopy to Coulomb scattering. 
 This prospect is applied below in  Secs.~\ref{sec.231221.2} and \ref{sec.240322.5}.

The previous conclusion is independent of the sign and strength of $ee'$, so that one can transit from different types of spectra, and its validity always holds. Therefore, one can confidently  conclude that the interaction kernel $V^{(J,\ell)}(p^2)$, introduced in Eq.~\eqref{231221.3}, with $g(p^2)$ given in Eq.~\eqref{240121.2}, is properly constructed.  As a result, the $T$-matrix in PWAs given in Eq.~\eqref{240122.1} from the calculated $V^{(J,\ell)}(p^2)$ is the adequate or physical one.  For instance, for nonrelativistic Coulomb scattering the $S$-matrix is well-known from Quantum Mechanics \cite{Kang:1962}, and $V^{(J,\ell)}$ up to ${\cal O}(e^2)$ is shown to be properly calculated by one of the authors in Ref.~\cite{Oller:2022tmo}. Similarly, it is also shown how the exact  PWAs are reproduced with increasing accuracy as the order in the calculation of $V^{(J,\ell)}$ increases.

Our discussion here establishes a general procedure that, from the practical point of view, avoids the explicit construction of Chung's and, more generally, of FK states in order to handle with the IR divergences. The unitary and IR-finite $S$ matrix  can be obtained from the $S$ matrix directly obtained by applying standard Feynman diagrams in QED  by its multiplication with Weinberg's exponential factor. The latter is generally denoted by  $\exp(-\Phi_{\beta\alpha})$ for the scattering process $\alpha\to \beta$. Precisely, when taking only the imaginary part of $\Phi_{\beta\alpha}$ we have the redefinition of the $S$ matrix, cf. Eq.~\eqref{240317.4}, while the real part of $\Phi_{\beta\alpha}$ would account for the removal of IR divergences accomplished by the introduction of the Chung's or FK states. Interestingly, this new point of view in applying Weinberg's results in \cite{Weinberg:1965nx} can be easily translated to gravity as well \cite{Choi:2017bna}. We  will further elaborate on this method in \cite{Oller:2024neq}.

\subsection{$J=1/2$ in $S$-wave}
\label{sec.240202.1}

\begin{figure}[h]
  \begin{center}
    \includegraphics[width=0.7\textwidth,angle=0]{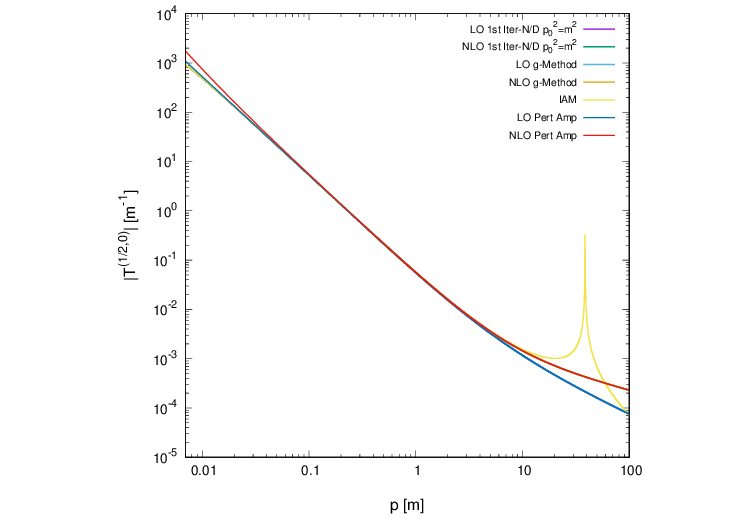}
    \caption{{\small $|T^{(\frac{1}{2},0)}(p^2)|$ as a function of $p$ in units of $m^{-1}$ and $m$, respectively, in log scale.  We show the results from the LO and NLO $g$-method and 1st-iterated $N/D$ methods,  the IAM, and the LO and NLO QED perturbative PWAs. The types of lines are explained in the keys of the figure. The IAM result shows a fake resonant structure for large $p$, which does not correspond to any pole in the 2nd RS.} \label{fig.240202.1}}
    \end{center}
\end{figure}

We show in Fig.~\ref{fig.240202.1} the modulus of $T^{(\frac{1}{2},0)}(p^2)$ as a function of $p>0$ up to $100 m$ in log scale. Energy units are taken such that $m=1$. There we show both the unitarized PWA with different methods and the perturbative QED calculations at LO and NLO. The curve corresponding to each case is indicated by the keys in the figure. Namely, the 1st-iterated $N/D$ method for unitarizing the LO and NLO QED amplitudes correspond to the magenta and green lines, respectively. The  $g$-method for unitarizing the  LO and NLO QED amplitudes are the light blue and orange lines, in this order. The IAM is the yellow line, and the LO and NLO QED perturbative PWAs are the blue and red lines, respectively.

Regarding the perturbative amplitudes we see that they depart from each other for small momenta, $p\lesssim 0.1~m$, and for large ones, $p\gtrsim 6~m$. It is interesting to discuss the reasons involved. For the small momenta  the point is that for $|p|={\cal O}( m\alpha)$ Coulomb scattering becomes nonperturbative, with $im\alpha$ the binding momentum of the hydrogen atom, and the perturbative expansion breaks down as $p\to 0$. The interested reader can consult Fig.~1 of Ref.~\cite{Oller:2022tmo} where this point is explicitly shown for nonrelativistic scattering. For the large $p$ part the issue there is, as already pointed out in Sec.~\ref{sec.240128.3}, that $T_1^{(\frac{1}{2},0)}(p^2)$ vanishes as $1/p$ for $p\to \infty$, while  $T_2^{(\frac{1}{2},0)}$ is asymptotically dominated in this limit by the contribution from the dressed electromagnetic vertex of the electron, because despite it also decreases like $1/p$ its size is relatively enhanced by  factors $\ln^2 (p/m)^2$ [coming from the function $Li_2$ in Eq.\eqref{230901.1}]. This is not a particularity of $J=1/2$ in $S$-wave, but it applies to any PWA. In this way, for $p\sim m/\alpha$ this ultraviolet enhancement is large enough to compensate the extra factor of $\alpha$ of $T_2^{(J,\ell)}$ relative to $T_1^{(J,\ell)}$, and $T_2^{(J,\ell)}$ starts becoming larger in absolute value than $|T_1^{(J,\ell)}|$ as $p$ increases.

Concerning the unitarization methods we would like to stress two points:  i) The different unitarization results follow closely the perturbative QED amplitudes, such that they cannot typically be distinguished in the figure, except for small $p\lesssim 5 m\alpha$. It is worth noting that there the results from the three unitarization methods overlap each other, both by unitarizing the LO and NLO perturbative QED amplitudes, and they cannot be distinguished in the scale of the figure. However, the blue and red lines corresponding to the LO and NLO QED perturbative amplitudes, respectively, can be distinguished between them and with respect to the unitarized results for the small momentum region.
ii) The figure clearly shows the problem of the IAM at large momenta, as already anticipated in Sec.~\ref{sec.240128.3}. The peak occurs at $p\simeq 38.5~m$ because of the cancelation of $T_1^{(\frac{1}{2},0)}-\Re T_2^{(\frac{1}{2},0)}$ at this crossover point. The peak in Fig.~\ref{fig.240202.1} does not correspond to any resonance manifested as a pole in the complex $p^2$ plane in the second Riemann sheet, but it is just an artifact of the method. 

\begin{figure}[h]
  \begin{center}
    \includegraphics[width=0.7\textwidth,angle=0]{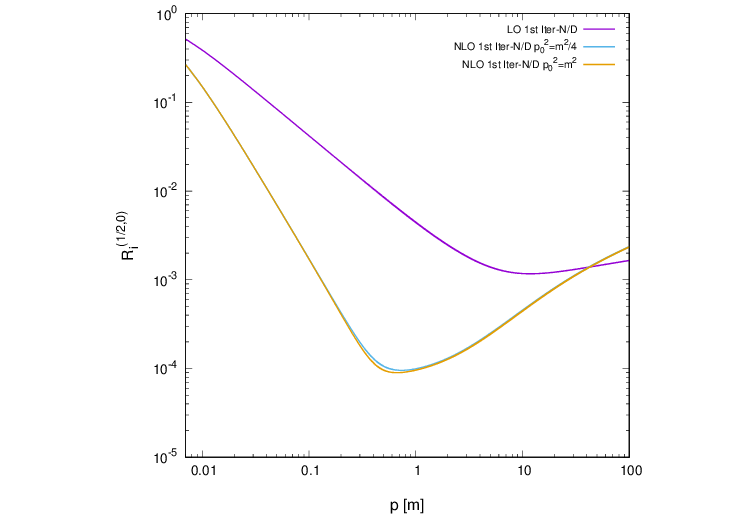}
    \caption{{\small We plot in log scale  $R_i$ defined in Eq.~\eqref{240207.1} as a function of $p$ (in units of $m$) for the unitarization with the 1st-iterated $N/D$ method of the LO and NLO QED amplitudes. For the latter  case we plot the results for  $p_0^2=m^2$ and $m^2/4$. The types of lines are explained in the keys of the figure.} \label{fig.240207.1}}
    \end{center}
\end{figure}

In order to ascertain the impact of the unitarization of the perturbative QED PWAs $T^{(\frac{1}{2},0)}_i(p^2)$, we define the ratio $R_i^{(J,\ell)}$ as
\begin{align}
  \label{240207.1}
  R^{(J,\ell)}_{i}&=\left|\frac{T^{(J,\ell)}_{{\rm nd};i}}{\displaystyle{\sum_{j=1}^i T^{(J,\ell)}_{j}}}-1\right|~.
\end{align}
This is shown in Fig.~\ref{fig.240207.1} for the unitarization of the LO ($i=1$) and NLO ($i=2$) perturbative QED $S$-wave  by employing the first-iterated $N/D$ method. We also show the results for two values of the subtraction point $p_0^2$ at $m^2$ and $m^2/4$. Unitarization corrections to the LO QED PWA at typical  momenta  $p={\cal O}(m)$ are ${\cal O}(\alpha^2)$ in absolute terms, and  ${\cal O}(\alpha)$ in relative ones. This sets the size  of the magenta line in Fig.~\ref{fig.240207.1} for momenta $p={\cal O}(m)$.  For the unitarization of the NLO QED PWA the corrections  with respect to the NLO QED PWA are of ${\cal O}(\alpha^3)$ and ${\cal O}(\alpha^2)$ in absolute and relative terms, respectively. This  sets the  size of the unitarization corrections for the blue and orange lines in Fig.~\ref{fig.240207.1} for typical momenta of ${\cal O}(m)$. In more detail, for the NLO case 
 we can write
\begin{align}
  \label{240209.1}
  R_2&=\left|\frac{\cV_1^{(J,\ell)}+\cV_2^{(J,\ell)}-(T_1^{(J,\ell)}+T_2^{(J,\ell)})D^{(J,\ell)}}{(T_1^{(J,\ell)}+T_2^{(J,\ell)})D^{(J,\ell)}}\right|={\cal O}(\alpha^3)~,
\end{align}
because of the matching procedure in Eq.~\eqref{240122.2}.

However, the corrections are much larger  for $p={\cal O}( m\alpha)$ because QED is nonperturbative there and, therefore, the largest corrections are precisely expected at such momenta.\footnote{Notice that in addition to the nominal global power in $\alpha$ corresponding to $T_1^{(J,\ell)}$ and $T_2^{(J,\ell)}$, the momenta themselves carry powers of $\alpha$ when $p={\cal O}(m\alpha)$. } On the other side, for large momenta $p={\cal O}(m/\alpha)$, the unitarity corrections to the NLO QED PWAs grow and become comparable to the corrections of the LO QED PWA. This is due to the aforementioned faster decrease  for $p\to\infty$ of the LO amplitude at higher momentum compared with the NLO QED one. It is also worth stressing that the left dependence of the results on $p_0^2$  due to the unitarization of $\cV_{\rm it}^{(\frac{1}{2},0)}$ is already barely visible in Fig.~\ref{fig.240207.1}, and it can be neglected in front of the rest of contributions.

\subsection{$J=3/2$ and  $J=1/2$ in $P$-waves}
\label{sec.240202.2}

\begin{figure}[h]
  \begin{center}
\begin{tabular}{ll}
    \hspace{-1.5cm} \includegraphics[width=0.6\textwidth,angle=0]{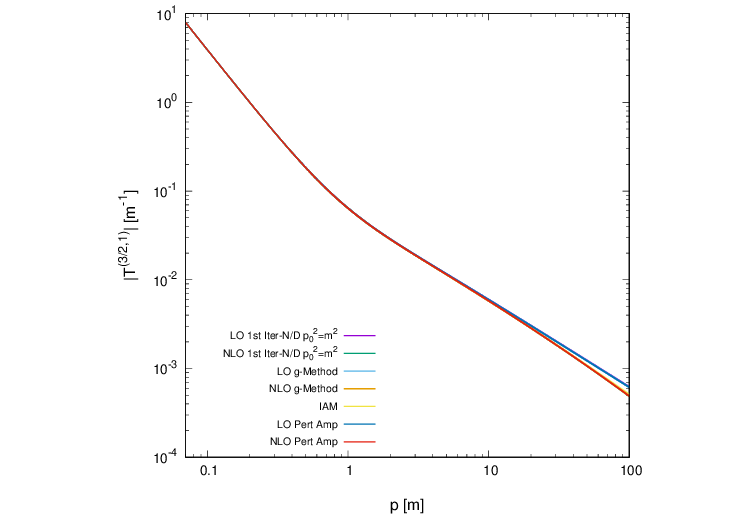} &
   \hspace{-2.5cm} \includegraphics[width=0.6\textwidth,angle=0]{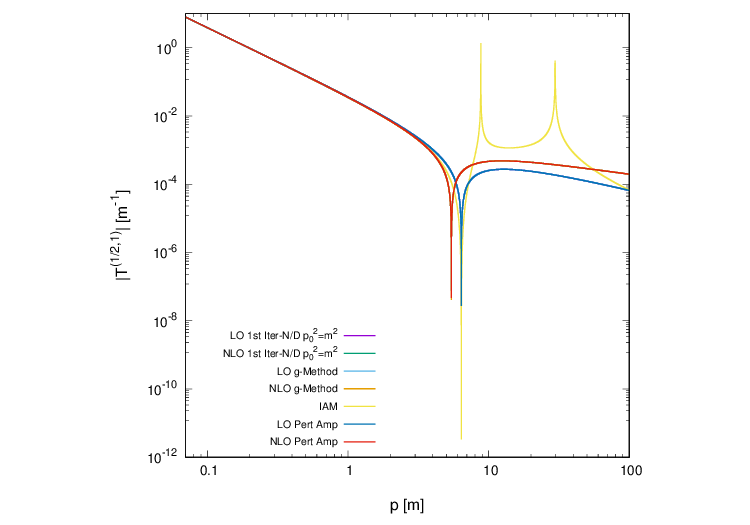}
\end{tabular}
    \caption{{\small $|T^{(3/2,1)}(p^2)|$ (left panel) and $|T^{(1/2,1)}(p^2)|$ (right panel) are plotted in log scale as a function of $p$ in units of $m^{-1}$ and $m$, respectively.  We show the LO and NLO $g$- and 1st-iterated $N/D$ methods,  the IAM and the LO and NLO QED perturbative PWAs.  The types of lines are explained in the keys of the figure, being the same as in Fig.~\ref{fig.240202.1}.} \label{fig.240208.1}}
    \end{center}
\end{figure}

For $P$-wave scattering there are two possible values for the total angular momentum, $J=3/2$ and $J=1/2$.  
We proceed along similar lines as for the $S$-wave.  First, we show  in Fig.~\ref{fig.240208.1} the modulus of $T^{(J,1)}(p^2)$  for $J=3/2$ (left panel) and $J=1/2$ (right panel), as resulting from the application of the unitarization methods and the pure perturbative calculations. We follow the same convention for the type of lines as in Fig.~\ref{fig.240202.1}. The basic features that one can observe in the Fig.~\ref{fig.240208.1} for the $P$ waves are  similar to those for the $S$ wave. The $g$-method and the 1st-iterated $N/D$ method provide unitarized results that cannot be distinguished from the perturbative PWAs in the scale of the figures. The LO and NLO QED PWAs can only be distinguished in the large momentum region because of the faster vanishing for large of $p$ of the LO result. For $J=1/2$ the difference between the red and blue lines is more visible in the log scale used in the figure because the displacement in the position of the zero of the LO QED PWA  at $6.4 m$ and the one in the real part of the NLO QED PWA at $5.4 m$.\footnote{Let us indicate that its imaginary part does not vanish because it is given by unitarity, and $T_1^{(\frac{1}{2},1)}$ is not zero at $p\approx 5.4 m$ but at $p\approx 6.4m$. It is however very small because $T_1^{(\frac{1}{2},1)}$ is suppressed due to its nearby zero.} The application of the IAM in the $J=1/2$ case is handicapped because $T_1^{(\frac{1}{2},0)}-\Re T_2^{(\frac{1}{2},0)}$ vanishes at two points with momenta equal to $8.7m$ and $29.6m$. This is not the case for $J=3/2$ since $T_1^{(\frac{3}{2},1)}$ and $T_2^{(\frac{3}{2},1)}$ have opposite signs, and $T_2^{(\frac{3}{2},1)}$ remains very small in comparison with $T_1^{(\frac{3}{2},1)}$ in the range of momentum shown. 

\begin{figure}[h]
  \begin{center}
    \begin{tabular}{ll}
    \hspace{-1.5cm}  \includegraphics[width=0.6\textwidth,angle=0]{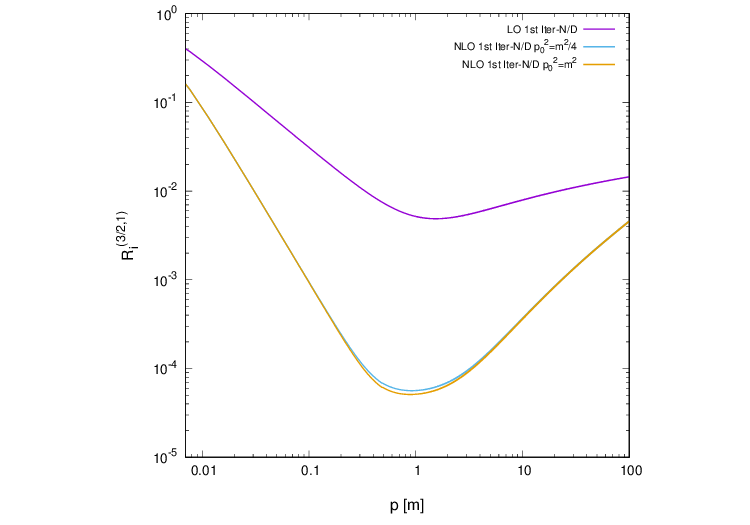} &
   \hspace{-2.5cm}   \includegraphics[width=0.6\textwidth,angle=0]{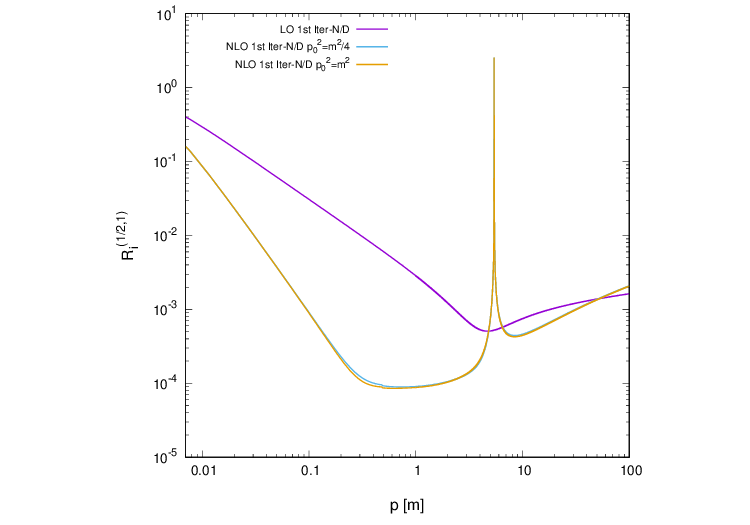}
    \end{tabular}
    \caption{{\small We plot in log scale $R_i^{(3/2,1)}$ (left panel) and $R_i^{(1/2,1)}$ (right panel) as functions of $p$ (in units of $m$) for the unitarization with the 1st-iterated $N/D$ method of the LO and NLO QED amplitudes. For every case we plot the results for two subtractions points $p_0^2=m^2$ and $m^2/4$. The types of lines are explained in the keys of the figure, being the same as in Fig.~\ref{fig.240207.1}.} \label{fig.240208.2}}
    \end{center}
\end{figure}

The ratios $R_i^{(J,1)}$, cf. Eq.~\eqref{240207.1}, are shown in Fig.~\ref{fig.240208.2}, with the lines corresponding to the same casuistry as in Fig.~\ref{fig.240207.1}. From Fig.~\ref{fig.240208.2} we see that the relative size of  the unitarization corrections for typical momenta of ${\cal O}(m)$ are as expected, namely, of ${\cal O}(\alpha)$ and ${\cal O}(\alpha^2)$ for the unitarization of the LO and NLO QED PWAs, in this order. 
 The peak of the relative unitarization corrections in the $J=1/2$ NLO QED PWA just reflects the almost zero of $T_2^{(\frac{1}{2},1)}$ at $p\approx 5.4 m$. However,  $R^{(\frac{1}{2},1)}_1$ is smooth when $T_1^{(\frac{1}{2},1)}$ crosses its zero at $p\approx 5.4 m$ because $\cV_1^{(J,\ell)}=T_1^{(J,\ell)}$, and  the zero at $T_1^{(\frac{1}{2},1)}$ drops in the ratio between the unitarized  and the perturbative LO QED PWA. As in the $S$-wave case, the change in the results from the two values of $p_0^2$ is very small and it can only be disentangled in the figure for $p$ close to $m$. 

\section{The bound-state equation for the different unitarization methods}
\label{sec.231221.2}
\setcounter{equation}{0}   
\def\theequation{\arabic{section}.\arabic{equation}}

\jo{In Quantum Mechanics, the full S matrix for nonrelativisitc Coulomb scattering is exactly solvable.}  In terms of it Ref.~\cite{Oller:2022tmo} gave the exact expression for $V_{\rm nr }^{(\ell)}(p)$, the nonrelativistic counterpart of $V^{(J,\ell)}$ of Eq.~\eqref{231221.3}. \jo{We have followed these lines above in Sec.~\ref{sec.240128.5} giving $V_{\rm nr}^{(\ell)}(p)$ in Eq.~\eqref{240123.1}}. 
The expansion of $V_{\rm nr}^{(\ell)}(p)$ 
in powers of $\alpha$ breaks down  for $|p|<p_{\rm critical}$, with $p_{\rm critical}$ the largest momentum at which the denominator in Eq.~\eqref{240123.1} vanishes. The values of such critical momenta are given in Table~\ref{tab.240123.1} for small $\ell\leq 5$.       For large values of $\ell$ an approximate expression  for $p_{\rm critical}$ can be obtained by expanding the denominator in Eq.~\eqref{240123.1} in powers of $\ell^{-1}$. Its asymptotic behavior for $\ell\gg 1$ is 

    \begin{table}
      \begin{center}
        \begin{tabular}{|r|llllll|}
          \hline
          $\ell$                            & 0   & 1    & 2    & 3    & 4   & 5 \\
          \hline
          &&&&&&\\
          $\frac{p_{\rm critical}}{m\alpha}$ &0.18 & 0.43 & 0.62 & 0.81 & 0.96 & 1.09\\
          &&&&&&\\
          \hline
          \end{tabular}
        \caption{{\small Largest root of the denominator in Eq.~\eqref{240123.1} for $\ell\leq 5$.}
          \label{tab.240123.1}}
        \end{center}
      \end{table}

      \begin{align}
        \label{240123.2}
\Gamma(1+\ell-i\frac{m\alpha}{p})+\Gamma(1+\ell+i\frac{m\alpha}{p})\longrightarrow \sqrt{2\pi l}\exp\left(\ell[-1+\ln \ell]\right)\ell^{-\frac{im \alpha}{p}}\left(1+\ell^{\frac{2im\alpha}{p}}\right)~,
      \end{align}
      from where we deduce that
      \begin{align}
        \label{240123.4}
p_{\rm critical}\to  \frac{2 m\alpha \ln \ell}{\pi}~.
      \end{align}
   Let us notice   that  for $\ell=5$  this formula gives the approximate result of $p_{\rm critical}\approx 1.02 m\alpha$, already rather close to the numerical result in Table~\ref{tab.240123.1}.  We should compare $p_{\rm critical}$ with the modulus of the binding momentum of the nonrelativistic hydrogen atom with  principal quantum number $n$, $im\alpha/n$, such that the angular momentum $\ell$ is within the range $0\leq \ell\leq n-1$. 
      Because the binding momentum decreases in modulus with $\ell$ while $p_{\rm critical}$ increases, as it is clear from Table~\ref{tab.240123.1} and Eq.~\eqref{240123.4}, it follows that $|p_b|>p_{\rm critical}$ only for $\ell<2$.\footnote{In this statement we take into account that QED corrections to the nonrelativistic Quantum Mechanics values of the binding momentum are suppressed by extra powers of $\alpha$ and are small for the physical value of the fine structure constant.}

      For  $P$ wave one needs at least $n=2$, and $1/2$ is quite close to the radius of convergence $p_{\rm critical}/m\alpha=0.43$ for $\ell=1$ in Table~\ref{tab.240123.1}, such that one really needs high orders in the expansion of $V^{(1)}_{\rm nr}$. It is straightforward to check from the expansion of Eq.~\eqref{240123.1} that $V^{(1)}_{\rm nr}$ becomes negative for $p^2<0$ and $|p|\ll m$  (and then able to give rise to a bound state) only when including contributions of ${\cal O}(\alpha^5)$, though higher orders are still necessary to reach convergence. This would require more than 4 loops  in QED, certainly beyond the scope of this study based on unitarized NLO QED. For $S$ wave, in principle, one could study $n\leq 5$ as $p_{\rm critical}=0.18$, cf. Table~\ref{tab.240123.1}. However, the unitarized LO QED amplitude only gives rise to the fundamental state of the hydrogen atom,  and higher-order QED calculations are needed to reach accurateness in the result. Indeed, $p_b$ for the fundamental state  is reproduced with a precision of a few $\permil$ only after including the ${\cal O}(\alpha^3)$ contribution in the expansion of $V_{{\rm nr}}^{(0)}$ (see Table~1 of Ref.~\cite{Oller:2022tmo}). The calculation is more and more demanding as $n$ increases. Nevertheless, it is a compelling  prospect for  future studies the possibility of studying  within this  new formalism  the Lamb shift effect, corresponding to the energy difference between the $^2S_{1/2}$ and $^2P_{1/2}$ energy levels by unitarizing  high-loop calculations in QED. In this respect, a two-loop calculation would be of interest since it is at this order when the self-energy of the electron inside the iteration diagram first appears. This is a key effect for the Lamb shift, as it is well known \cite{Bethe:1947id,Mandl:1985bg,Weinberg:1995mt}.



      \bigskip
      
Next, we explain a procedure to solve the bound-state equation for the $g$-method and the IAM. The $g$-method can be  applied to unitarize the LO and NLO QED PWAs, while  the IAM for being applied requires at least the NLO QED PWAs. We will not treat separately the bound-state equation from the IAM because, as shown below,  
the bound-state equation for the $g$-method  
with a perturbative treatment of the NLO contributions encompasses that of the IAM. 


\bigskip
The bound-state equation from the knowledge of the LO QED PWAs is
\begin{align}
  \label{231224.1}
  \frac{1}{V_1^{(J,\ell)}(p^2)}-i\frac{p}{4\pi}=0~.
\end{align}
In order to handle with  transcendental equations we typically proceed perturbatively expanding the solution in powers of $\alpha$, being the nonrelativistic result  the  leading contribution. This process is illustrated below with explicit examples. 


The total result at NLO QED contains the vacuum polarization, vertex diagrams plus the Chung's diagrams, and the once-time iteration, that also includes the redefinition of the $S$ matrix up to ${\cal O}(\alpha^2)$ to make it IR finite.
Since the NLO QED result is ${\cal O}(\alpha^2)$, its contribution to $V^{(J,\ell)}(p^2)$  can be calculated by applying Eq.~\eqref{231221.7} and  is also ${\cal O}(\alpha^2)$. The partial-wave projection of the vacuum polarization and dressed vertex diagram, including Chung's diagrams, has no  algebraic expression, though it can be done numerically. However, their expansion at low momentum can be algebraically worked out order by order in a power series in $p^2$, which is precisely what we need in order to implement a  perturbative treatment  for solving the bound-state equation.  All these terms, $V_{\vc}^{(J,\ell)}$ and $V_{\rm vx}^{(J,\ell)}$  start at  ${\cal O}(\alpha^2 p^0)$, cf. Eq.~\eqref{240131.1}. In contrast, the LO QED amplitude is ${\cal O}(\alpha/p^2)$ and  for $p={\cal O}(m\alpha)$, as it is the case for the binding momentum, it counts as ${\cal O}(\alpha^{-1})$. Therefore, the contributions from the vacuum polarization and dressed vertex diagram  to the binding momentum count three extra powers of $\alpha$, and they are ${\cal O}(\alpha^4)$.   Higher order terms in the low-momentum expansion would contribute with extra powers of $\alpha^2$, giving rise to ${\cal O}(\alpha^6)$ contributions that we do not discuss.

The once-iterated diagram  $V_{\rm it}^{(J,\ell)}$ depends in momentum as  ${\cal O}(\alpha^2 p^{-1})$. This is a pure relativistic contribution because the leading nonrelativistic contribution to the first-iterated scattering amplitude is purely imaginary for real $p$ and it is removed when calculating $V_2^{(J,\ell)}$, as shown in \cite{Oller:2022tmo}. For momentum of ${\cal O}(m\alpha)$ it then counts as ${\cal O}(\alpha)$, two powers higher than the LO QED amplitude that counts as ${\cal O}(\alpha^{-1})$ in the same momentum region. Therefore, 
the contribution to  the binding momentum from the first-iterated diagram  is ${\cal O}(\alpha^{3})$. 
  Thus, from the unitarization of the NLO QED scattering amplitude, the leading corrections to the nonrelativistic result for the binding momentum are ${\cal O}(\alpha^3)$,  and stem from the relativistic corrections to the nonrelativistic  LO amplitude and from the once-iterated amplitude.

Since, as discussed, $V_2^{(J,\ell)}$ contributes perturbatively to the binding momentum, the bound-state equation Eq.~\eqref{231224.1} with the NLO QED contribution included can be expanded on $V_2^{(J,\ell)}$ and it becomes then
\begin{align}
  \label{240108.1}
\frac{1}{V_1^{(J,\ell)}(p^2)+V_2^{(J,\ell)}(p^2)}-i\frac{p}{4\pi}=  \frac{1}{V_1^{(J,\ell)}}-\frac{V_2^{(J,\ell)}}{(V_1^{(J,\ell)})^2}+{\cal O}(\alpha)-i\frac{p}{4\pi}=0~.
\end{align}
This extra contribution proportional to $V_2^{(J,\ell)}$ starts at  ${\cal O}(\alpha^0)$. If the previous equation is multiplied  by $V_1^{(J,\ell)}(p^2)^2$ one realizes,  after taking into account Eq.~\eqref{231221.7} to determine $V_2^{(J,\ell)}$ and the expression of $g(p)$ in Eq.~\eqref{240121.2}, that it can also be written as $T_1^{(J,\ell)}-T_2^{(J,\ell)}=0$. Thus, Eq.~\eqref{240108.1} is the same bound-state equation as  for the IAM case (and this is why we do not treat it separately).

However, Eq.~\eqref{240108.1} cannot be so straightforwardly applied to the bound-state problem and it requires some amends. 
 The problem is that  the once-iterated amplitude contributing to  $V^{(J,\ell)}_2(p^2)$ is proportional to $1/\sqrt{p^2}$, so that it has a LHC for $p^2<0$ that  overlaps with the bound-state binding momentum $p_b$. As a result, the direct application of Eq.~\eqref{240108.1} drives to a  pole in the first Riemann sheet with $p_b$ having a positive imaginary part but also a non-vanishing real part. This last feature is unacceptable from the analytical point of view. 
  This issue is solved by the 1st-iterated $N/D$ method that keeps the right analytical structure of a PWA, as already discussed in Sec.~\ref{sec.240128.4}. The general bound-state equation for calculating the binding momentum $p_b$ within the 1st-iterated $N/D$ method is
 \begin{align}
   \label{240210.1}
   D^{(J,\ell)}(p_b^2)&=0~.
 \end{align}
 This equation is then applied to  $D_{{\rm LO}}^{(J,\ell)}$ and $D_{{\rm LO}}^{(J,\ell)}+D_{{\rm NLO}}^{(J,\ell)}$, Eqs.~\eqref{240131.4b} and \eqref{240210.2}, respectively. Since $D^{(J,\ell)}(p^2)$ only has RHC, the problem in the secular equation due to the overlapping of the binding momentum with the LHC is avoided by construction.

Finally, it is important to realize that for the bound-state problem not all the contributions to $V_n^{(J,\ell)}$, with $n\geq 3$, can be treated perturbatively in comparison with $V_1^{(J,\ell)}$, as discussed in Ref.~\cite{Oller:2022tmo}. To show it let us consider the expansion  $V_{\rm nr}^{(\ell)}$, Eq.~\eqref{240123.1} in powers of $\alpha$, namely, $V_{{\rm nr}}^{(\ell)}=\sum_{j=0}V_{{\rm nr};2j+1}\alpha^{2j+1}$, since it only involves odd powers of $\alpha$. Therefore, for $p={\cal O}(m\alpha)$ the functions $V_{{\rm nr;}2j+1}^{(\ell)}(p^2)$, with  $j\geq 1$, are not suppressed compared to $V_{{\rm nr};1}^{(\ell)}(p^2)$ (the expansion in powers of $\alpha$ of Eq.~\eqref{240123.1} is actually an expansion in powers of $\alpha m/p$, the so-called Sommerfeld parameter). 


\section{Binding momentum for the ground state in the $J=1/2$ and $\ell=0$ PWA}
\setcounter{equation}{0}   
\def\theequation{\arabic{section}.\arabic{equation}}
\label{sec.240322.5}

In this section $J=1/2$ and $\ell=0$. Firstly, we analyze the results for the binding momentum from the unitarized LO QED PWA, and then we move on and  consider the NLO QED PWA.  

\subsection{Bound state from the unitarized LO QED PWA}
\label{sec.240212.1}

The LO QED result plus the contribution from the $S$ matrix redefinition, that removes the infrared divergences, is
\begin{align}
  \label{231228.1}
  T^{(\frac{1}{2},0)}_{1}(p^2)&=\frac{2\pi \alpha(m-E(1-2\ln a))}{p^2}~.
\end{align}
The bound-state equation Eq.~\eqref{231224.1} is now
\begin{align}
\label{231228.2}
\frac{p^2}{2\pi\alpha(m-(1-2\ln a)E)}-i\frac{p}{4\pi}=0~.
\end{align}
Its solution for the binding momentum $p_b$ is 
\begin{align}
  \label{240123.5}
  p_{b}&=im\alpha \frac{1+|1-2\ln a|\sqrt{1-\alpha^2 (1-\ln a)\ln a}}{2\left(1-\alpha^2(1-\ln a)\ln a\right)+\alpha^2/2}\\
  &=im\alpha \ln a+\frac{1}{4}im\alpha^3(1-2\ln a)\ln^2a
+\frac{im \alpha^5\ln^3 a}{16}(2+\ln a(-7+6\ln a))+{\cal O}(\alpha^7)~.\nn
\end{align}
In the pure nonrelativistic case the previous result reduces to  $p_b=im\alpha \ln a$ \cite{Oller:2022tmo}. 

Next, we consider the application of the first-iterated $N/D$ method for unitarizing $V_1^{(\frac{1}{2},0)}$. For that we evaluate algebraically the function $D^{(\frac{1}{2},0)}(p^2)$, Eq.~\eqref{240131.4b}, that  explicitly reads
\begin{align}
  \label{240123.6}
  D^{(\frac{1}{2},0)}_{{\rm LO}}(p^2)&
  =  1+C^{(\frac{1}{2},0)}-\frac{\alpha(p^2+p_0^2)}{\pi}\int_0^\infty dk \frac{m-(1-2\ln a)\sqrt{m^2+k^2}}{(k^2-p^2)(k^2+p_0^2)}\\
&=1+C^{(\frac{1}{2},0)}-i\alpha\frac{m-(1-2\ln a)E}{2p}
-\frac{\alpha(1-2\ln a)}{2\pi p}E\ln\frac{E+p}{E-p}\nn\\
&+\frac{\alpha}{2\pi p_0}\left(\pi m+2(1-2\ln a)\sqrt{p_0^2-m^2}{\rm arctanh}\sqrt{1-\frac{m^2}{p_0^2}}\right)~.\nn
\end{align}
Here $C^{(J,\ell)}$ is a subtraction constant that we expand in powers of $\alpha$ as
\begin{align}
  \label{240210.3}
 C^{(J,\ell)}&=\sum_{n=1}C_n^{(J,\ell)}\alpha^n~,
\end{align}
where $C_1^{(J,\ell)}$ and $C_2^{(J,\ell)}$ are going to be fixed in the unitarization process of the LO and NLO QED amplitudes, respectively, by following Sec.~\ref{sec.240128.4}.  The expansion in powers of $p$ of Eq.~\eqref{240123.6} is
\begin{align}
  \label{240210.4}
  D^{(\frac{1}{2},0)}_{{\rm LO}}(p^2)&=1-i\frac{m\alpha \ln a}{p}+C^{(\frac{1}{2},0)}+\frac{\alpha}{\pi p_0}\left(\frac{m\pi}{2}+(1-2 \ln a)\left[\sqrt{p_0^2-m^2}{\rm arctanh}\sqrt{1-\frac{m^2}{p_0^2}}-p_0\right]\right)\\
  &+i\frac{\alpha(1-2\ln a)p}{4m}+{\cal O}(p^2)~.\nn
\end{align}
By requiring that its real part goes to 1 for $p\to 0$ we then have that
\begin{align}
\label{240210.6}
  C_1^{(\frac{1}{2},0)}&=-\frac{1}{\pi p_0}\left(\frac{m\pi}{2}+(1-2 \ln a)\left[\sqrt{p_0^2-m^2}{\rm arctanh}\sqrt{1-\frac{m^2}{p_0^2}}-p_0\right]\right)~.
  \end{align}
Taking this expression into $D_{{\rm LO}}^{(\frac{1}{2},0)}(p^2)$ it follows that 
\begin{align}
\label{240210.7}
D_{{\rm LO}}^{(\frac{1}{2},0)}(p^2)&=1-i\alpha\frac{m-(1-2\ln a)E}{2p}+\frac{\alpha(1-2\ln a)}{2\pi p}\left[2p-E\ln\frac{E+p}{E-p}\right]~.
\end{align}

The $D_{{\rm LO;nr}}^{(\frac{1}{2},0)}$ function calculated by employing the nonrelativistic version of $V_1^{(0,\frac{1}{2})}$, denoted above by $V_{{\rm nr}}^{(0)}$, is
\begin{align}
  \label{240210.5}
D_{{\rm LO;nr}}^{(0)}(p^2)&=1-\frac{ 2m\alpha\ln a}{\pi}\int_0^\infty \frac{dk}{k^2-p^2}
=1-i\frac{m\alpha \ln a}{p}~,
\end{align}
in agreement with \cite{Oller:2022tmo}. Notice that the integration in Eq.~\eqref{240210.5} is convergent and it does not require of any subtraction. In the limit $p\to 0$ we have from Eqs.~\eqref{240210.7} and \eqref{240210.5} that
\begin{align}
  \label{240717.1}
D_{{\rm LO}}^{(\frac{1}{2},0)}(p^2)&\to  D_{{\rm LO;nr}}^{(0)}(p^2)+{\cal O}(p)~.
\end{align}

The bound-state equation $D_{\rm LO}^{(\frac{1}{2},0)}(p_b^2)=0$ is a transcendental equation and we proceed  to solve it in a power series expansion in $\alpha$. For that, it is 
convenient to multiply Eq.~\eqref{240210.7} by $1/V_1^{(\frac{1}{2},0)}$, and the resulting secular equation is 
\begin{align}
  \label{240210.8}
  -i\frac{p}{4\pi}+\frac{p^2}{2\pi\alpha(m-E[1-2\ln a])}+\frac{(1-2\ln a)p}{4\pi^2(m-E[1-2\ln a])}\left(2p-E\ln\frac{E+p}{E-p}\right)=0~.
\end{align}
Except for the term $-ip/4\pi$ this equation only contains even powers of $p$ if expanded around $p=0$. The leading order contribution is $p_{b;1}=im\alpha\ln a$, by constructions the same result as in the nonrelativistic treatment. Next, to get the ${\cal O}(\alpha^2)$ correction we write the binding momentum $p_b=p_1+\alpha^2\kappa_2$, and expand Eq.~\eqref{240210.8} up to ${\cal O}(\alpha^2)$. One then obtains that $\kappa_2=0$. For the first non-vanishing corrections we write $p_b=p_1+\alpha^3 \kappa_3$, expand Eq.~\eqref{240210.8} up to ${\cal O}(\alpha^3)$ and the results is 
\begin{align}
  \label{240210.9}
\kappa_3&=i\frac{m(1-2\ln a)\ln^2 a}{4}~.
\end{align}
Therefore,
\begin{align}
  \label{240123.11b}
  p_b=im\alpha\ln a+im\alpha^3\frac{(1-2\ln a)\ln^2 a}{4}+{\cal O}(\alpha^4)~.
\end{align}

In connection with the Dirac equation for the binding momentum $p_D^{(J,n)}$ for the hydrogen atom,
\begin{align}
  \label{240123.12}
  p^{(J,n)}_D&=im\alpha\left[\alpha^2+\left(
 n-J-\frac{1}{2}+\sqrt{(J+\frac{1}{2})^2-\alpha^2}\right)^2\right]^{-\frac{1}{2}}~,  
\end{align}
with $n$  the principal quantum number with $J+\frac{1}{2}\leq n$, we notice that the iteration of the LO QED amplitude should represent the same physics. For the case of the lowest energy level of the hydrogen atom  $n=1$ and one has that 
$p_D^{(\frac{1}{2},1)}=im\alpha$, as in the nonrelativistic case. This result differs from Eq.~\eqref{240123.11b} by ${\cal O}(\alpha^3)$. 
 The difference, of course, stems from the need to know $V^{(J,\ell)}$ to all orders, similarly as for the nonrelativistic case, treated in detail in Ref.~\cite{Oller:2022tmo}. There, it was shown that once higher orders terms in the expansion in powers of $\alpha$ of  $V_{\rm nr}^{(0)}$ are taken into account from Eq.~\eqref{240123.1}, the exact value $p_b=im\alpha$ is obtained, despite that at LO one has  $p_b=im\alpha\ln a$ with $\ln a=\gamma_E$. One can see from Table~1 of \cite{Oller:2022tmo} that the largest correction to $p_b$ comes from the ${\cal O}(\alpha^3)$  contribution to $V_{\rm nr}^{(0)}$, while higher order terms in its expansion gives contributions less than a few per mil. 
It is certainly worth stressing that the suppression of these higher orders in the binding momentum is purely numerical, and does not arise because higher powers of $\alpha$ are involved. 

The ${\cal O}(\alpha^3)$ contribution found in Eq.~\eqref{240123.11b} to $p_D^{(\frac{1}{2},1)}=i m\alpha$ is connected to the relativistic correction to $D_{{\rm LO;nr}}^{(0)}$ for $p={\cal O}(m\alpha)$ in Eq.~\eqref{240717.1}, since the latter is also two orders higher in the $\alpha$ counting than the LO contribution. Our freedom to choose $C_1^{(\frac{1}{2},0)}$ as in Sec.~\ref{sec.240128.4}, cf. Eq.~\eqref{240210.6}, defines our renormalization scheme when applying the once-iterated $N/D$ method. Of course, one could choose other scheme and, as a result, $\cV_2^{(\frac{1}{2},0)}$ would be different in order to guarantee that the perturbative QED amplitude is reproduced order by order. In this respect, we can choose $C_1^{(\frac{1}{2},0)}$ so as to reproduce the exact Dirac result for the binding momentum by adding to $\alpha C_1^{(\frac{1}{2},0)}$, Eq.~\eqref{240210.6}, the amount
\begin{align}
  \label{240717.2}
  \delta C^{(\frac{1}{2},0)}&=\frac{m\alpha}{4\pi^2\left(1-\sqrt{1-\alpha^2}(1-2\ln a)\right)}\left(\pi+(1-2\ln a)2\left[\alpha+\sqrt{1-\alpha^2}\left\{\frac{\pi}{2}-\arcsin\alpha\right\}\right]
  \right)~.
\end{align}



\subsection{Bound state from the unitarized NLO QED PWA}
 \label{sec.240123.2}

 Consider first the application of the $g$-method, such that $V_2^{(\frac{1}{2},0)}(p^2)$ according to Eq.~\eqref{231221.7} is given by
 \begin{align}
   \label{240211.1a}
V_2^{(\frac{1}{2},0)}= T_2^{(\frac{1}{2},0)}
   -i\alpha^2\pi\frac{(m-(1-2\ln a)E)^2}{p^3}~.
 \end{align}
   The  partial contribution to Eq.~\eqref{240211.1a} from the once-iterated diagram plus the $S$-matrix renormalization at ${\cal O}(\alpha^2)$ is  
\begin{align}
  \label{231231.1}
  V_{\rm it}^{(\frac{1}{2},0)}(p^2)&=\frac{\fa^2\pi^2}{\sqrt{p^2}}~.
\end{align}
Let us notice that this contribution has a branch-point singularity at $p^2=0$ and a LHC for $p^2<0$. As a result, its use in Eq.~\eqref{240122.1} drives to incorrect analytical properties of a PWA (in comparison with the right analytical properties of a triangular loop \cite{Eden:1966dnq}). This issue manifests in the calculation of the solution to the bound-state equation,
\begin{align}
  \label{240211.2}
1-i\alpha\frac{(m-E_b(1-2\ln a))}{2p_b}-i\frac{\pi\alpha^2}{4}=0~,
\end{align}
 since then $\displaystyle{p_b=im\alpha\ln a\left(1+\frac{\alpha^2(1-2\ln a)\ln a}{4}\right)-\alpha^3 m\pi\frac{\ln a}{4}+{\cal O}(\alpha^5)}$ has also a real  together  a positive imaginary part, which is  unacceptable.

 We now proceed to apply the first-iterated $N/D$ method to unitarize the NLO QED PWA. We have to fix $\cV_2^{(\frac{1}{2},0)}$ by employing Eq.~\eqref{240128.3} which implies that 
 \begin{align}
   \label{240211.1}
   \cV_2^{(\frac{1}{2},0)}&
   =T_2^{(\frac{1}{2},0)}
   -i\alpha^2\pi\frac{(m-(1-2\ln a)E)^2}{p^3}
   +\alpha^2\frac{(1-2\ln a)(m-(1-2\ln a)E)}{p^3}
   \left[2p-E\ln\frac{E-p}{E+p}\right]~.
   \end{align}

 $V_{{\rm it}}^{(\frac{1}{2},0)}(p^2)$ is also part of $\cV_2^{(\frac{1}{2},0)}$ and now it contributes to $D^{(\frac{1}{2},0)}(p^2)$ as, cf. Eq.~\eqref{240213.1},  
\begin{align}
  \label{241123.14}
-\frac{\alpha^2(p^2+p_0^2)}{4}\int_0^\infty \frac{dk^2}{(k^2-p^2)(k^2+p_0^2)}=\frac{\alpha^2}{4}\ln\frac{-p^2}{p_0^2}~,
\end{align}
so that it has only RHC and then  does not pose any problem on analyticity of the $T$ matrix. 

The contribution from  Eq.~\eqref{241123.14} to the binding momentum is obtained by multiplying it by $- p_b$  and evaluate the result with $p\to im\alpha \ln a$ [this is clear  from the secular equation Eq.~\eqref{240210.1}].  It is an ${\cal O}(\alpha^3)$ contribution given by
\begin{align}
  \label{240211.3}
\delta p_{b;{\rm it}}&=-\frac{i}{2}m\alpha^3 \ln a\, \ln\!\left(\!\frac{m\alpha \ln a}{p_0}\!\right)+{\cal O}(\alpha^5)~.
\end{align}
We set $p_0=m$, 
and vary $p_0^2$ between $m^2/4$ and $m^2$, which implies a relative uncertainty of $\frac{\alpha^2}{2}\ln 2$ in $\cV_2^{(\frac{1}{2},0)}$. For the seeking of the binding momentum this source of uncertainty is  suppressed in front of the infrared enhanced log-factor $\frac{\alpha^2}{2}\ln \frac{|p_b|}{m}\sim \frac{\alpha^2}{2}\ln \alpha$ with $p_b\approx im\alpha$. The relative error here is around a 14\%.


We next consider the contributions from the one-loop vacuum polarization ($\cV_{\rm vc}^{(\frac{1}{2},0)}$), the vertex corrections plus Chung's diagrams ($\cV_{\rm vx}^{(\frac{1}{2},0)}$), and the last term to the right in Eq.~\eqref{240211.1} ($\cV_{D}^{(\frac{1}{2},0)}$), equal to  $V_1^{(\frac{1}{2},0)}(\Re D_{{\rm LO}}^{(\frac{1}{2},0)}-1)$.  $\cV_{\rm vc}^{(\frac{1}{2},0)}$ and $\cV_D^{(\frac{1}{2},0)}$  admit a low-momentum expansion in powers of  $p^2$, and tend to a constant for $p\to 0$. For $\cV_{\rm vx}^{(\frac{1}{2},0)}$ an expansion in powers of $p^2$ is also possible modulo $\ln p$. One then has, 

\begin{align}
\label{231219.2}
\cV_{\rm vc}^{(\frac{1}{2},0)}(p^2)&=\frac{8\fa^2}{15m}
  +{\cal O}(p^2)~,\\
\cV_{\rm vx}^{(\frac{1}{2},0)}(p^2)&=-\frac{\fa^2}{24m}
\left(  28-3\pi +16  \ln\frac{m^4 a^4}{256 p^4} \right)
+{\cal O}(p^2)~,\nn\\
\cV_D^{(\frac{1}{2},0)}&=-\frac{4\alpha^2}{3m}(1-2\ln a)\ln a+{\cal O}(p^2)~.\nn
\end{align}

When inserted in the calculation of the function $D^{(\frac{1}{2},0)}(p^2)$ a subtraction is needed because they vanish as $1/p$(modulo powers of $\ln p$) for $p\to\infty$. Within the iterated $N/D$ method we can proceed as we did concerning $D_{LO}^{(\frac{1}{2},0)}(p^2)$ and Eq.~\eqref{241123.14} from $\cV_{{\rm it}}^{(\frac{1}{2},0)}$, so that the constant term that adds to the 1 in $D^{(J,\ell)}(p^2)$, cf. Eq.~\eqref{240210.6}, can be removed. This is then reshuffled as a contribution to the $\cV^{(\frac{1}{2},0)}(p^2)$ function by the matching procedure explained in Sec.~\ref{sec.240128.4}, such that the perturbative QED PWA can be reproduced at any order  in the expansion in powers of $\alpha$. The removal of this constant term is equivalent to take the subtraction at $p_0=0$ with zero subtraction constant when calculating their contribution to $D^{(\frac{1}{2},0)}(p^2)$, denoted by $\delta D^{(\frac{1}{2},0)}_{{\rm vc;vx;}D}$. Thus, we have
\begin{align}
  \label{240211.5}
\delta D^{(\frac{1}{2},0)}_{{\rm vc;vx;}D}(p^2)&=-\frac{p^2}{4\pi^2}\int_0^\infty dk^2\frac{k\{\cV_{{\rm vc}}^{(\frac{1}{2},0)}(k^2)+\cV_{{\rm vx}}^{(\frac{1}{2},0)}(k^2)+\cV_D^{(\frac{1}{2},0)}(k^2)\}}{k^2(k^2-p^2)}
\end{align}
Eq.~\eqref{240211.5} contributes to the binding momentum by
\begin{align}
  \label{240211.6}
  \delta p_{b}&=-p_{b;1}\delta D^{(\frac{1}{2},0)}_{{\rm vc;vx;}D}(p_{b;1}^2)+{\cal O}(\alpha^6)~,
\end{align}
and it is ${\cal O}(\alpha^4)$. This can be manifestly shown by introducing the dimensionless variable $y$  such that  $k=m\alpha\ln a\, y$, together with the dimensionless function $W^{(\frac{1}{2},0)}=m\cV^{(\frac{1}{2},0)}/\alpha^2$. Then, we can rewrite Eq.~\eqref{240211.5} for $p=im\alpha\ln a$ as
\begin{align}
  \label{240211.6}
\delta p_b&=-i\frac{m\alpha^4 \ln^2 a}{2\pi^2}\int_0^\infty dy\frac{\left(W_{{\rm vc}}^{(\frac{1}{2},0)}+W_{{\rm vx}}^{(\frac{1}{2},0)}+W_D^{(\frac{1}{2},0)}\right)}{y^2+1}~.
\end{align}
These integrals are calculated numerically, and we have checked that they perfectly agree with the algebraic results that one obtains by applying the $g$-method
\begin{align}
  \label{240211.8}
  \delta p_{b;{\rm vc}}&=-im\alpha^4\gamma_E^2\frac{2}{15\pi}=-im\alpha^4\gamma_E^2\, 4.2\, 10^{-2}~,\\
  \delta p_{b;{\rm vx}}&=im\alpha^4\gamma_E^2\frac{1}{96\pi}\left(28-3\pi+64\ln\frac{a}{4\alpha\ln a}\right)=im\alpha^4\gamma_E^2\,1.0~,\nn\\
 \delta p_{b;D}&=-im\alpha^4\gamma_E^3 \frac{1-2\gamma_E}{3\pi}
   =-im\alpha^4\gamma_E^2\, 9.4\,10^{-2}~,\nn
\end{align}
plus ${\cal O}(\alpha^6)$ for all cases. 
\section{Conclusions and outlook}
\def\theequation{\arabic{section}.\arabic{equation}}
\setcounter{equation}{0}   
\label{sec.240322.6}

In this study we have unitarized  in partial wave amplitudes the one-loop scattering amplitude in QED of a relativistic electron by an external Coulomb potential. This work is an extension of Refs.~\cite{Blas:2020dyg,Oller:2022tmo}  that unitarized  nonrelativistic Coulomb scattering.  The approach of Refs.~\cite{Blas:2020dyg, Blas:2020och} has been  extended to deal with infrared divergences that go beyond the generalized divergent Coulomb phase already accounted for by the method of these references. We have then applied the redefinition of the electron asymptotic states by accounting for a cloud of real soft photons, first introduced by Chung in Ref.~\cite{Chung:1965zza}, and afterwards generalized by Faddeev-Kulish \cite{Kulish:1970ut}. The resulting QED scattering amplitudes that we have calculated up to NLO are then infrared finite. 

Based on them, we have calculated infrared finite partial-wave amplitudes, first calculated perturbatively and then unitarized by applying three different unitarization methods: the algebraic $N/D$ method, renamed $g$-method for brevity, the Inverse Amplitude Method (IAM), and the first-iterated $N/D$ method. In this way, we have presented explicit results for the $S$- and $P$-wave scattering.  The results from the different unitarization methods typically agree, though for $|p| \gtrsim m/\alpha$ the IAM fails in those partial-wave amplitudes for which the real part of the $T$ matrix crosses zeroes.  This deficiency at high momentum is similar to the well-known issue for the straightforward application of this method around an Adler zero in meson-meson scattering \cite{GomezNicola:2007qj,Salas-Bernardez:2020hua,Escudero-Pedrosa:2020rwb}. 

We have also discussed the secular equation for the calculation of the binding momenta by applying the different unitarization methods. 
However, it is only the first-iterated $N/D$ method that is suitable for looking for bound states which position lie on top of the left-hand cut, as it is the case here with the perturbative QED  scattering amplitude calculated to the one-loop level. We have then evaluated the binding momentum for the fundamental state from the unitarized perturbative QED partial-wave amplitudes calculated at leading and up to next-to-leading orders. 

As an outlook, we propose that this method could be applied to study properties of the hydrogen bound states by unitarizing higher-order perturbative QED scattering amplitudes. This is a novel approach  that does not require to use quantum mechanical wave functions for the electron, and it is based entirely on the knowledge of the $S$ matrix.  It can be applied for bound states with $\ell\leq 1$ and principal quantum number up to $n=5$ in $S$ wave and  $n=2$ for $P$ waves. Similarly, this unitarization method  could be  applied to study the interplay between strong and electromagnetic interactions in processes of interest, like hadronic or nuclear low-energy scattering of charged particles, electromagnetic corrections to hadronic and nuclear bound states and resonances, etc. For the near future, we are interested in extending the results
 of Refs.~\cite{Blas:2020dyg,Blas:2020och} by unitarizing  graviton-graviton  scattering from a one-loop calculation in the low-energy effective-field theory of gravity. As discussed in Sec.~\ref{sec.240316.2}, this can be done by employing the results of Ref.~\cite{Weinberg:1965nx} in a way that would make it unnecessary to explicitly introduce the states analogous to the Faddeev-Kulish ones for gravity  \cite{Oller:2024neq}. 
 Of special interest will be to look for  the graviball resonance pole found in these references from the unitarized leading-order graviton-graviton scattering amplitudes.

\bigskip
\bigskip

\section*{Acknowledgements}

JAO would like to acknowledge partial financial support to the Grant PID2022-136510NB-C32 funded
by \\MCIN/AEI/10.13039/501100011033/ and FEDER, UE, and to the EU Horizon 2020 research and innovation program, STRONG-2020 project, under grant agreement no. 824093. MP acknowledges financial support from the ANII-FCE-175902 project.

\appendix

\section{QED one-loop results}
\label{sec.240322.2}
\def\theequation{\Alph{section}.\arabic{equation}}
\setcounter{equation}{0}\label{App:QED}

In this appendix we provide a concise summary of QED results up to one loop. 
The results are first expressed as Feynman integrals, then simplified to Passarino-Veltman integrals, and finally presented analytically. All the integrals are evaluated in dimensional regularization.  
A photon mass, denoted by $\lambda$, appears explicitly only in the terms that give rise to infrared divergences, otherwise is set to zero.  
Some definitions are repeated here again for consistency in the appendix. In the following $q=p'-p$, $t=q^2$ and $\sigma=\sigma(t)=\sqrt{1-4m^2/t}$.

\subsection{Electron self energy}
\begin{figure}
\centering
 \includegraphics[width=0.15\textwidth]{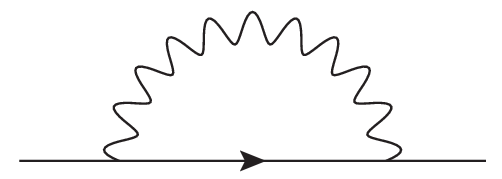}
     \caption{\label{Fig:electronselfenergy} One-loop diagram to the electron self energy.}
\end{figure}
The one-loop diagram for the electron self energy is depicted in Fig.~\ref{Fig:electronselfenergy}, which expression is denoted as $-i \Sigma (\slashed{p})$. It  takes the form, 
\begin{align}
-i \Sigma (\slashed{p})
&=-e^2\mu^{-2\epsilon}\int\frac{d^dk}{(2\pi)^d} \frac{(2-d)(\slashed{p}-\slashed{k})+d m}{\left(k^2-\lambda^2+i\ve \right)\left((p-k)^2-m^2+i \ve\right)}~,
\end{align}
where $d=4+2\ep$, $\mu$ represents the dimension of the coupling constant, $p$ is the electron incoming momentum and $k$ is the photon one. 
Thus, 
\begin{align}
-i \Sigma (\slashed{p})=-e^2\left(\frac{(d-2)\slashed{p}}{2p^2}\left[A(m^2)-A(\lambda^2)+\left(\lambda^2-p^2-m^2\right)B(\lambda^2,m^2,p^2)\right]+d\hspace{1mm}m B(\lambda^2,m^2,p^2)\right)~,
\end{align} 
where the integrals $A$ and $B$ are defined in Sec.\ref{Sec:MasterIntegrals}, together with their divergent ($A_d$ and $B_d$) and finite ($A_f$ and $B_f$) parts.

In order to have the pole in the electron propagator at the physical mass with residue equal to one, the renormalization proceeds by requiring that 
\begin{align}
    \delta Z_m=-\frac{\Sigma(\slashed{p}=m)}{m} \quad \& \quad
\delta Z_2=\frac{d \Sigma(\slashed{p})}{d\slashed{p}}\Big|_{\slashed{p}=m}~. 
\end{align}
Therefore, 
\begin{align}
\delta Z_m 
&=\frac{e^2}{(4\pi)^2}\left(\frac{3}{\hat{\ep}}+3\ln\frac{m^2}{\mu^2}-4\right)~,
\end{align}
where here $\lambda^2=0$ has been set,   and 
\begin{align}
\label{240412.3}
\frac{1}{\hat{\ep}}&=\frac{1}{\ep}+\gamma_E-\ln 4\pi~,
    \end{align}
with $\gamma_E$ the  Euler constant.  In turn, 
\begin{align}
\delta Z_2
&=\frac{e^2}{(4\pi)^2}\left(\frac{1}{\hat{\ep}}+\ln\frac{m^2}{\mu^2}+2\ln\frac{m^2}{\lambda^2}-4\right)~,
\end{align} 
where the infrared divergence as well as the ultraviolet one appear in this renormalization factor.

\subsection{Vacuum polarization}

We compute here the correction  coming from diagram (d) of Fig.~\ref{Fig:todos}, properly renormalized with diagram (e). The expression for the one-loop amplitude given by the  diagram (d) reads
\begin{equation}
  \bar u_{\sigma'}(\boldsymbol{p'})(-i e \gamma^\mu)u_{\sigma}(\boldsymbol{p})\left(\frac{-ig_{\mu\nu}}{q^2-\lambda^2}\right)i\Pi^{\nu\rho}(q) \mathcal{A}_\rho(q)~,
\end{equation}
where
\begin{align}
    i\Pi^{\nu\rho}(q)
    &=-de^2\mu^{-2\ep}\int \frac{d^d k}{(2\pi)^d}\frac{(-k^2+k\cdot q+m^2)g^{\nu\rho}+2k^\nu k^\rho-k^\nu q^\rho-q^\nu k^\rho}{(k^2-m^2+i\ve)((k-q)^2-m^2+i\ve)}~,
    \end{align}
which, after the reduction to master integrals, has the form $\Pi^{\nu\rho}(q)=\Pi(q^2)\left(q^2 g^{\nu\rho}-q^\nu q^\rho\right)$, with
\begin{align}
\Pi(q^2)
=i e^2 \frac{d}{(d-1)q^2} \left((2-d)A(m^2)+\left(2m^2+(d-2)\frac{q^2}{2}\right)B(m^2,m^2,q^2)\right).
\end{align}
The effect of taking into account the counterterm reduces to subtract the expression evaluated at vanishing momentum, so the counterterm is
\begin{equation}
    \delta Z_3=\Pi(0)=\frac{e^2}{24 \pi^2}\left(\frac{2}{\hat{\ep}}+2\ln\frac{m^2}{\mu^2}+1\right)~.
\end{equation}
 It  leads us to the amplitude
\begin{align}
\bar u_{\sigma'}(\boldsymbol{p'})(-i e \gamma^\mu \mathcal{A}_\mu(q))u_{\sigma}(\boldsymbol{p}) \frac{q^2}{q^2-\lambda^2}\Pi_R(q^2),
\end{align}
with $\Pi_R(q^2)=\Pi(q^2)-\Pi(0)$, corresponding to Eq.(\ref{Eq:A1}) .

\subsection{Vertex modification of the  electron-photon coupling}

The expression for the sum of the diagrams (b) and (c) of Fig.~\ref{Fig:todos} can be explicitly written as $\displaystyle{\bar{u}_{\sigma'}\left(\boldsymbol{p}^{\prime}\right) (-ie\Gamma^\mu) u_{\sigma}(\boldsymbol{p})A_\mu(q)}$, with
\begin{align}
\label{240412.2}
\Gamma^\mu&=-i e^2 \int \frac{d^dk}{(2\pi)^d}\frac{\gamma^\alpha (\slashed{p'}-\slashed{k}+m)\gamma^\mu(\slashed{p}-\slashed{k}+m)\gamma_\alpha}{((p'-k)^2-m^2+i\ve)((p-k)^2-m^2+i\ve)(k^2-\lambda^2+i\ve)}-ie^2\delta Z_1~.
\end{align}

The former diagrams  are often  written in terms of two form factors:
\begin{align}
\Gamma^\mu(p',p)=
F_1(q)\gamma^\mu+\frac{i\sigma^{\mu\nu}}{2m}q_\nu F_2(q)~,
\end{align}
with $\sigma^{\mu\nu}=\frac{i}{2}[\gamma^\mu,\gamma^\nu]$. 

The analysis is performed using a general $\mu$-index for the potential that we should consider equal to $0$ for the Coulomb potential. The final result of the renormalized vertex is: 
\begin{align}
\label{240412.5}
 F_1(q)&=
ie^2\left(2(t-2m^2)\mathcal{C}(\lambda^2,m^2,t)+\frac{4A_f(m^2)}{t-4m^2}+\frac{B_f(m^2,m^2,t)(3t-8m^2)}{t-4m^2}+\frac{4B_f(0,m^2,m^2)(2m^2-t)}{t-4m^2}+\frac{2m^2-t}{8\pi^2(t-4m^2)}
\right)\nn\\
&+\delta Z_2^f~,
\end{align}
with $\delta Z_2^f=\delta Z_2-e^2/(16\pi^2\hat{\ep})$, and 
\begin{align}
\label{240413.1}
F_2(q)&=-\frac{4ie^2}{t-4m^2}\left(A_f(m^2)+m^2 B_f(m^2,m^2,t)-2m^2 B_f(0,m^2,m^2)+\frac{im^2}{16\pi^2}
\right)~.
\end{align}
 Let us notice that the infrared divergences in Eq.~\eqref{240412.5} stem only from the triangle integral 
$\mathcal{C}(\lambda^2,m^2,t)$ and $\delta Z_2^f$, while  $F_2(q)$ has no infrared divergences. We also note that $F_1(q)$  vanishes linearly in $q^2$ for $q^2\to0$. 
Renormalization takes place by imposing the following condition on the total vertex function,  $\Lambda^\mu(p',p)=\gamma^\mu+\Gamma^\mu(p',p)$, 
\begin{align}
\bar u (\boldsymbol{p}^{\prime})\Lambda^\mu(p,p')u(\boldsymbol{p})\Big|_{\slashed{p}_1=\slashed{p}_2=m}=\bar u (\boldsymbol{p}^{\prime})\gamma^\mu u(\boldsymbol{p})\Big|_{\slashed{p}_1=\slashed{p}_2=m}
\end{align}
The latter implies that 
\begin{align}
\bar u(\boldsymbol{p})\Gamma^\mu(p,p)u(\boldsymbol{p})\Big|_{\slashed{p}=m}=(\delta Z_e+\frac{1}{2}\delta Z_3) \bar u(\boldsymbol{p})\gamma^\mu u(\boldsymbol{p}) =0 ~,
\end{align}
in agreement with the Ward identity. Equation~\eqref{Eq:A2} results by substituting in the expressions given for $F_1(q)$ and $F_2(q)$ the master integrals in the appendix~\ref{Sec:MasterIntegrals} and $\delta Z_2^f$. 




\subsection{Iterated diagram}
The amplitude for the iteration of the Coulomb potential corresponding to diagram (f) in Fig.\ref{Fig:todos}  is
\begin{align}
\label{Eq.Iterado}
&-i(ee')^2 (2\pi)\delta(E'-E)\bar u_{\sigma'}(\vp')\int \frac{d^3q}{(2\pi)^3}\frac{E \gamma^0+\vec{q}.\vec{\gamma}+m}{((\vp'-\boldsymbol{q})^2+\lambda^2+i\ve)((\vp-\boldsymbol{q})^2+\lambda^2+i\ve)(\vp^2-\boldsymbol{q}^2+i\ve)} u_\sigma(\vp)\\
    &\equiv -i(ee')^2 (2\pi)\delta(E'-E)\bar u_{\sigma'}(\vp')\left((E \gamma^0+m)I+\frac{\vec{\gamma}.(\vp+\vp')}{2}J\right) u_{\sigma}(\vp)~,\nn
\end{align}
where
\begin{equation}
\label{240413.6}
 I=   \int \frac{d^3q}{(2\pi)^3}\frac{1}{((\vp'-\boldsymbol{q})^2+\lambda^2+i\ve)((\vp-\boldsymbol{q})^2+\lambda^2+i\ve)(\vp^2-\boldsymbol{q}^2+i\ve)}~,
\end{equation} and
\begin{equation}
\label{240413.7}
    \frac{\vp'+\vp}{2}J=\int \frac{d^3q}{(2\pi)^3}\frac{\vq}{((\vp'-\boldsymbol{q})^2+\lambda^2+i\ve)((\vp-\boldsymbol{q})^2+\lambda^2+i\ve)(\vp^2-\boldsymbol{q}^2+i\ve)}~,
\end{equation}
which are the same loop functions as introduced by Dalitz in Ref.~\cite{Dalitz:1951ah} except for a global factor $(2\pi)^{-3}$. However, we have now evaluated them using dimensional regularization. 
 After some manipulations we obtain:
\begin{align}
\label{240413.2}
I&=\frac{-1}{8\pi p\sin\frac{\theta}{2}\sqrt{4p^4\sin^2\frac{\theta}{2}+4p^2\lambda^2+\lambda^4}}
\Big(
\arctan\frac{\lambda p\sin\frac{\theta}{2}}{\sqrt{4p^4\sin^2\frac{\theta}{2}+4p^2\lambda^2+\lambda^4}}
+\frac{i}{2}\ln\frac{\sqrt{4p^4\sin^2\frac{\theta}{2}+4p^2\lambda^2+\lambda^4}+2p^2\sin\frac{\theta}{2}}{\sqrt{4p^4\sin^2\frac{\theta}{2}+4p^2\lambda^2+\lambda^4}-2p^2\sin\frac{\theta}{2}}
\Big)~,\\
\label{240413.3}
J&=-\frac{1}{2\vp^2\cos^2\frac{\theta}{2}}\Big(
{\cal A}_3(\lambda^2,\vq^2)+{\cal B}_3(\lambda^2,\vp^2)-(2\vp^2+\lambda^2) I\Big)\nn\\
&=\frac{(2\vp^2+\lambda^2)I}{2\vp^2\cos^2\frac{\theta}{2}}-\frac{\arctan\frac{p\sin\theta/2}{\lambda}}{16\pi|\vp|^3\cos^2\frac{\theta}{2}\sin\frac{\theta}{2}}+\frac{i}{16\pi|\vp|^3\cos^2\frac{\theta}{2}}\ln\left(1-\frac{2i|\vp|}{\lambda}\right)~.
\end{align}
The master integrals in three dimensions ${\cal A}_3$ and ${\cal B}_3$ are evaluated in Sec.~\ref{Ap.Master3dim}.  Both functions $I$ and $J$ calculated here are in agreement with their computation in Ref.~\cite{Dalitz:1951ah}. 
By using the Dirac equation  in  Eq.(\ref{Eq.Iterado}) we can rewrite it as
\begin{align}
&-ie^4 (2\pi)\delta(E'-E)\bar u_{\sigma'}(\vp')\left((E \gamma^0+m)I+(E\gamma^0-m))J\right) u_\sigma(\vp)\nonumber\\
&=-ie^4 (2\pi)\delta(E'-E)\bar u_{\sigma'}(\vp')\left(E \gamma^0 (I+J)+m (I-J)\right) u_\sigma(\vp)~.
\end{align}

As discussed in Secs.~\ref{sec.240318.1} and \ref{sec.140413.1}, one needs to carefully take the limit $\lambda\to 0$ for the iterated diagram $A_3$ in order to properly handle the infrared divergences and its cancellation within partial-wave amplitudes. This is a point obviated  in Ref.~\cite{Dalitz:1951ah} since it was interested in the  Born series associated to the Coulomb potential. The limit $\lambda\to 0$ can be taken straightforwardly for the combination $I-J$  because of the factor $\displaystyle{1-\frac{1}{\cos^2\theta/2}}=-\tan^2\frac{\theta}{2}$ that vanishes for $\theta\to 0$. The resulting expression is the term between square brackets multiplied by $m$ in Eq.~\eqref{Eq:A3}.

The infrared divergent part for the sum $I+J$ is more delicate. In the term $\sqrt{4p^4\sin^2\frac{\theta}{2}+4p^2\lambda^2+\lambda^4}$, which repeatedly appears in the expression for $I$ in Eq.~\eqref{240413.2}, we can neglect $\lambda^4$ in front of $p^2\lambda^2$, but the latter cannot always be neglected compared with $4p^4\sin^2\theta/2$, as the latter vanishes for $\theta\to 0$.  However,  $\displaystyle{\lambda p \sin\frac{\theta}{2}/\sqrt{4p^4\sin^2\frac{\theta}{2}+4p^2\lambda^2+\lambda^4}\to 0}$ for $\lambda\to 0$ because this ratio vanishes in any case, either when $\lambda^2\ll p^2\theta^2$ or $p^2\theta^2\ll \lambda^2$. For the last two terms in the expression for $J$ in Eq.~\eqref{240413.3} its asymptotic behavior for $\lambda\to 0$  is straightforward, and it reads 
\begin{align}
\label{240413.8}
-\frac{\arctan\frac{p\sin\theta/2}{\lambda}}{16\pi|\vp|^3\cos^2\frac{\theta}{2}\sin\frac{\theta}{2}}+\frac{i}{16\pi|\vp|^3\cos^2\frac{\theta}{2}}\ln\left(1-\frac{2i|\vp|}{\lambda}\right)\xrightarrow[\lambda\to 0]{} -\frac{1}{32|\vp|^3\cos^2\frac{\theta}{2}\sin\frac{\theta}{2}}+\frac{i}{16\pi|\vp|^3\cos^2\frac{\theta}{2}}\left(\ln\frac{2|\vp|}{\lambda}-i\frac{\pi}{2}\right)
\end{align}
Then, the term between square brackets multiplied by $E \gamma^0$ in Eq.~\eqref{Eq:A3} results. 

We have detected some typos in Ref.~\cite{Dalitz:1951ah}, which we list here with reference to the numbered equations in \cite{Dalitz:1951ah}: $m$ should be changed to $-m$ in the numerator of the second line of the equation before (2.3) and in the first line of this equation.  In addition, a global minus sign is missing in the expression for $I$ in Eq.~(2.6) in the limit $\lambda\to 0$. We stress that the final expression for $M$ in the second line of Eq.~(2.3) is correct.


\subsection{Master Integrals}
\label{Sec:MasterIntegrals}

In this part of the Appendix we summarize the definitions and analytical results for master integrals in $d=4+2\epsilon$ dimensions. To begin with, the tadpole integral that is represented with $A(m^2)$ has the expression:
\begin{align}
A(m^2)=\mu^{-2\epsilon}\int\frac{d^dq}{(2\pi)^{d/2}}\frac{1}{q^2-m^2+i\ve}= -\frac{i m^2}{(4\pi)^2}\left(\frac{1}{\hat{\ep}}-1+\ln\left(\frac{m^2}{ \mu^2}\right)\right)~,
\end{align}
from where we can separate the divergent and finite parts of it in order to define $\displaystyle{A(m^2)=\frac{A_d(m^2)}{\hat{\ep}}+A_f(m^2)}$: 
\begin{align}
\label{240412.3}
    A_d(m^2)= -\frac{im^2}{(4\pi)^2}\hspace{1cm}\& \hspace{1cm} A_f(m^2)= -\frac{im^2}{(4\pi)^2}\left(\ln\left(\frac{m^2}{\mu^2}\right)-1\right)~.
\end{align}
A similar procedure is used to describe the one loop sunset diagram,
\begin{align}
B(m_1^2,m_2^2,p^2)=\mu^{-2\epsilon}\int\frac{d^dq}{(2\pi)^{d/2}}\frac{1}{q^2-m_1^2+i\ve}\frac{1}{(q-p)^2-m_2^2+i\ve}=\frac{B_d}{\hat{\epsilon}}+B_f(m_1^2,m_2^2,p^2)~,
\end{align}
where $B_f$ is finite. It then results that
\begin{align}
    \label{240412.6}
B_d&=-\frac{i}{(4\pi)^2}=\frac{A_d(m^2)}{m^2}~,\\
B_f(\lambda^2,m^2,p^2)&=\frac{i}{(4\pi)^2}\left(2-\ln\frac{\lambda m}{\mu^2}+\frac{m^2-\lambda^2}{2 p^2}\ln\frac{\lambda^2}{m^2}-\frac{h}{2p^2}\left[\ln\frac{h+p^2+m^2-\lambda^2}{h-p^2-m^2+\lambda^2}+\ln\frac{h+p^2-m^2+\lambda^2}{h-p^2+m^2-\lambda^2}\right]\right)~,\nn\\
h&=\sqrt{(p^2)^2+\lambda^4+m^4-2m^2 p^2-2m^2\lambda^2-2p^2\lambda^2}~,\nn
\end{align}
with $h$ the square root of the K\"allen function. 
For the case of equal masses, $B_f$ simplifies as 
\begin{align}
B_f(m^2,m^2,p^2)=\frac{i}{16\pi^2}\left(2-\ln\left(\frac{m^2}{\mu^2}\right)+ \sigma(p^2) \ln\left(\frac{\sigma(p^2)-1}{\sigma(p^2)+1}\right)\right)~.
\end{align}
Other particular results derived from Eq.~\eqref{240412.6} are
\begin{align}
\label{240412.7}
B_f(0,m^2,m^2)&=\frac{i}{(4\pi)^2}\left( 2-\ln \frac{m^2}{\mu^2}\right)~,\\
\left.\frac{\partial B_f(\lambda^2,m^2,p^2)}{\partial p^2}\right|_{p^2=m^2}&=-\frac{i}{(4\pi m)^2}\left(1-\ln\frac{m}{\lambda}\right)+{\cal O}(\lambda)~.\nn
\end{align}

It is generally possible to give analytical expressions for bubble integrals, 
 while it is not as straightforward for triangle integrals 
 in arbitrary configuration. In the case of one loop QED we need to compute only,
\begin{align}
\mathcal{C}(\lambda^2,m^2,t)=\mu^{-2\epsilon}\int\frac{d^dk}{(2\pi)^{d/2}}\frac{1}{k^2-\lambda^2}\frac{1}{(k-p_1)^2-m^2}\frac{1}{(k-p_2)^2-m^2}\Big|_{p_1^2=p_2^2=m^2}.
\end{align}
This integral is UV finite but IR divergent when $\lambda$ goes to zero. 
In the kinematics when $t<0$, the result can be expanded in powers of $\lambda$ as
\begin{align}
\mathcal{C}(\lambda^2,m^2,t)&=\frac{i}{(4 \pi)^2}\frac{\ln\lambda^2/\mu^2}{t\sigma(t)}\ln\left(\frac{\sigma(t)-1}{\sigma(t)+1}\right)-\frac{i}{32 \pi^2  t \sigma(t)}\Big[\ln
   \left(-\frac{t \sigma(t)^2}{m^2}\right)
   \ln \left(\frac{\sigma(t)-1
   }{\sigma(t) +1}\right)+2 \text{Li}_2\left(\frac{\sigma(t)
   +1}{2 \sigma(t) }\right)\\
   &-2 \text{Li}_2\left(\frac{\sigma(t) -1}{2 \sigma(t)
   }\right) \Big]+\mathcal{O}(\lambda)~.\nn
\end{align}
This expression is explicit for $t<0$, but it is still valid for $0<t<4m^2$ using the analytical continuation of the functions.

\subsubsection{Master integrals in Euclidean three dimensions}
\label{Ap.Master3dim}
Master integrals also appear in three dimensions in the Euclidean space, in particular when computing the iterated diagram. For that case, apart from $I$ given in Eq.~\eqref{240413.2}, the expressions needed are the 
bubble integrals 
\begin{align}
\label{240413.4}
{\cal A}_3(\lambda^2,\vp^2)&=\int\frac{d^3 q}{(2\pi)^3}\frac{1}{(\vq^2+\lambda^2+i\ve)((\vq+\vp)^2+\lambda^2+i\ve)}=\frac{1}{4\pi|\vp|}\arctan\frac{|\vp|}{2\lambda}~,\\
{\cal B}_3(\lambda^2,\vp^2)&=\int\frac{d^3q}{(2\pi)^3}\frac{1}{((\vq-\vp)^2+\lambda^2+i\ve)(\vp^2-\vp^2+i\ve)}=\frac{-i}{8\pi |\vp|}\ln\left(1-2i\frac{|\vp|}{\lambda}\right)~.\nn
\end{align}
By passing, let us notice that  all of them are UV finite, as it must be by  simple counting of  powers of $\vq$ in the numerator and denominator inside the integral symbol.


 \section{No infrared-divergent phase for $n=m$ in Ref.~\cite{Weinberg:1965nx}}
\label{sec.240413.2} \def\theequation{\Alph{section}.\arabic{equation}}
\setcounter{equation}{0}

We briefly discuss that the integral $J_{nn}$ in Ref.~\cite{Weinberg:1965nx} is purely real, so that only $n\neq m$ should be considered for the infrared-divergent phase, as indicated in Eqs.~\eqref{240317.2}. 
The integral $J_{nn}$ introduced in   Ref.~\cite{Weinberg:1965nx} is given by\footnote{We have adapted its expression to our convention for the Minkowski metric, with $g_{00}=1$ and $g_{ij}=-\delta_{ij}$.}
\begin{align}
\label{240413.9}
J_{nn}&=i\int^\Lambda \frac{d^4 q}{(q^2-\lambda^2+i\ve)(p_n\cdot q+i\eta_n\ve)^2}~.
\end{align}
For definiteness, let us take  the particle $n$  as outgoing, so that $\eta_n=1$ (the procedure is analogous for $\eta_n=-1$ corresponding to an incoming particle). Then, the poles in the complex $q^0$-plane are locate at 
\begin{align}
\label{240413.10}
q^0&=\pm w(q)\mp i\ve~,\\
q^0&=\boldsymbol{u}_n\cdot \boldsymbol{v}+i\ve~,\nn
\end{align}
with 
\begin{align}
\label{240413.11}
w(q)&=\sqrt{\vq^2+\lambda^2}~,~\boldsymbol{v}=\frac{\boldsymbol{q}}{w(q)}~,~
\boldsymbol{u}_n=\frac{\boldsymbol{p}_n}{E_n}~.
\end{align}
Closing the integration contour in the complex $q^0$-plane by a semicircle of infinite radius 
 running along the lower half plane, we can write $J_{nn}$ as
\begin{align}
\label{240413.12}
J_{nn}&=\frac{\pi}{E_n^2}\int^\Lambda
\frac{d^3 q}{w(q)^3(1-\boldsymbol{u}_n\cdot \boldsymbol{v})^2}~.
\end{align}
Since $|\boldsymbol{u}_n|<1$ and $|\boldsymbol{v}|<1$ the previous integral exists and is purely real.

 \section{Partial-wave amplitudes in the $\ell S J$ basis}

\def\theequation{\Alph{section}.\arabic{equation}}
\setcounter{equation}{0}
\label{App:PWA}

The series expansion of $|\vp,\sigma s\ra$ in the $\ell s J$-basis follows by inverting Eq.~\eqref{240320.4}, and it  yields
\begin{align}
\label{240320.5}
|\vp, \sigma s\rangle & =\sqrt{4 \pi} \sum_{l, m} Y_{\ell}^{m}(\hat{\vp})^{*}\left|\ell m, \sigma s\right\rangle =\sqrt{4\pi}\sum_{l, m,J,\mu}Y_{\ell}^{m}(\hat{\vp})^{*} C(m \sigma \mu \mid \ell s J)|J \mu, \ell s\rangle~.
\end{align}

Following chapter~2 of Ref.~\cite{Oller:2019rej}, the  expression for the calculation of a PWA in our case reads

\begin{align}
\label{Eq:pwa1}
\left\langle J^{\prime} \mu^{\prime}, \ell s|T| J \mu, \ell s\right\rangle&=\frac{\delta_{\mu^{\prime} \mu} \delta_{J^{\prime}J}}{2 J+1} \sqrt{\frac{2 \ell+1}{4 \pi}}\sum_{\sigma,\sigma^{\prime}=-s}^s\sum_{m=-\ell}^\ell \int d \hat{\vp}\, C\left(m \sigma^{\prime} \sigma \mid \ell s J\right)  C\left(0 \sigma \sigma \mid \ell s J\right) Y_{\ell}^{m}\left(\hat{\vp}\right)^{\ast} \left\langle\vp, \sigma^{\prime} s|T| p \hat{\vz}, \sigma s\right\rangle~.
\end{align}

Nonetheless, this expression can be further simplified using a rotation around the $z$ axis that takes $\vp$ in the integrand to the $\widehat{xz}-$plane, $R_{z}(-\phi) \vp=\vp_{x z}$, that acts in the kets as
\begin{align}
R_{z}(\phi)^{\dagger}\left|\vp, \sigma^{\prime} s\right\rangle
= \sum_{\sigma^{\prime \prime}} D_{\sigma^{\prime \prime}\sigma^{\prime}}^{(s)}(-\phi, 0,0)\left|\vp_{x z}, {\sigma}^{\prime \prime} s\right\rangle~,
\end{align}

where $D^{(s)}_{\sigma^{\prime \prime}\sigma^{\prime}}(\alpha,\beta,\gamma)$ are the matrix elements of the representation of the rotation group with spin $s$  over the ket space, and $\alpha$, $\beta$ and $\gamma$ are the Euler angles. Thus, 
\begin{align}
D_{{\sigma}^{\prime \prime} \sigma^{\prime}}^{(s)}(-\phi, 0,0)=\left\langle{\sigma}^{\prime \prime} s\left|e^{i J_{z} \phi}\right| \sigma^{\prime} s\right\rangle=\delta_{{\sigma}^{\prime \prime} \sigma^\prime }e^{i \sigma^{\prime} \phi}
\end{align}

Applying rotational invariance, we get that

\begin{align}
\label{240321.1}
& \left\langle \vp, \sigma^{\prime} s|T| p \hat{\vz}, \sigma s\right\rangle=\left\langle\vp, \sigma^{\prime} s\left|R_z(\phi) T R_{z}(\phi)^{\dagger}\right| p \hat{\vz}, \sigma s\right\rangle 
 =\left\langle \vp_{xz}, \sigma^{\prime} s|T| p\hat{\vz},  \sigma s\right\rangle e^{i \phi\left(\sigma-\sigma^{\prime}\right)}~.
\end{align}

 Next, we make use of the standard properties of transformation of a spherical harmonics under a rotation \cite{rose.170517.1,Oller:2019rej}:
\begin{align}
\label{240321.2}
Y_\ell^m(\hat{\vp})=Y_\ell^m(R_z(\phi)\hat{\vp}_{xz})=e^{i m\phi}Y_\ell^m(\hat{\vp}_{xz})~.
    \end{align}
Taking Eqs.~\eqref{240321.1} and \eqref{240321.2} into  Eq.~\eqref{Eq:pwa1} the $\phi$ dependence in the integrand drops out because of $m+\sigma'=\sigma$, and the  $\phi$ integral becomes trivial. With $\mu'=\mu$ and $J'=J$ we can then rewrite Eq.~\eqref{Eq:pwa1} as 

\begin{align}
\label{240321.3}
\left\la J\mu,\ell s\left|T\right|J\mu,\ell s\right\ra&=\frac{\sqrt{\pi(2\ell+1)}}{2J+1}\sum_{\sigma',\sigma=-s}^s\sum_{m=-\ell}^\ell \int_{-1}^{1}d\cos\theta\,Y_\ell^m(0,\theta)C(m\sigma'\sigma|\ell sJ)C(0\sigma\sigma|\ell sJ)\left\la \vp_{xz},\sigma' s\left|T\right|p\hat{\vz},\sigma s\right\ra~.
\end{align}
Let us note that $Y_\ell^m(0,\theta)$ is real. 
To end with Eq.~\eqref{231216.1} a further simplification is implemented by restricting the sum over $m$ to values $m\geq 0$. Since $s=1/2$ the index $m=\pm 1$ or 0. For the latter, $\sigma'=-\sigma$, and the same result is obtained for $\sigma=\pm 1/2$ whatever sign is taken. Similarly, for $m=-1$ we have the contribution with $\sigma'=\sigma=-1/2$ and the same result is obtained as with $m=1$ and $\sigma=\sigma'=1/2$. It therefore follows Eq.~\eqref{231216.1}. 

In this last process we have made use that 
\begin{align}
\label{240321.5}
Y_\ell^{-m}(0,\theta)&=(-1)^m Y_\ell^m(0,\theta)~,\\
C(-m-\sigma'-\sigma|\ell sJ)&=(-1)^{\ell+s-J}C(m\sigma'\sigma|\ell sJ)~.\nn
\end{align}

To show the equality of the results for opposite signs in the spin projections, let us notice that symmetry under time reversal, driven by the antiunitary operator $\theta_t$, allows to flip the sign of the third components of spin. This symmetry implies that
\begin{align}
\label{240321.6}
\la \vp,\sigma's|T|p\hat{\vz},\sigma s\ra&=(\theta_t| \vp,\sigma's\ra)^\dagger \theta_t T|p\hat{\vz},\sigma s\ra^*~.
\end{align}
The transformation under $\theta_t$ of the vector kets is
\begin{align}
\label{240321.7}
\theta_t|\vp,\sigma' s\ra&=\xi_s i^{2\sigma'}|-\vp,-\sigma' s\ra~,
\end{align}
where $\xi_s$ is a phase factor independent of $\sigma$, and a similar transformation rule applies to  $\theta_t|p\hat{\vz},\sigma s\ra$. For the $T$-matrix operator we have that $\theta_t T \theta_t^{-1}=T^\dagger$. Additionally, we also implement invariance under parity to reverse the signs of the momentum of every state after the $\theta_t$ transformation:
\begin{align}
\label{240321.8}
P|\vp,\sigma s\ra&=|-\vp,\sigma s\ra~,
\end{align}
since the intrinsic parity of the electron is $+1$. 

Using  Eqs.~\eqref{240321.5}-\eqref{240321.8} into Eq.~\eqref{240321.3}  for the matrix elements involving negative values of $\sigma$, we end with 
\begin{align}
\label{240321.9}
\la J\mu,\ell s|T|J\mu,\ell s\ra&=\frac{\sqrt{\pi(2\ell+1)}}{2J+1}\int_{-1}^{+1} d\cos\theta\left[
Y_\ell^0(0,\theta)C(0\frac{1}{2}\frac{1}{2}|\ell sJ)^2
\left(\la \vp_{xz},\frac{1}{2}s|T|p\hat{\vz},\frac{1}{2}s\ra+\la p\hat{\vz},\frac{1}{2}s|T|\hat{\vp}_{xz},\frac{1}{2}s\ra\right)\right.\\
&\left.+
Y_\ell^1(0,\theta)C(1-\frac{1}{2}\frac{1}{2}|\ell sJ)C(0\frac{1}{2}\frac{1}{2}|\ell sJ)
\left(\la \vp_{xz},-\frac{1}{2}s|T|p\hat{\vz},\frac{1}{2}s\ra+\la p\hat{\vz},-\frac{1}{2}s|T|\hat{\vp}_{xz},\frac{1}{2}s\ra\right)\right]~.\nn
\end{align}
Now, for the momenta $\vp_{xz}$ and $p\hat{\vz}$ the associated Dirac spinors are real. Similarly, the Dirac-matrix structures in the matrix elements for the amplitudes $A_0$, $A_1$, $A_2$,  and $A_3$ also give rise to real matrices, including $i(p'-p)^\nu \sigma_{\mu\nu}$ for $A_2$. If we designate by $\Gamma$ any of those Dirac-matrix structures, namely, $\Gamma(\vp_{xz},p\hat{\vz})=\left\{I,\gamma^0,i(p'-p)^\nu \sigma_{\mu\nu}\right\}$, one can check that it is verified that
\begin{align}
\bar{u}_{\sigma'}(\vp_{xz})\Gamma(\vp_{xz},p\hat{\vz}) u_\sigma(p\hat{\vz})=\left({u}_{\sigma'}(\vp_{xz})^T\gamma^0\Gamma(\vp_{xz},p\hat{\vz}) u_\sigma(p\hat{\vz})\right)^T={u}_{\sigma}(p\hat{\vz})^T \gamma^0 \Gamma(p\hat{\vz},\vp_{xz})  u_{\sigma'}(\vp_{xz})~,
\end{align}
since in the Dirac representation for the $\gamma$ matrices, $\gamma^0$ and $\Gamma$ are symmetric and, furthermore, $\Gamma(\vp_{xz},p\hat{\vz})\gamma^0=\gamma^0\Gamma(p\hat{\vz},\vp_{xz})$. From here it follows that the matrix elements between brackets in Eq.~\eqref{240321.9} give the same result, and a global factor of 2 results.  Finally, to end with Eq.~\eqref{231216.1},  $Y_\ell^m(0,\theta)$ is expressed  in terms of the associated Legendre polynomials $P_\ell^m(\theta)$, by making use of the relation
\begin{align}
\label{240321.4}
Y_\ell^m(0,\theta)&=\sqrt{\frac{2\ell+1}{4\pi}\frac{(\ell-m)!}{(\ell+m)!}}P_\ell^m(\theta)~.
    \end{align}

    To end with this appendix we collect the basics Dirac algebra expressions used along the manuscript.

\bigskip
{\bf Dirac spinor:}
\bigskip

\begin{align}
  \label{231216.2}
u_\sigma(\vp)^T&=\sqrt{m+E(p)}\left(\chi_\sigma,\frac{\vp\cdot \boldsymbol{\sigma}}{m+E(p)}\chi_\sigma\right)~,
  \end{align}
with $\chi_\sigma$ a standard Pauli spinor.

  \bigskip
{\bf Matrix elements:}
\bigskip

As said, we take the Dirac representations for the $\gamma$ matrices, and $\sigma^{\mu\nu}=\frac{i}{2}[\gamma^\mu,\gamma^\nu]$. 

  \begin{align}
  \label{240629.1}
\bar{u}_\sigma(\vp_{xz})\gamma^0u_\sigma(\vp_z)&=E+m+(E-m)\cos\theta~,\\
\bar{u}_{-\frac{1}{2}}(\vp_{xz})\gamma^0u_\frac{1}{2}(\vp_z)&=(E-m)\sin\theta~,\nn\\
\bar{u}_\sigma(\vp_{xz})u_\sigma(\vp_z)&=E+m+(m-E)\cos\theta~,\nn\\
\bar{u}_{-\frac{1}{2}}(\vp'_{xz})u_\frac{1}{2}(\vp_z)&=(m-E)\sin\theta~,\nn\\
i\bar{u}_\frac{1}{2}(\vp_{xz})(p'_{xz}-p_z)^\nu\sigma_{0\nu}u_\frac{1}{2}(\vp_z)&=4p^2\sin^2\frac{\theta}{2}\cos\frac{\theta}{2}\left(\cos\frac{\theta}{2}+\sin\frac{\theta}{2}\right)~,\nn\\
i\bar{u}_{-\frac{1}{2}}(\vp'_{xz})(p'_{xz}-p_z)^\nu\sigma_{0\nu}u_\frac{1}{2}(\vp_z)&=4p^2\sin^3\frac{\theta}{2}\left(\cos\frac{\theta}{2}+\sin\frac{\theta}{2}\right)~.\nn
\end{align}


\bibliographystyle{unsrt}
\bibliography{references}

\begin{thebibliography}{10}

\bibitem{Weinberg:1965nx}
Steven Weinberg.
\newblock {Infrared photons and gravitons}.
\newblock {\em Phys. Rev.}, 140:B516--B524, 1965.

\bibitem{Mott1929TheSO}
Nevill~Francis Mott.
\newblock The scattering of fast electrons by atomic nuclei.
\newblock {\em Proceedings of The Royal Society A: Mathematical, Physical and
  Engineering Sciences}, 124:425--442, 1929.

\bibitem{Dalitz:1951ah}
R.~H. Dalitz.
\newblock {On higher Born approximations in potential scattering}.
\newblock {\em Proc. Roy. Soc. Lond. A}, 206:509--520, 1951.

\bibitem{McKinley:1948zz}
William~A. McKinley and Herman Feshbach.
\newblock {The Coulomb Scattering of Relativistic Electrons by Nuclei}.
\newblock {\em Phys. Rev.}, 74:1759--1763, 1948.

\bibitem{Yennie:1954zz}
D.~R. Yennie, D.~G. Ravenhall, and R.~N. Wilson.
\newblock {Phase-Shift Calculation of High-Energy Electron Scattering}.
\newblock {\em Phys. Rev.}, 95:500--512, 1954.

\bibitem{Baier:1972jm}
V.~N. Baier and V.~M. Katkov.
\newblock {On bremsstrahlung in the collision of high-energy particles in a
  magnetic field}.
\newblock {\em Dokl. Akad. Nauk Ser. Fiz.}, 207:68--70, 1972.

\bibitem{Sikora:2018zda}
B.~Sikora, V.~A. Yerokhin, N.~S. Oreshkina, H.~Cakir, C.~H. Keitel, and
  Z.~Harman.
\newblock {Theory of the two-loop self-energy correction to the $g$ factor in
  nonperturbative Coulomb fields}.
\newblock {\em Phys. Rev. Res.}, 2(1):012002, 2020.

\bibitem{Delto:2023kqv}
Maximilian Delto, Claude Duhr, Lorenzo Tancredi, and Yu~Jiao Zhu.
\newblock {Two-loop QED corrections to the scattering of four massive leptons}.
\newblock arXiv:2311.06385 [hep-ph].

\bibitem{Banerjee:2021qvi}
Pulak Banerjee, Tim Engel, Nicolas Schalch, Adrian Signer, and Yannick Ulrich.
\newblock {M\o{}ller scattering at NNLO}.
\newblock {\em Phys. Rev. D}, 105(3):L031904, 2022.

\bibitem{Bonciani:2005im}
R.~Bonciani and A.~Ferroglia.
\newblock {Two-loop Bhabha scattering in QED}.
\newblock {\em Phys. Rev. D}, 72:056004, 2005.

\bibitem{Fleischer:2006ht}
J.~Fleischer, J.~Gluza, A.~Lorca, and T.~Riemann.
\newblock {One-loop photonic corrections to Bhabha scattering in
  d=4-2\ensuremath{\varepsilon} dimensions}.
\newblock {\em Eur. Phys. J. C}, 48:35--52, 2006.

\bibitem{Engel:2022kde}
Tim Engel.
\newblock {\em {Muon-Electron Scattering at NNLO}}.
\newblock PhD thesis, Zurich U., 2022.

\bibitem{Khalil:2017cqm}
Abdullah Khalil and W.~A. Horowitz.
\newblock {Next-to-leading order corrections to the elastic scattering of an
  electron off of a static scattering center}.
\newblock {\em J. Phys. Conf. Ser.}, 802(1):012004, 2017.

\bibitem{Khalil:2017ons}
Abdullah Khalil.
\newblock {NLO Rutherford Scattering and the Kinoshita-Lee-Nauenberg Theorem}.
\newblock Master's thesis, Cape Town U., 2017.

\bibitem{Oller:2020guq}
Jos\'e~Antonio Oller.
\newblock {Unitarization Technics in Hadron Physics with Historical Remarks}.
\newblock {\em Symmetry}, 12(7):1114, 2020.

\bibitem{Blas:2020dyg}
Diego Blas, Jorge Martin~Camalich, and Jos\'e~Antonio Oller.
\newblock {Unitarization of infinite-range forces: graviton-graviton
  scattering}.
\newblock {\em JHEP}, 08:266, 2022.

\bibitem{Blas:2020och}
D.~Blas, J.~Martin~Camalich, and J.~A. Oller.
\newblock {Scalar resonance in graviton-graviton scattering at high-energies:
  The graviball}.
\newblock {\em Phys. Lett. B}, 827:136991, 2022.

\bibitem{Oller:2022tmo}
Jos\'e~Antonio Oller.
\newblock {Unitarizing non-relativistic Coulomb scattering}.
\newblock {\em Phys. Lett. B}, 835:137568, 2022.

\bibitem{landau.170517.1}
R.H. Landau.
\newblock {\em {Quantum mechanics. Vol. 2: A second course in quantum theory}}.
\newblock WILEY-VCH, Weinheim, 1995.

\bibitem{Delgado:2022qnh}
Rafael~L. Delgado, Antonio Dobado, and Dom\`enec Espriu.
\newblock {Seeking for resonances in unitarized one-loop graviton-graviton
  scattering}.
\newblock {\em Phys. Rev. D}, 107(4):044073, 2023.

\bibitem{Bloch:1937pw}
F.~Bloch and A.~Nordsieck.
\newblock {Note on the Radiation Field of the electron}.
\newblock {\em Phys. Rev.}, 52:54--59, 1937.

\bibitem{Kulish:1970ut}
P.~P. Kulish and L.~D. Faddeev.
\newblock {Asymptotic conditions and infrared divergences in quantum
  electrodynamics}.
\newblock {\em Theor. Math. Phys.}, 4:745, 1970.

\bibitem{Ware:2013zja}
John Ware, Ryo Saotome, and Ratindranath Akhoury.
\newblock {Construction of an asymptotic S matrix for perturbative quantum
  gravity}.
\newblock {\em JHEP}, 10:159, 2013.

\bibitem{Choi:2017bna}
Sangmin Choi, Uri Kol, and Ratindranath Akhoury.
\newblock {Asymptotic Dynamics in Perturbative Quantum Gravity and BMS
  Supertranslations}.
\newblock {\em JHEP}, 01:142, 2018.

\bibitem{Choi:2017ylo}
Sangmin Choi and Ratindranath Akhoury.
\newblock {BMS Supertranslation Symmetry Implies Faddeev-Kulish Amplitudes}.
\newblock {\em JHEP}, 02:171, 2018.

\bibitem{Prabhu:2022zcr}
Kartik Prabhu, Gautam Satishchandran, and Robert~M. Wald.
\newblock {Infrared finite scattering theory in quantum field theory and
  quantum gravity}.
\newblock {\em Phys. Rev. D}, 106(6):066005, 2022.

\bibitem{Prabhu:2024zwl}
Kartik Prabhu and Gautam Satishchandran.
\newblock {Infrared finite scattering theory: Scattering states and
  representations of the BMS group}.
\newblock arXiv:2402.00102 [hep-th].

\bibitem{Prabhu:2024lmg}
Kartik Prabhu and Gautam Satishchandran.
\newblock {Infrared finite scattering theory: Amplitudes and soft theorems}.
\newblock arXiv:2402.18637 [hep-th].

\bibitem{Campiglia:2016hvg}
Miguel Campiglia and Alok Laddha.
\newblock {Subleading soft photons and large gauge transformations}.
\newblock {\em JHEP}, 11:012, 2016.

\bibitem{Campiglia:2014yka}
Miguel Campiglia and Alok Laddha.
\newblock {Asymptotic symmetries and subleading soft graviton theorem}.
\newblock {\em Phys. Rev. D}, 90(12):124028, 2014.

\bibitem{Ashtekar:2018lor}
Abhay Ashtekar, Miguel Campiglia, and Alok Laddha.
\newblock {Null infinity, the BMS group and infrared issues}.
\newblock {\em Gen. Rel. Grav.}, 50(11):140--163, 2018.

\bibitem{Hirai:2020kzx}
Hayato Hirai and Sotaro Sugishita.
\newblock {IR finite S-matrix by gauge invariant dressed states}.
\newblock {\em JHEP}, 02:025, 2021.

\bibitem{Himwich:2020rro}
Elizabeth Himwich, Sruthi~A. Narayanan, Monica Pate, Nisarga Paul, and Andrew
  Strominger.
\newblock {The Soft $\mathcal{S}$-Matrix in Gravity}.
\newblock {\em JHEP}, 09:129, 2020.

\bibitem{Hannesdottir:2019umk}
Holmfridur Hannesdottir and Matthew~D. Schwartz.
\newblock {Finite $S$ matrix}.
\newblock {\em Phys. Rev. D}, 107(2):L021701, 2023.

\bibitem{Chung:1965zza}
Victor Chung.
\newblock {Infrared Divergence in Quantum Electrodynamics}.
\newblock {\em Phys. Rev.}, 140:B1110--B1122, 1965.

\bibitem{Oller:1998zr}
J.~A. Oller and E.~Oset.
\newblock {N/D description of two meson amplitudes and chiral symmetry}.
\newblock {\em Phys. Rev. D}, 60:074023, 1999.

\bibitem{Oller:1999me}
J.~A. Oller.
\newblock {The Case of a $W W$ dynamical scalar resonance within a chiral
  effective description of the strongly interacting Higgs sector}.
\newblock {\em Phys. Lett. B}, 477:187--194, 2000.

\bibitem{Lehmann:1972kv}
H.~Lehmann.
\newblock {Chiral invariance and effective range expansion for pion pion
  scattering}.
\newblock {\em Phys. Lett. B}, 41:529--532, 1972.

\bibitem{Dobado:1989qm}
A.~Dobado, Maria~J. Herrero, and Tran~N. Truong.
\newblock {Unitarized Chiral Perturbation Theory for Elastic Pion-Pion
  Scattering}.
\newblock {\em Phys. Lett. B}, 235:134--140, 1990.

\bibitem{Oller:1997ng}
Jos\'e~Antonio Oller, E.~Oset, and J.~R. Pelaez.
\newblock {Nonperturbative approach to effective chiral Lagrangians and meson
  interactions}.
\newblock {\em Phys. Rev. Lett.}, 80:3452--3455, 1998.

\bibitem{Oller:2019opk}
Jos\'e~Antonio Oller.
\newblock {Coupled-channel approach in hadron\textendash{}hadron scattering}.
\newblock {\em Prog. Part. Nucl. Phys.}, 110:103728, 2020.

\bibitem{Oller:2019rej}
Jos\'e~Antonio Oller.
\newblock {\em {A Brief Introduction to Dispersion Relations}}.
\newblock SpringerBriefs in Physics. Springer, 2019.

\bibitem{GomezNicola:2007qj}
A.~Gomez~Nicola, J.~R. Pelaez, and G.~Rios.
\newblock {The Inverse Amplitude Method and Adler Zeros}.
\newblock {\em Phys. Rev. D}, 77:056006, 2008.

\bibitem{Salas-Bernardez:2020hua}
Alexandre Salas-Bern\'ardez, Felipe~J. Llanes-Estrada, Juan Escudero-Pedrosa,
  and Jose~Antonio Oller.
\newblock {Systematizing and addressing theory uncertainties of unitarization
  with the Inverse Amplitude Method}.
\newblock {\em SciPost Phys.}, 11(2):020, 2021.

\bibitem{Escudero-Pedrosa:2020rwb}
Juan Escudero-Pedrosa, Felipe~J. Llanes-Estrada, Jos\'e~Antonio Oller, and
  Alexandre Salas-Bern\'ardez.
\newblock {Assessment of systematic theory uncertainties in IAM unitarization}.
\newblock {\em Nucl. Part. Phys. Proc.}, 312-317:82--86, 2021.

\bibitem{Pineda:1998kn}
A.~Pineda and J.~Soto.
\newblock {Potential NRQED: The Positronium case}.
\newblock {\em Phys. Rev. D}, 59:016005, 1999.

\bibitem{Haidar:2019kcp}
M.~Haidar, Z-X Zhong, V.~I. Korobov, and J-Ph. Karr.
\newblock {Nonrelativistic QED approach to the fine- and hyperfine-structure
  corrections of order $m\alpha^6$ and $m\alpha^6(m/M)$ : Application to the
  hydrogen atom}.
\newblock {\em Phys. Rev. A}, 101(2):022501, 2020.

\bibitem{Peskin:1995ev}
Michael~E. Peskin and Daniel~V. Schroeder.
\newblock {\em {An Introduction to quantum field theory}}.
\newblock Addison-Wesley, Reading, USA, 1995.

\bibitem{Passarino:1978jh}
G.~Passarino and M.~J.~G. Veltman.
\newblock {One Loop Corrections for e+ e- Annihilation Into mu+ mu- in the
  Weinberg Model}.
\newblock {\em Nucl. Phys. B}, 160:151--207, 1979.

\bibitem{Weinberg:1995mt}
Steven Weinberg.
\newblock {\em {The Quantum theory of fields. Vol. 1: Foundations}}.
\newblock Cambridge University Press, 6 2005.

\bibitem{Yennie:1961ad}
D.~R. Yennie, Steven~C. Frautschi, and H.~Suura.
\newblock {The infrared divergence phenomena and high-energy processes}.
\newblock {\em Annals Phys.}, 13:379--452, 1961.

\bibitem{Gaharia:2019xlh}
David Gaharia.
\newblock {Asymptotic Symmetries and Faddeev-Kulish states in QED and Gravity}.
\newblock Master's thesis, Stockholm U. (main), 2019.

\bibitem{Oller:2000ma}
J.~A. Oller, E.~Oset, and A.~Ramos.
\newblock {Chiral unitary approach to meson meson and meson - baryon
  interactions and nuclear applications}.
\newblock {\em Prog. Part. Nucl. Phys.}, 45:157--242, 2000.

\bibitem{Albaladejo:2010tj}
M.~Albaladejo, J.~A. Oller, and L.~Roca.
\newblock {Dynamical generation of pseudoscalar resonances}.
\newblock {\em Phys. Rev. D}, 82:094019, 2010.

\bibitem{Zwanziger:1974jz}
Daniel Zwanziger.
\newblock {Scattering Theory for Quantum Electrodynamics. 1. Infrared
  Renormalization and Asymptotic Fields}.
\newblock {\em Phys. Rev. D}, 11:3481, 1975.

\bibitem{Giddings:2009gj}
Steven~B. Giddings and Rafael~A. Porto.
\newblock {The Gravitational S-matrix}.
\newblock {\em Phys. Rev. D}, 81:025002, 2010.

\bibitem{Lehmann:1958ita}
H.~Lehmann.
\newblock {Analytic properties of scattering amplitudes as functions of
  momentum transfer}.
\newblock {\em Nuovo Cim.}, 10(4):579--589, 1958.

\bibitem{martin.200705.1}
A.~D. Martin and T.~D. Spearman.
\newblock {\em Elementary Particle Theory}.
\newblock North-Holland Publishing Company, Amsterdam, 1970.

\bibitem{Kang:1962}
Ik-Ju Kang and L.~M. Brown.
\newblock {Higher Born Approximations for the Coulomb Scattering of a Spinless
  Particle}.
\newblock {\em Phys. Rev.}, 128:2828, 1962.

\bibitem{Weinberg:1991um}
Steven Weinberg.
\newblock {Effective chiral Lagrangians for nucleon - pion interactions and
  nuclear forces}.
\newblock {\em Nucl. Phys. B}, 363:3--18, 1991.

\bibitem{Oller:1997ti}
J.~A. Oller and E.~Oset.
\newblock {Chiral symmetry amplitudes in the S wave isoscalar and isovector
  channels and the $\sigma$, f$_0$(980), a$_0$(980) scalar mesons}.
\newblock {\em Nucl. Phys. A}, 620:438--456, 1997.
\newblock [Erratum: Nucl.Phys.A 652, 407--409 (1999)].

\bibitem{Oller:2000fj}
J.~A. Oller and Ulf~G. Meissner.
\newblock {Chiral dynamics in the presence of bound states: Kaon nucleon
  interactions revisited}.
\newblock {\em Phys. Lett. B}, 500:263--272, 2001.

\bibitem{Oller:2005ig}
Jose~A. Oller, Joaquim Prades, and Michela Verbeni.
\newblock {Surprises in threshold antikaon-nucleon physics}.
\newblock {\em Phys. Rev. Lett.}, 95:172502, 2005.

\bibitem{Dobado:1996ps}
A.~Dobado and J.~R. Pelaez.
\newblock {The Inverse amplitude method in chiral perturbation theory}.
\newblock {\em Phys. Rev. D}, 56:3057--3073, 1997.

\bibitem{Oller:1998hw}
Jos\'e~Antonio Oller, E.~Oset, and J.~R. Pelaez.
\newblock {Meson meson interaction in a nonperturbative chiral approach}.
\newblock {\em Phys. Rev. D}, 59:074001, 1999.
\newblock [Erratum: {\it ibid} 60, 099906 (1999); {\it ibid} 75, 099903
  (2007)].

\bibitem{Eden:1966dnq}
Richard~John Eden, Peter~V. Landshoff, David~I. Olive, and John~Charlton
  Polkinghorne.
\newblock {\em {The analytic S-matrix}}.
\newblock Cambridge Univ. Press, Cambridge, 1966.

\bibitem{Tricomi85}
F.~G. Tricomi.
\newblock {\em {Integral Equations}}.
\newblock Dover Publications, Inc., New York, USA, 1985.

\bibitem{Hernandez:1984zzb}
E.~Hernandez and A.~Mondragon.
\newblock {Resonant states in momentum representation}.
\newblock {\em Phys. Rev. C}, 29:722--738, 1984.

\bibitem{Bazhanov:1977fa}
V.~V. Bazhanov, G.~P. Pronko, L.~D. Solovev, and Yu.~Ya. Yushin.
\newblock {Small Angle Scattering in Quantum Electrodynamics}.
\newblock {\em Teor. Mat. Fiz.}, 33:218--230, 1977.

\bibitem{Oller:2024neq}
J.~A. Oller and Marcela Pel\'aez.
\newblock {Unitarization of the one-loop graviton-graviton scattering
  amplitudes and study of the graviball}.
\newblock arXiv:2407.16538 [hep-th].

\bibitem{Bethe:1947id}
H.~A. Bethe.
\newblock {The Electromagnetic shift of energy levels}.
\newblock {\em Phys. Rev.}, 72:339--341, 1947.

\bibitem{Mandl:1985bg}
Franz Mandl and Graham Shaw.
\newblock {\em {Quantum Field Theory}}.
\newblock John Wiley \& Sons, 1985.

\bibitem{rose.170517.1}
M.~E. Rose.
\newblock {\em Elementary Theory of Angular Momentum}.
\newblock John Wiley \& Sons, Inc. New York., 1957.

\end{thebibliography}

\end{document}